\documentclass[aps,prd,twocolumn,showpacs,superscriptaddress,nofootinbib,10pt]{revtex4-1}

\usepackage{amsfonts}
\usepackage{latexsym}
\usepackage{yfonts}
\usepackage{amsmath}
\usepackage{amssymb}
\usepackage{upgreek}
\usepackage{bm}
\usepackage{dcolumn}
\usepackage{epsfig}
\usepackage{graphicx}
\usepackage[latin1]{inputenc}
\usepackage{latexsym}
\usepackage{rotating}
\usepackage{hyperref}

\newcommand\be{\begin{equation}}
\newcommand\ba{\begin{eqnarray}}
\newcommand\ee{\end{equation}}
\newcommand\ea{\end{eqnarray}}

\newcommand{\ACMC}{{\mbox{\tiny ACMC}}}
\newcommand{\AHAR}{{\mbox{\tiny AH}}}
\newcommand{\APO}{{\mbox{\tiny apo}}}
\newcommand{\BL}{{\mbox{\tiny BL}}}
\newcommand{\GG}{{\mbox{\tiny G}}}

\newcommand{\GW}{{\mbox{\tiny GW}}}
\newcommand{\HAR}{{\mbox{\tiny H}}}
\newcommand{\HEL}{{\mbox{\tiny Hel}}}
\newcommand{\INC}{{\mbox{\tiny inc}}}
\newcommand{\KERR}{{\mbox{\tiny K}}}
\newcommand{\LSO}{{\mbox{\tiny LSO}}}
\newcommand{\MBH}{{{\bullet}}}
\newcommand{\met}{\mbox{g}}
\newcommand{\MIN}{{\mbox{\tiny min}}}
\newcommand{\ORB}{{\mbox{\tiny Orbital}}}
\newcommand{\PERI}{{\mbox{\tiny peri}}}
\newcommand{\PM}{{\mbox{\tiny PM}}}
\newcommand{\REG}{{\mbox{\tiny R}}}
\newcommand{\RR}{{\mbox{\tiny RR}}}
\newcommand{\SCO}{{\star}}
\newcommand{\SF}{{\mbox{\tiny SF}}}
\newcommand{\STF}{{\mbox{\tiny STF}}}
\newcommand{\TOT}{{\mbox{\tiny Tot}}}
\newcommand{\TT}{{\mbox{\tiny TT}}}

\begin{document}
\title{New Kludge Scheme for the Construction of Approximate Waveforms for Extreme-Mass-Ratio Inspirals}

\author{Carlos F.~Sopuerta}
\affiliation{Institut de Ci\`encies de l'Espai (CSIC-IEEC), 
Facultat de Ci\`encies, Campus UAB, Torre C5 parells, 
Bellaterra, 08193 Barcelona, Spain.}

\author{Nicol\'as Yunes}
\affiliation{Department of Physics, Montana State University, Bozeman, MT 59717, USA.}
\affiliation{Department of Physics and MIT Kavli Institute, 77 Massachusetts Avenue, Cambridge, MA 02139, USA.}
\affiliation{Princeton University, Physics Department, Princeton, NJ 08544, USA.}

\date{\today}

\begin{abstract} 
We introduce a new kludge scheme to model the dynamics of generic extreme mass-ratio inspirals 
(stellar compact objects spiraling into a spinning supermassive black hole) and to produce the gravitational
waveforms that describe the gravitational-wave emission of these systems.  
This scheme combines tools from different techniques in General Relativity: It uses a a multipolar, post-Minkowskian expansion 
for the far-zone metric perturbation (which provides the gravitational waveforms, here taken up to mass hexadecapole and current octopole order)
and for the local prescription of the self-force (since we are lacking a general prescription for it);
a post-Newtonian expansion for the computation of the multipole moments in terms of the
trajectories; and a BH perturbation theory expansion when treating the 
trajectories as a sequence of self-adjusting Kerr geodesics.
The orbital evolution is thus equivalent to solving the geodesic equations with {\emph{time-dependent}} 
orbital elements, as dictated by the multipolar post-Minkowskian radiation-reaction prescription.
To complete the scheme, both the orbital evolution and wave generation require to map the Boyer-Lindquist 
coordinates of the orbits to the harmonic coordinates in which the different multipolar post-Minkowskian quantities have been derived, 
a mapping that we provide explicitly in this paper.   
This new kludge scheme is thus a combination of approximations that can be used to model generic inspirals of
systems with extreme mass ratios to systems with more moderate mass ratios, and hence can provide valuable
information for future space-based gravitational-wave observatories like the 
Laser Interferometer Space Antenna and even for advanced ground detectors.  
Finally, due to the local character in time of our multipolar post-Minkowskian self-force, this scheme can be
used to perform studies of the possible appearance of transient resonances in generic inspirals.
\end{abstract}


\maketitle

\section{Introduction}
\label{intro}

Gravitational waves (GWs) hold the promise to provide detailed information about astrophysical bodies that are 
obscure in the electromagnetic spectrum, such as binary black hole (BH) systems. Moreover, such waves will allow 
for the first studies of the nature of the gravitational interaction and of the validity of General Relativity (GR) 
in the strongest regimes~\cite{lrr-2006-3,Yunes:2009ke}. An accurate modeling of such GWs is essential for 
the extraction and characterization of weak signals buried in detector noise. This is because waveform templates 
act as an optimal linear filter that maximizes the signal-to-noise ratio (SNR) in the presence of stochastic noise. 
The absence of such templates for certain GW sources renders suboptimal any GW search strategy. Therefore, the 
construction and modeling of GWs to construct accurate templates for data analysis is of paramount importance in 
the blossoming of GW astrophysics.   

One of the staple GW sources of the planned space-based observatory {\em Laser Interferometer Space Antenna}
(LISA)~\cite{Danzmann:2003tv,*Danzmann:2003ad,*Prince:2003aa,lisa} that require accurate templates for their
detection and analysis are extreme-mass-ratio inspirals (EMRIs)~\cite{AmaroSeoane:2007aw}. These events 
consist of a small compact object (SCO), such as a stellar-mass BH or neutron 
star (NS), spiraling in a generic orbit into a spinning, (super)massive black hole (MBH), whose evolution is 
GW dominated.  In such inspirals, the SCO spends up to millions of cycles in close orbits around the MBH, possibly 
with large pericenter velocities and eccentricities, sampling the strong gravitational field of the MBH. 

Many astrophysical scenarios predict the existence of such EMRIs. One such scenario postulates that the 
SCO exchanges energy and angular momentum with other stars in a stellar core/cusp near 
a MBH at a galactic center, via two-body relaxation and dynamical 
friction~\cite{AmaroSeoane:2007aw}. If so, the SCO can be swung sufficiently close to the MBH to be gravitationally 
captured~\cite{1995ApJ...445L...7H,*1997MNRAS.284..318S,AmaroSeoane:2007aw}, at which point it would slowly
inspiral until being swallowed by the MBH. Of course, such a scenario is 
complicated by mass segregation~\cite{2006ApJ...649...91F,*2006ApJ...645L.133H,*2006RPPh...69.2513M}, triaxial 
density profiles~\cite{2002ApJ...567..817H,*2005ApJ...629..362H},  resonant 
relaxation~\cite{1996NewA....1..149R,*2006ApJ...645.1152H}, etc. Other channels of EMRI formation include binary 
tidal separation~\cite{2005ApJ...631L.117M} (where a binary is disrupted with one component captured by the 
MBH) and massive star capture or production in accretion discs~\cite{2003astro.ph..7084L,*2007MNRAS.374..515L}, 
where the stellar-mass BH is directly formed in the accretion disk of the MBH.  It turns out that the EMRI 
GWs produced by EMRIs in each astrophysical mechanism have distinct characteristic that may be use to
distinguish them from EMRI observations~\cite{Yunes:2011ws,*2011PhRvD..84b4032K}.

The inspiral is by far the dominant GW phase for EMRI data analysis purposes. This can be understood rather 
easily (see, e.g.~\cite{Finn:2000sy,Yunes:2010zj}). The number of cycles accumulated in the inspiral scales with the inverse of the 
mass ratio $m_{\SCO}/M_{\MBH}$, while in the plunge and merger it scales with the MBH mass $M_{\MBH}$ only. 
Since the mass ratio for EMRIs is in the range ${\cal{O}}(10^{-4})$-${\cal{O}}(10^{-6})$, the characteristic duration 
and cycle accumulation during the inspiral phase is several orders of magnitude larger than during plunge and merger. 
In turn, the SNR is in the range ${\cal{O}}(10)$-${\cal{O}}(10^{2})$ for EMRIs at realistic 
distances and it scales linearly with the total number of cycles.   Thus, the contribution of the plunge and 
merger to the SNR is reduced by a factor of ${\cal{O}}(m_{\SCO}/M_{\MBH})$ relative to the inspiral 
contribution. In addition, there is no detectable ringdown for EMRIs, as the SCO barely perturbs the 
background geometry as it crosses the MBH's event horizon. Since the SCO is not disrupted in EMRI plunges, 
the SCOs internal structure is erased or {\emph{effaced}} almost completely, without affecting the inspiral signal.

A positive consequence of the large number of EMRI GW cycles is that these waves carry a detailed map of the
MBH geometry, so that we expect to determine the EMRI physical parameters with high precision~\cite{Barack:2003fp}.  
This information will be very useful, in particular to test the spacetime geometry of 
MBHs~\cite{Collins:2004ex,*Glampedakis:2005cf,*Barack:2006pq,*Yunes:2008gb,*Brink:2008xx,*Vigeland:2009pr,*Hughes:2010xf,Sopuerta:2010zy} 
and even alternative theories of gravity (see, e.g.~\cite{Hughes:2006pm,Schutz:2009zz,Sopuerta:2010zy,Babak:2010ej,*Vigeland:2011ji,*Gair:2011ym}).  
Moreover, given that the expected even rate is in the range $10-10^3$ EMRIs$/$yr~\cite{Gair:2004ea,2006ApJ...645L.133H},  
EMRI observations will allow us to understand better the stellar dynamics near galactic nuclei, 
populations of stellar BHs, etc. (for a review see~\cite{AmaroSeoane:2007aw}), and it also possible that
they will tell us about cosmology~\cite{MacLeod:2007jd,*Gair:2008bx}.

Although one is left to model the inspiral phase of EMRIs, this task remains a gargantuan endeavor for several 
reasons.  First, the accuracy requirements for EMRI templates are much more 
stringent than for comparable-mass binaries. For detection and parameter estimation one usually demands an 
absolute accuracy of better than $1$ and SNR$^{-1}$ radians in the GW phase respectively over the entire 
time of observation. In a $1$ yr inspiral, a typical EMRI can have $10^{5}$ cycles in-band, which then 
translates into a relative radian accuracy of ${\cal{O}}(10^{-6})$ and ${\cal{O}}(10^{-8})$ for detection 
and parameter estimation respectively. 
It is important to note that these precision requirements are just simple estimates that do not take into account
data analysis strategies that may relax them (see, e.g.~\cite{Gair:2004ea}).
In contrast, the above relative measure becomes of 
${\cal{O}}(10^{-2}/\mbox{SNR})$ for ground-based data analysis of comparable-mass plunge-merger-ringdowns with 
current numerical relativity simulations, because these accumulate only ${\cal{O}}(10)$ GW cycles in these phases.
In between EMRIs and comparable-mass inspirals, there are Intermediate-Mass-Ratio Inspirals (IMRIs), which can be potential sources for both
space-based detectors (where an Intermediate-Mass BH (IMBH) inspirals into a MBH) and advanced ground-based
detectors (where a SCO inspirals into an IMBH).

The extreme mass ratios involved in the problem also lead to the appearance of two different spatial and
time scales.  The two different spatial scales are represented by the very different sizes of the MBH and
the SCO, $m_{\SCO}/M_{\MBH}\ll 1$, whereas the different time scales are the orbital one and the one associated
with the {\em radiation-reaction} effects, $T^{}_{\ORB}/T^{}_{\RR} \sim m_{\SCO}/M_{\MBH}\ll 1$.
An illustration of how this complicates the EMRI problem is the recent work of Lousto and Zlochower~\cite{Lousto:2010ut,*Nakano:2011pb}, 
who evolved the first $1:100$ mass-ratio binary over the last two orbits before merger and plunge; this simulation took 
approximately $100$ days of computational time using full numerical relativity.  This means that with present
numerical relativity techniques, full numerical simulations are out of the question for EMRI modeling. 

Another key reason for the difficulty of EMRI modeling is their intrinsic strong-relativistic nature.  In
the interesting part of the EMRI dynamics, the SCO is moving around the strong-field region of the MBH,
acquiring large pericenter velocities and even sampling regions inside the ergosphere, leading to 
large relativistic $\Gamma$-factors over tens of thousands of GW cycles. Approximate techniques employed in the comparable-mass 
regime, such as low-velocity expansions in the post-Newtonian (PN) approximation, are then ill-suited for EMRIs.
So neither numerical relativity nor PN theory are, for very different reasons, suitable schemes to model EMRIs.

A better-suited framework that exploits the extreme mass ratios involved is BH perturbation theory, where one treats 
the SCO as a small perturbation of the MBH background geometry.   In this context, the inspiral can be described as 
the action of a {\em self-force}. This local vector force is made out of the regularized metric perturbations generated by the SCO,
after eliminating divergences due to the particle description of the SCO. The SCO's motion is then
governed by the MiSaTaQuWa equation of motion equation, derived in~\cite{Mino:1997nk,Quinn:1997am}
(see also Sec.~\ref{self-force-emris}).   The MiSaTaQuWa equation is considered   
the foundation of a self-consistent scheme to describe EMRIs in an \emph{adiabatic} way by coupling it to the 
partial differential equations that describe the perturbations produced by the SCO.  For recent
discussions on these issues see~\cite{Gralla:2008fg,Pound:2009sm,Pound:2010pj} and for general reviews
see~\cite{Glampedakis:2005hs,Poisson:2004lr,Barack:2009ux}.  

At present, the gravitational self-force has 
been computed for the case of a nonrotating MBH using time-domain techniques~\cite{Barack:2009ey,*Barack:2010tm}
(see also~\cite{Barack:2010ny,Barack:2011ed,Tiec:2011bk} for the study of the physical consequences of the self-force)
and progress is being made towards calculations for the more astrophysically relevant case of a spinning 
MBH~\cite{Shah:2010bi}.  In the meantime, a number of new techniques in the frequency and time domains are
being developed to produce accurate and efficient self-force 
calculations~\cite{Sopuerta:2005rd,*Barack:2007jh,*Barack:2007we,*Vega:2007mc,*Lousto:2008mb,*Canizares:2008dp,*Canizares:2009ay,*Field:2009kk,*Canizares:2010yx,*Thornburg:2010tq,*Canizares:2011ft,*Warburton:2011hp,*Dolan:2011dx}.  
In any case, given the amount of cycles required for EMRI GWs and 
the present complexity of self-force calculations, we cannot expect to generate complete
waveform template banks by means of full self-force calculations.  Instead, the goal of these studies
should be to understand all the details of the structure of the self-force so that we can formulate
efficient and precise algorithms to create the waveforms needed for LISA data analysis, perhaps
complementing some of the existent approximating schemes that we review below.

\subsection{Existent EMRI Waveform Models}
In parallel to the efforts to make progress in the self-force program, there have also been some
efforts to build certain less-reliable approximation schemes to model EMRIs. These are very useful,
for instance, for parameter estimation studies. We compare and contrast these in Table~\ref{model-table}, 
ordered by level of complexity from simplest (top) to most complex (bottom), where we also include the
new Kludge scheme at the bottom for comparison.

\begin{table*}[htb]
\begin{tabular}{cccc}
\hline\hline
Scheme Name  & Orbital Motion & Radiation-Reaction &  Waveform Generation \\
\hline
Peters \& Mathews~\cite{Peters:1963ux,Peters:1964zz} & Keplerian Ellipses &  Newtonian Order & Multipolar decomposition ($l=2$) \\
Analytic Kludge~\cite{Barack:2003fp} & Keplerian Ellipses & Low-Order PN & Multipolar decomposition ($l=2$)\\
Numerical Kludge~\cite{Babak:2006uv} & Kerr Geodesics & Calibrated Low-Order PN & Multipolar decomposition ($l\leq3$) \\
EOB~\cite{Yunes:2009ef,AmaroSeoane:2010ub,Yunes:2010zj} & Kerr Geodesics & Calibrated, Resummed 5.5PN & Resummed 5.5PN ($l\leq8$)\\
Teukolsky-based~\cite{Hughes:1999bq,Hughes:2001jr} & Kerr Geodesics & Averaged Teukolsky & Adiabatic Teukolsky ($l \lesssim 60$) \\
This work & Kerr Geodesics & Local Multipolar Post-Minkowskian~~ & Multipolar decomposition ($l\leq4$) \\
\hline\hline
\end{tabular}
\caption{\label{model-table} Comparison of existent modeling schemes.}
\end{table*}

The simplest model is that of Peters and Mathews~\cite{Peters:1963ux,Peters:1964zz}, 
in which the SCO is assumed to be moving on Keplerian ellipses. The orbital elements of 
this ellipse evolve according to leading-order (Newtonian), dissipative radiation-reaction, i.e.~the 
quadrupole formula for the loss of energy and angular momentum. Waveforms are then computed also to 
leading order via the quadrupole formula~\cite{Misner:1973cw}. 

A better model was introduced by Barack and Cutler~\cite{Barack:2003fp}, the so-called {\emph{Analytical 
Kludge}} waveform model.  This model is based on the Peters \& Mathews~\cite{Peters:1963ux,*Peters:1964zz} 
model, but it is enhanced via different PN formulas in order to account for all the relativistic effects, 
both dissipative and conservative, present in a generic EMRI event.  It has the advantage that the orbital 
evolution is decoupled from the evolution of the additional relativistic effects (which evolve effectively
in the radiation-reaction time scale), and hence EMRI waveforms, also prescribed via the quadrupole 
formula~\cite{Misner:1973cw}, can be computed very fast.  For this reason, it has become the method of 
choice for many LISA parameter estimation and data analysis studies (see, e.g.~\cite{Arnaud:2007jy}).

A more sophisticated approach was proposed by Babak, et.~al.~\cite{Babak:2006uv} and is sometimes referred 
to as the {\emph{Numerical Kludge}} waveform model. In this setup, the orbital motion is given by geodesics
around a Kerr BH and the {\em radiative} effects are prescribed via PN evolution equations for orbital elements
(from 2PN expressions for the fluxes of energy and angular momentum) calibrated to more accurate 
Teukolsky fluxes with $45$ fitting parameters~\cite{Gair:2005ih}. The waveforms are then modeled again via 
a multipolar expansion~\cite{Thorne:1980rm}, but this time taken to next-to-leading-order (quadrupole plus 
octopole). 

Recently, a new hybrid scheme has been proposed by 
Yunes, et.~al.~\cite{Yunes:2009ef,AmaroSeoane:2010ub,Yunes:2010zj} based on effective-one-body (EOB) 
techniques~\cite{Buonanno:1998gg,*Buonanno:2000ef,*Damour:2009sm}. In this approach, the SCO-MBH, two-body system is mapped to 
an effective one-body system: a Kerr BH perturbed by a small effective object. The orbital motion is obtained 
by solving the Hamilton equations for the Hamiltonian of the effective system. When neglecting conservative 
self-force corrections, this reduces to solving the geodesic equations in the Kerr background. 
The radiation-reaction comes from the short-wavelength approximation of 
Isaacson's~\cite{Isaacson:1968ra,*Isaacson:1968gw}, 
where the waveform is constructed from an orbit-averaged, but resummed PN expression~\cite{Damour:1997ub}. This 
radiation-reaction force is then enhanced through the addition of very high PN order point-particle 
results~\cite{Tanaka:1997dj,Mino:1997bx} and expressions that account for the flux of energy and angular 
momentum into the MBH's horizon~\cite{Shibata:1994jx,Mino:1997bx}. Although shown to be accurate for 
equatorial, circular orbits~\cite{Yunes:2009ef} relative to Teukolsky waveforms, the EOB scheme has 
not yet been tested for eccentric or inclined orbits.  

Another approach is the \emph{Teukolsky}-based scheme of Hughes and 
others~\cite{Poisson:1993vp,Hughes:1999bq,Hughes:2001jr,Glampedakis:2002ya,Drasco:2005kz,Drasco:2005is,Hughes:2005qb,LopezAleman:2003ik,Khanna:2003qv,Burko:2006ua,Sundararajan:2007jg,Sundararajan:2008zm} 
have developed.  In this model, one prescribes the inspiral as a sequence of  
{\em slowly}-changing geodesics. The mapping between them is given by the orbit-averaged evolution of orbital elements, 
which in turn is obtained by a balance law relating averaged fluxes at the boundaries of spacetime 
and at the location of the SCO. These averaged fluxes at a given geodesic or point in orbital phase 
space are computed by solving the Teukolsky equation at that point. Therefore, the construction of 
any single waveform requires the mapping of the entire orbital phase space, which in turn is 
computationally prohibitive for truly generic EMRIs. Moreover, this scheme has numerical 
(and conceptual) difficulties when modeling EMRIs in regimes of spacetime where the evolution 
deviates from adiabaticity, such as in or close to plunge, around rapidly spinning MBHs 
($a/M_{\MBH} > 0.9$), or highly eccentric inspirals.

\subsection{The New Kludge Scheme}

In this paper, we devise a new kludge approximation scheme (relative to numerical kludge) that combines seemingly disparate ingredients 
from BH perturbation theory and the multipolar post-Minkowskian formalism of Blanchet and Damour~\cite{Blanchet:1984wm}. We shall here combine two different 
approximation schemes, BH perturbation theory (which assumes only that the mass-ratio is small)
and the multipolar post-Minkowskian formalism (which assumes the gravitational field strength is small), together
with other ingredients related to the choice of coordinate system and waveform construction
method. 

In the new kludge scheme, the orbital motion is prescribed as a spacetime trajectory that is piecewise
geodesic (with respect to the MBH geometry) and such that the different geodesic intervals are 
connected via the SCO's local, self-acceleration (due to its own gravity in the presence of the MBH).  
In this sense, each geodesic interval can be chosen arbitrarily small. This is in 
contrast to the Teukolsky approach (see, e.g.~\cite{Hughes:1999bq}) where the mapping between 
sequences is given by the averaged (over several orbits) GW energy-momentum fluxes and balance laws.   
This fact, that the mapping is given purely in terms of quantities {\emph{local}} to the 
SCO's worldline and not by nonlocal balance laws, is a distinctive feature of the new kludge scheme.  
The implementation of this idea uses the evolution of geodesics with {\emph{varying orbital elements}}; 
the energy, angular momentum in the spin direction, and Carter constants are then functions of time 
governed by the self-force. 

Such an evolution scheme is a direct implementation of the {\emph{osculating orbits}} method, 
proposed by Pound and Poisson~\cite{Pound:2007th} and Pound~\cite{Pound:2009sm}. This method 
professes that at each point of the SCO's nongeodesic worldline there is a unique geodesic 
that lies tangent to it. Therefore, the worldline is simply an interpolation between these 
tangent geodesics. Such a scheme hinges on a fundamental assumption of EMRI modeling: 
{\emph{the adiabatic approximation}}, which assumes the deviation vector between adjacent 
tangent geodesics is small, i.e.~the radiation-reaction time scale is much longer than all 
other timescales, particularly the orbital one. As Gralla and Wald~\cite{Gralla:2008fg} 
explained, the adiabatic approximation only holds quasilocally around some small neighborhood 
of proper time at each point of the SCO's worldline. This problem can be circumvented, however, 
if at each point on the worldline the deviation vector is recomputed, which is exactly the basis 
of the method of osculating orbits~\cite{Pound:2007th,Pound:2009sm}.

In the new kludge scheme, the evolution of orbital elements is prescribed by the SCO's 
self-acceleration, but this quantity is not known exactly (numerically or otherwise) for generic orbits around a 
spinning MBH. In view of this, we model the self-force via a {\emph{radiative approximation}}, 
i.e.,~through the time-asymmetric part of the radiation field, given by the ``half-retarded 
minus half-advanced'' Green function~\cite{Mino:2003yg}. This can be implemented with a 
high-order multipolar, post-Minkowskian expansion, which assumes gravitational radiation is small relative 
to the gravitational field of the background, i.e.~an expansion in powers of Newton's 
gravitational constant $G$. 

The new kludge, local self-force prescription is completed once we say how the multipole moments
depend on the orbital trajectories. This can be achieved by asymptotically matching a PN and a post-Minkowskian
solution~\cite{Burke:2010wb,Blanchet:1987wq,Blanchet:1989ki,Damour:1990ji,Blanchet:1993ng,2002PhRvD..65l4020P}. 
In this paper, we use only the leading-order expressions for these multipoles, although in the future it 
would be trivial to including higher PN corrections. As such, we are not consistently keeping a given PN order, 
but instead using leading-order expressions for the multipole moments, while keeping several such moments 
in the expansions. 

The use of the radiative, multipolar post-Minkowskian expansion is only because 
of the lack of a more precise self-force. Clearly, this radiative approximation neglects the 
conservative part of the self-force, which could be important in GW modeling~\cite{Pound:2007th}. 
Once the full self-force becomes available, however, one could easily employ it instead of its 
post-Minkowskian expansion. Our set-up is general and easily adaptable to other, more precise 
expressions for the self-force. 

Once the orbital evolution has been prescribed, one can construct the GWs 
again through multipolar, post-Minkowskian expressions in terms of a sum over multipole moments. Since 
the mapping between Boyer-Lindquist and harmonic coordinates is known, there are no coordinate 
issues to relate the trajectories obtained from the orbital evolution to the trajectories that 
enter the definition of the multipole moments. We here employ an expansion to second-order 
in the multipole moments, including both the mass hexadecapole and the current octopole, thus 
keeping contributions one order higher than traditional kludge waveforms.   

Let us emphasize that the new kludge waveforms are only {\emph{approximate}} gravitational-wave solutions,
meant to be used for qualitative descoping studies and investigations of resonant behavior. That is, the waveforms
constructed could be used to determine the accuracy to which parameters could be extracted given a detection with
new space-based gravitational-wave observatories as a function of the detector. Moreover, one could also study how
the resonances found by Flanagan and Hinderer~\cite{Flanagan:2010cd} affect such parameter estimation and detection. 
Having said that, improvements would have to be implemented before our kludge waveforms can be used 
as realistic templates in data analysis. As we will see, the new kludge scheme is amenable to such improvements,
which will be the focus of future work. 

\subsection{Comparison to Standard Kludge Waveforms}

The main advantage of the new kludge scheme is the fact that the radiation-reaction effects 
are described in terms of a {\emph{local}} self-force.  This means that the inspiral
description does not need to average certain gravitational-wave fluxes over a number of 
periods/orbits like is traditionally done in kludge implementations. 
Instead, the self-force is prescribed through a multipolar, post-Minkowskian expansion 
(e.g.~the quantity ${\cal{A}}^{}_{\RR}$ mentioned above) that contains time-derivatives of 
the system multipole moments in a nonaveraged form.

When implementing such a nonaveraged and local scheme, it is critical to use an exact mapping between 
Boyer-Lindquist coordinates, used in the integration of the geodesic equations of motion, 
and harmonic coordinates, employed in the calculation of the multipolar,
post-Minkowskian self-force.  This eliminates gauge issues that plague kludge waveforms due  
to the neglect of such a mapping (i.e.~kludge schemes simply use Boyer-Lindquist like Cartesian coordinates in the 
multipolar decomposition of the GWs). 

The use of a local self-force provides the freedom to choose how often to apply radiation-reaction effects in the numerical implementation 
of the dynamics (trajectory and waveform construction) of the new kludge scheme.  The two extremes are: (i) We can apply
the self-force at every single time step, which corresponds to the case of a continuous
local self-force, or (ii) we can store the information about the self-force and apply it
after a certain period of time, mimicking the averaging procedure of other schemes.

These two {\em extreme} ways of using the new kludge scheme are in correspondence with two
very relevant potential applications.  Whereas the type of application (ii) can be used
to try to generate efficiently EMRI gravitational-wave templates for parameter estimation
and data analysis development purposes, the type of application (i) seems to be very well
fitted for studying local phenomena in the dynamics of EMRIs.  In this sense, an important
application of our scheme would be to study the transient resonances that 
Flanagan and Hinderer~\cite{Flanagan:2010cd} have recently reported in generic EMRI
orbits (eccentric and inclined) when the fundamental orbital frequencies become commensurate.
The new kludge scheme can be in principle used to study such behavior and shed some light on
the relevance that it may have for future EMRI detection with LISA-like detectors.

Let us clearly state, however, that the new kludges presented here are not intended to model
comparable-mass systems, for which the effective-one-body framework has proven the most 
successful~\cite{Buonanno:1998gg,Buonanno:2000ef}. Instead, we here focus on EMRIs only and we
are interested in studying local phenomena that no other kludge scheme can currently handle. Perhaps
in the future, one could also use effective-one-body tools for such studies, but this would require a tested
framework capable of handling completely generic orbits. Work along this lines is currently underway~\cite{Yunes:2009ef,AmaroSeoane:2010ub,Yunes:2010zj}. 

\subsection{Executive Summary of Main Results}

In this paper we present the new kludge scheme, introducing one by one each of its ingredients,
from the form of the equations of motion for the inspiral to the GW construction, 
including the details of the approximations that we use to construct
the local self-force that drives the inspiral.  From a technical point of view, one 
of the main challenges of the numerical implementation of our scheme is the 
computation of high-order time derivatives (of the mass and current multipole moments),
which are crucial for the estimation of the radiation-reaction effects (the multipolar post-Minkowskian
self-force involves up to eight-order time derivatives of the trajectory in harmonic
coordinates) and the GW construction (since we are using up to the
mass hexadecapole and current octopole multipoles in the calculation, we require up to fourth-order time derivates of the
trajectory in harmonic coordinates).  The computation of these time derivatives is very challenging,
forcing us to implement numerical techniques adapted to the properties of EMRIs dynamics. 
The key point is to use the fact that geodesic orbits have, in the generic case of eccentric and inclined trajectories,
three fundamental frequencies.  This then allows us to fit any quantity that needs differentiating to a 
multiple Fourier series, using a standard least-squares technique. Numerical derivatives of such quantities  
can then be obtained simply by analytically differentiating the Fourier expansion.  Numerical experimentation
has shown that this technique works remarkably well, even for the highest-order
time derivatives that the new kludge requires.

To illustrate the capabilities of our scheme we show in this paper results from
evolutions for different types of orbits: circular equatorial, eccentric equatorial, 
circular inclined, and the generic eccentric inclined orbits.   Using these evolutions we
study different aspects of the new kludge scheme. First, we consider the impact of harmonic coordinates in the 
trajectories and waveform observables in comparison with using other coordinate systems. 
We find that not properly accounting for this transformation can lead to huge errors in the amplitude
and phase of the waveform, eg.~up to errors of order a factor of 2 in the total accumulated cycles after a 
$1$ yr evolution. Second, we study the impact of the different radiation-reaction potentials in the resulting
waveforms. Although these corrections have a smaller impact than the proper use of coordinates, they are
still large for strong-field EMRIs. For example, including higher-order terms to the Burke-Thorne potential 
leads to corrections of order $10^{4}$ radians in a two month evolution. Third, we investigate the use of a quadrupole
waveform prescription versus a more accurate hexadecapole-octopole prescription. We find no change in the
resulting waveform phases, but an amplitude correction of less than $5\%$. Fourth, we  consider the
time-evolution of different orbital parameters when we apply radiation-reaction effects locally
in time through our multipolar, post-Minkowskian self-force. Although there are some orbital parameters whose time evolution
is not affected, other parameters like the inclination angle can acquire small oscillations with period equal to the orbital
period. Such small oscillations are not captured if the self-force is orbit-averaged.

Finally, we perform some tests and comparisons with results in the literature to validate the new kludge
numerical implementation.  In particular we test the prediction that the inclination 
angle remains almost constant during the evolution, a test that our scheme successfully passes. 

\subsection{Notation and Organization of the Paper}

Throughout this paper we use the metric signature $(-,+,+,+)$ and geometric units in which $G = c = 1$.
The MBH geometry, whose metric we denote by $\met^{\KERR}_{\alpha\beta}$, is determined by its mass 
$M^{}_{\MBH}$ and (magnitude of the) spin angular momentum $S_{\MBH} = M^{}_{\MBH} a$, 
with dimensionless, Kerr spin parameter $a/M^{}_{\MBH}$ ($-1\leq a/M^{}_{\MBH}\leq 1$). 
The SCO is parameterized only in terms of its mass $m^{}_{\SCO}$ since we neglect its spin and other 
internal properties.  The binary system's parameters are the mass ratio $q \equiv m_{\SCO}/M_{\MBH}$ and the total mass  
$M_{\TOT} = m_{\SCO} + M_{\MBH}$.  The reduced mass of the system is therefore 
$\mu \equiv m_{\SCO} M_{\MBH}/M_{\TOT}$, while the symmetric mass ratio is $\eta \equiv \mu/M_{\TOT} = q/(1 + q)^{2}$. 
The mass difference is denoted by $\delta m \equiv M_{\MBH} - m_{\SCO}$.

The SCO orbit can be parameterized in terms of the constants of geodesic motion
${\cal{I}}^{A} = (E,L_{z},Q/C)$, which stand for the SCO's energy normalized with respect to  
$m_{\SCO}$, the $z$-component of the angular momentum normalized to $m_{\SCO}^{2}$, and the Carter constant also 
normalized to $m_{\SCO}^{2}$ ($C$ and $Q$ stand for two definitions of the 
Carter constant that we use in this paper).  Alternatively, we will also parameterize the SCO orbit in terms
of the orbital elements ${\cal O}^{A} = (e,p,\iota/\theta^{}_{\INC})$, which are also constants of the
geodesic motion, where $e$ is the orbit eccentricity,
$p$ is the semilatus rectum, and $\iota$ and $\theta_{\INC}$ are two measures of the orbit inclination.
We present the mappings between ${\cal{I}}^{A}$ and ${\cal{O}}^{A}$ in Appendix~\ref{app-const-of-motion}.
The SCO spacetime trajectory is denoted via $z^{\mu}(\tau)$, where $\tau$ is proper time and thus, its four-velocity is the 
unit timelike vector $u^{\mu} \equiv d z^{\mu}/d\tau$.  

Post-Newtonian orders always refer to a {\emph{relative}} ordering scheme (instead of an absolute one), such 
that the $N$-th PN order term refers to one of the form $A = A_{\rm Newtonian} [1 + \ldots + {\cal{O}}(v^{2N}/c^{2N})]$, 
where $A_{\rm Newtonian}$ is the leading order contribution.  We shall commonly drop the factor of 
$1/c$ when referring to PN expansions. 

Greek letters in index lists are used to denote indices on the $4$-dimensional spacetime, while Latin 
letters in the middle of the alphabet $i,j,k,\ldots$ denote spatial indices only. 
Covariant differentiation is denoted using the symbol $\nabla^{}_\mu$, while  partial derivatives with respect to 
the coordinate $x^{\mu}$ are denoted as $\partial^{}_{\mu} B^{}_{\nu}$ or $B^{}_{\nu,\mu}$. 
We denote symmetrization and antisymmetrization with parenthesis and square brackets around the 
indices respectively, such as $A^{}_{(\mu\nu)}\equiv [A^{}_{\mu\nu} + A^{}_{\nu\mu}]/2$ and 
$A^{}_{[\mu\nu]} \equiv [A^{}_{\mu\nu} - A^{}_{\nu\mu}]/2$. 

We use two main sets of coordinate systems: Boyer-Lindquist coordinates $x^{\mu}_{\BL}$ and 
harmonic coordinates $x^{\mu}_{\HAR}$.  Other systems of coordinates that we also consider 
are asymptotic-Cartesian mass-centered (ACMC) coordinates, $x^{\mu}_{\ACMC}$, and approximate harmonic 
coordinates, $x^{\mu}_{\AHAR}$.  
Retarded time is denoted in harmonic coordinates via 
$t^{}_{r} \equiv t^{}_{\HAR} - r^{}_{\HAR}$, where 
$r^{}_{\HAR} \equiv (x_{\HAR}^{2} + y_{\HAR}^{2} + z_{\HAR}^{2})^{1/2}$. 
When we refer to the Kerr metric, we sometimes use the label $K$,
e.g. $\met_{\alpha\beta}^{\KERR}$ and $\met^{\alpha\beta}_{\KERR}$.  In some situations, it is crucial
to specify in which coordinate system the metric has to be written in a certain equation, like
in a coordinate transformation.  In those situations, we also incorporate in the metric (or related
objects) a label associated with the coordinate system, e.g. $\met^{\KERR,\HAR}_{\alpha\beta}$
or $\met^{\KERR,\BL}_{\alpha\beta}$. As for angular coordinates, we shall find it convenient to sometimes 
perform multipolar decompositions as in~\cite{Thorne:1980rm}, with spin-weighted spherical harmonics 
${}^{}_{-2}Y^{\ell m}$ and symmetric and trace-free spherical harmonic tensors ${\cal{Y}}_{L}^{\ell m}$, 
where $L = (i_{1}, i_{2}, \ldots, i_{n})$ is a multi-index (see~\cite{Thorne:1980rm} for details
on the multi-index notation).  

The organization of this paper is as follows.
Section~\ref{self-force-emris} introduces the self-force approach to EMRI modeling and some basic 
details of the method of osculating orbits. 
Section~\ref{chimera} describes in detail the new kludge scheme, including the MBH geometry and the
properties of the geodesic orbits, the application to them of the method of osculating orbits,
the multipolar post-Minkowskian approximation to the self-force we use and the radiation-reaction potentials
from which it can be obtained, the mapping between Boyer-Lindquist and harmonic 
coordinates, and the multipolar post-Minkowskian waveform generation formalism that we use in our scheme. 
Section~\ref{num-implementation} explains the numerical implementation of the new kludge approach
and the main numerical techniques that we use.  This includes the algorithms for the integration
of the ODEs governing the local geodesic motion and the accurate estimation of time derivatives of 
several orders of the multipole moments.
Section~\ref{num-results} summarizes the different ways in which our scheme can
be used and presents several numerical results that illustrate its main features, 
using from circular-equatorial orbits to eccentric inclined orbits.
Section~\ref{discussion-conclusions} concludes and discusses several avenues for future research
applying the new kludge scheme.

We have attempted to present the main ingredients of our approach in the main body of 
the paper, relegating some details to the Appendices. 
Appendix~\ref{app-coeffs} gives explicit expressions of the different pieces of the multipolar post-Minkowskian
self-force in terms of the radiation-reaction and local potentials.  It also gives formulas to 
simplify the computation of the different spatial derivatives of the Kerr local potentials.
Appendix~\ref{app-blkerrinharmonic} provides complementary formulas related to the mapping between Boyer-Lindquist 
and harmonic coordinates.  In particular, we give expressions for the components of the Jacobian and
Hessian, and for the components of the covariant and contravariant Kerr metric tensor in harmonic coordinates.
Appendix~\ref{app-coord-details} constructs a system of asymptotically, mass-centered coordinates
and a system of approximate harmonic coordinates that we compare with the exact harmonic coordinates of
Sec.~\ref{coord-sec}. 
Appendix~\ref{app-Far-Field-Qs-and-Ks} performs a far-field expansion of the Kerr metric 
coefficients in the approximate harmonic coordinates of Appendix~\ref{app-coord-details}.
Appendix~\ref{app-const-of-motion} describes in detail how to implement the one-to-one mapping between 
the orbital elements ${\cal O}^{A} = (e,p,\iota/\theta_{\INC})$ and the constants 
of motion ${\cal I}^{A} = (E,L_{z},C/Q)$.
Appendix~\ref{fundamental-frequencies-and-periods} summarizes the main formulas for the 
computation of the fundamental frequencies and periods with respect to the
Boyer-Lindquist coordinate time.
Finally, Appendix~\ref{circular-nonequatorial} provides expressions for the coefficients that
determine the evolution of the radius and Carter constants of an inspiral through circular
nonequatorial geodesics.

\section{The Self-Force Approach to EMRIs}\label{self-force-emris}

In this section we review some of the basic and well-established concepts
related to the self-force as they are related to the new kludge approach
(see the review papers~\cite{Poisson:2004lr,Poisson:2011nh,Barack:2009ux} for details).

The foundations for the first-order perturbative description of EMRIs were laid down in 
the papers by Mino, Tanaka, and Sasaki~\cite{Mino:1997nk} and Quinn and Wald~\cite{Quinn:1997am}.  
The main result of these papers was the equation of motion of a massive pointlike object 
(describing the SCO) in the geometry of a MBH.  The stress-energy tensor of the SCO is 
then given by
\be
T^{\alpha\beta} = m^{}_{\ast} \int \frac{d\tau}{\sqrt{-{\met}}} \; 
\delta^{4}\left[x^{\mu} - z^{\mu}(\tau) \right] 
\frac{d z^{\alpha}}{d\tau} \frac{d z^{\beta}}{d\tau}\,,
\ee
where ${\met}$ denotes the determinant of ${\met}^{}_{\alpha \beta}$.  Then, the SCO
generates metric perturbations, $h^{}_{\alpha\beta}$, around the MBH background geometry that 
in the Lorenz gauge,
\be
\nabla^{}_{\mu}\tilde{h}^{\mu\nu} = 0\,,\quad
\tilde{h}^{\alpha\beta} \equiv h^{\alpha\beta}-\frac{1}{2}\met^{\alpha\beta}h\,,\quad
h \equiv \met^{\mu\nu}h^{}_{\mu\nu}\,, \label{boxh}
\ee
satisfy the linearized Einstein equations
\ba
{\square}\, \tilde{h}^{\alpha \beta} + 2\,{R}\,{}^{\alpha}{}_{\mu}{}^{\beta}{}_{\nu}\,\tilde{h}^{\mu \nu} 
= -16\, \pi\, m_{\SCO}\, T^{\alpha\beta}\,,  \label{evol-eq-hab}
\ea
where $\bar{R}^{\mu}{}_{\alpha \nu \beta}$ is the Riemann tensor of the MBH background geometry. 
However, according to this equation, the metric perturbations diverge at the particle location.
Then, the gravitational \emph{backreaction} on the particle motion, the self-force, is provided by the 
regularized part of the perturbations, say $h^{\REG}_{\alpha\beta}$, according to a Hadamard prescription 
given in~\cite{Mino:1997nk}.
Then, the equation of motion for an EMRI, the MiSaTaQuWa equation, is 
\be
\frac{d^{2}z^{\alpha}}{d \tau^{2}}+
{\Gamma}^{\alpha}_{\mu \nu} u^{\mu} u^{\nu} = m^{-1}_{\SCO} F_{\SF}^{\alpha}\,, \label{misataquwa1}
\ee
where the self-force, $F_{\SF}^{\alpha}$, is given by
\be
F_{\SF}^{\alpha} = - \frac{1}{2} m_{\SCO} \left({\met}^{\alpha \lambda} + u^{\alpha}u^{\lambda}\right)
u^{\mu}u^{\nu}
\left( 2 {\nabla}^{}_{\mu} h^{\REG}_{\nu\lambda} - {\nabla}^{}_{\lambda} 
h^{\REG}_{\mu \nu} \right)\,. \label{misataquwa2}
\ee
Therefore, the dynamics of EMRIs is determined by the coupled system of Eqs.~\eqref{misataquwa1},\eqref{misataquwa2},
and~\eqref{boxh} with a practical regularization scheme (like the \emph{mode sum scheme}~\cite{Barack:2001gx}).
A remarkable point is that Eqs.~\eqref{misataquwa1} and~\eqref{misataquwa2} can be rewritten as
geodesic equations of motion for a point particle in a perturbed geometry that only take into
account the regularized part of the metric perturbations, i.e.~geodesic in the geometry 
$\met^{}_{\alpha\beta}+h^{\REG}_{\alpha\beta}$~\cite{Detweiler:2000gt}.

But how do we evolve the trajectory of the SCO accounting for
the self-force in a self-consistent way?  As it was proposed in~\cite{Pound:2007th} (see also~\cite{Gralla:2009md}
and~\cite{Gair:2010iv} for a recent use of this technique), 
one can use a relativistic extension of the well-known method of \emph{osculating orbits}.
The idea is to consider always the geodesic tangent to the trajectory, so that the
motion transitions from one geodesic to the next.  Such smooth transition is
facilitated in EMRI by the fact that EMRI trajectories are very close to a (local) geodesic for a long time. 
This is because of the clean separation in EMRI time scales: the radiation-reaction time scale is much larger 
than the orbital (geodesic) one, except for the tiny fraction corresponding to the merger-plunge phase. 

The way to carry out this transition is to properly account for the time-evolution of the set of 
\emph{orbital elements} that completely characterize a geodesic orbits.
Following~\cite{Pound:2007th}, it is important to distinguish between two sets of orbital elements:
\emph{principal} orbital elements, in our case either ${\cal{I}}^{A} = (E,L_{z},Q/C)$ or 
${\cal O}^{A} = (e,p,\iota/\theta^{}_{\INC})$; and \emph{positional} orbital elements that 
determine the initial position in the geodesic as well as the geodesic initial spatial orientation. 
The radiation-reaction changes in the principal elements are due to the dissipative part
of the self-force, while radiation-reaction changes in the positional elements are due to
the conservative part of the self-force.  In this work we only consider dissipative effects
and hence we are only concerned with changes in the principal orbital elements.

The implementation of the method of osculating orbits consists in the translation of the 
fact that at any time there will be a geodesic trajectory, $z^{\mu}_{\GG}$, with orbital elements 
whose position and velocity at that time will coincide with those of the accelerated trajectory.
This can be written in the following way:
\ba
{z}^{\alpha}(\tau) & = & z^{\alpha}_{\GG}\left(\tau;{\cal{P}}^{A}(\tau);{\cal{I}}^{A}(\tau)\right)\,, \\
\frac{dz^{\alpha}}{d\tau}(\tau) & = & \frac{\partial z^{\alpha}_{\GG}}{\partial \tau}
\left(\tau;{\cal{P}}^{A}(\tau);{\cal{I}}^{A}(\tau)\right)\,,
\ea
where $\cal{P}^{A}$ denote the positional orbital elements. Although we have used here the constants
of motion as principal orbital elements, we could also have used ${\cal O}^{A}$.
Combining these osculation conditions with the equations of motion~\eqref{misataquwa1}, we can 
arrive at the following equations~\cite{Pound:2007th}:
\ba
&&\frac{\partial z^{\alpha}_{\GG}}{\partial\cal{P}^{A}}\; \frac{d\,{\cal{P}}^{A}}{d\tau} +
\frac{\partial z^{\alpha}_{\GG}}{\partial\cal{I}^{A}}\; \frac{d\,{\cal{I}}^{A}}{d\tau} = 0\,, 
\label{osculatory1} \\
&&\frac{\partial }{\partial\cal{P}^{A}}\left( \frac{\partial z^{\alpha}_{\GG}}{\partial \tau}\right) 
\frac{d\,{\cal{P}}^{A}}{d\tau} +
\frac{\partial }{\partial\cal{I}^{A}}\left( \frac{\partial z^{\alpha}_{\GG}}{\partial \tau}\right)
\frac{d\,{\cal{I}}^{A}}{d\tau} = a_{\SF}^{\alpha}\,, \label{osculatory2}
\ea
where $a^{\alpha}_{\SF}$ is the SCO self-acceleration, which is related to the self-force of
Eq.~\eqref{misataquwa2} by $a^{\alpha}_{\SF} = m^{-1}_{\SCO} F^{\alpha}_{\SF}$.  
This is due to the fact that ${\cal I}^{A}$ has been defined as the SCO constants of motion per unit mass.
From the inversion of these equations we can obtain the evolution of the different
orbital elements.  In this paper, we will focus exclusively on the dissipative effects of the
self-force which only affect to the principal orbital elements (either ${\cal I}^{A}$
or ${\cal O}^{A}$), i.e.~${d\,{\cal{P}}^{A}}/{d\tau}=0$ and 
we ignore the first term in Eqs.~\eqref{osculatory1} and~\eqref{osculatory2}.
We will refer to the principal orbital elements simply as orbital elements.

\section{The New Kludge Scheme}\label{chimera}

In this section we present all the details of the new kludge scheme. 
As explained in the previous sections, we separate the problem into two parts: that of constructing
the SCO's trajectory and that of building the waveform from these trajectories.  We begin then with
a description of the background MBH geometry and its associated geodesics. We then show how
to enhance the geodesic equation system to allow for the variation of the constants of the motion. 
The latter require knowledge of the self-acceleration, which we calculate in a multipolar, post-Minkowskian expansion. 
Finally, we present an explicit transformation between Boyer-Lindquist coordinates (used to evolve
the modified geodesic system) and harmonic coordinates, needed to generate waveforms in a multipolar
decomposition. 

\subsection{MBH Geometry and Geodesic Motion}\label{MBHgeometryandgeodesics}
The geometry of the MBH is modeled by the Kerr metric~\cite{Kerr:1963ud}, a vacuum stationary and 
axisymmetric spacetime that describes the final state of gravitational collapse, according to the
BH {\em no-hair} conjecture~\cite{Ruffini:1971rw} and uniqueness theorems (see, e.g.~\cite{Chandrasekhar:1992bo}).
In Boyer-Lindquist coordinates~\cite{Boyer:1966qh}, $(x^{\mu}_{\BL}) = (t,r,\theta,\phi)$, the line element
corresponding to the Kerr metric, $\met^{\KERR}_{\alpha\beta}$, is given by
\ba
ds^2 & = & -dt^2 + \frac{\rho^2}{\Delta}dr^2 + \rho^2d\theta^2 + (r^2+a^2)\sin^2\theta\, d\phi^2 
\nonumber \\
& + & \frac{2M_{\MBH}r}{\rho^2}(dt - a\sin^2\theta\, d\phi)^2 \,,
\ea
where $\rho^{2} \equiv r^{2} + a^{2} \cos^{2}{\theta}$ and $\Delta \equiv r^{2}-2M_{\MBH}r+a^{2} = 
r^{2} f + a^{2}$, with $f \equiv 1 - 2 M_{\MBH}/r$. For convenience, we also define the 
quantity $\Sigma^{2} \equiv (r^2 + a^2)^{2} - a^{2}\Delta\,\sin^{2}\theta\,$. 
The function $\Delta$ has two roots:
\be
r^{}_{\pm} \equiv M_{\MBH} \pm \sqrt{M_{\MBH}^2-a^2}\,.
\label{rpm}
\ee
The root $r^{}_{+}$ ($\geq r^{}_{-}$) coincides with the location of the event horizon.

The Kerr geometry is stationary, as described by the 
timelike Killing vector field ${\zeta}_{(t)}^{\alpha} =
\delta^{\alpha}_{t}$, and axisymmetric, as described by a spacelike Killing vector field ${\zeta}_{(\phi)}^{\alpha} = 
\delta^{\alpha}_{\phi}$.  It also well-known that the Kerr geometry has an additional symmetry described by a 2-rank 
Killing tensor, $\xi^{}_{\alpha \beta}$, given by
\be
\xi^{}_{\alpha \beta} = \Delta\, k^{}_{(\alpha} l^{}_{\beta)} + r^{2}\, {\met}^{\KERR}_{\alpha \beta}\,,
\label{killingtensor}
\ee
where $k^{\alpha}$ and $l^{\alpha}$ are the two null principal directions of the Kerr geometry
\be
k^{\alpha} = \left[\frac{r^{2}+a^{2}}{\Delta},-1,0,\frac{a}{\Delta} \right]\,,
~~
l^{\alpha} =  \left[\frac{r^{2}+a^{2}}{\Delta},1,0,\frac{a}{\Delta} \right]\,.
\ee

When the effect of the self-force is neglected (or equivalently, in the limit of zero mass for the SCO),
the SCO follows geodesics orbits of the MBH geometry.   The Boyer-Lindquist coordinates of a timelike
geodesic can be parameterized in terms of proper time as $z^{\mu}(\tau) = (t(\tau),r(\tau),\theta(\tau),\phi(\tau))$.
For a given geodesic, we can construct three geodesic constants of motion, corresponding to each of the three 
symmetries of the Kerr spacetime.  Stationarity leads to a conserved energy, ${\cal E}$, or equivalently an energy per unit mass
\be
\label{E-def}
E \equiv {\cal E}/m_{\SCO} \equiv - \zeta^{(t)}_{\alpha} {u}^{\alpha}\,.
\ee
Axial symmetry leads to a conserved component of the angular momentum vector (the one along the spin axis, which we choose
to be the $z$ axis)
\ba
\label{L-def}
L^{}_{z} \equiv {\cal L}^{}_{z}/m_{\SCO} \equiv \zeta^{(\phi)}_{\alpha} {u}^{\alpha}\,.
\ea
The Killing tensor symmetry~\eqref{killingtensor} leads to a conserved Carter constant, which
can be defined as follows
\be
Q \equiv {\cal Q}/m_{\SCO}^2 \equiv \xi^{}_{\alpha \beta} {u}^{\alpha} {u}^{\beta}\,,
\label{q_carter}
\ee
but also we will use the alternative definition
\be
C \equiv Q -  \left(L^{}_{z} - a E \right)^{2}\,. \label{c_carter}
\ee

The existence of these three symmetries makes the geodesic equations for $z^{\mu}(\tau)$ integrable and completely
separable~\cite{Carter:1968fa,*Carter:1968ks}.   The separation can be carried out using the definitions of 
these constants of motion and also the normalization condition of the four-velocity, $\met^{\KERR}_{\mu\nu}u^{\mu}u^{\nu}=-1$.  
The separated equations of motion for the components of $z^{\mu}(\tau)$ obey the following set of ordinary differential equations
(see~\cite{Bardeen:1972fi} for a detailed analysis of the physical properties of the solutions of these equations):
\be
\rho^{2}\, \frac{d{t}}{d\tau} = \frac{1}{\Delta}\left( \Sigma^{2} E - 2 M_{\MBH} a r {L}^{}_{z} \right)\,,  
\label{tdot-GR}
\ee
\be
\rho^{4}\, \left(\frac{d {r}}{d\tau} \right)^{2} = \left[ \left( r^{2} + a^{2} \right) E - a L^{}_{z} \right]^{2} - 
\left(Q+r^{2}\right)\Delta \,, \label{rdot-GR}
\ee
\be
\rho^{4}\, \left(\frac{d{\theta}}{d \tau} \right)^{2} = C - \cot^{2}{\theta}{L}^{2}_{z} 
- a^{2} \cos^{2}{\theta} \left(1 - {E}^{2} \right)\,,\label{thetadot-GR}
\ee
\be
\label{phidot-GR}
\rho^{2}\, \frac{d \phi}{d{\tau}} = \frac{1}{\Delta}\left[ 2M_{\MBH}\,a\,r{E} + \frac{{L}^{}_{z}}{\sin^{2}\theta} 
\left( \Delta -a^{2}{\sin^{2}{\theta}} \right)  \right]\,.
\ee

We are interested here in timelike bound and stable geodesics, what we call {\em orbits}.  These orbits,
apart from being characterized by the three constants of motion ${\cal I}^{A}$, can also be characterized
by the orbital elements ${\cal O}^{A}$, which can be defined in terms of the turning points of the
radial and polar motion (see Appendix~\ref{app-const-of-motion} for more details).  The turning points for the radial motion are 
just the minimum and maximum of $r$, also known as pericenter and apocenter, $r^{}_{\PERI}$ and $r^{}_{\APO}$, 
which can be used to introduce the concepts of semilatus rectum and eccentricity
\be 
r^{}_{\PERI} = \frac{pM^{}_{\bullet}}{1+e}\,, \qquad
r^{}_{\APO} = \frac{pM^{}_{\bullet}}{1-e}\,, \label{peri-and-apo}
\ee
or equivalently
\be
p = \frac{2\,r^{}_{\PERI}\,r^{}_{\APO}}{M^{}_{\bullet}(r^{}_{\PERI}+r^{}_{\APO})}\,, \qquad
e = \frac{r^{}_{\APO}-r^{}_{\PERI}}{r^{}_{\PERI}+r^{}_{\APO}}\,.
\ee
The turning point for the polar motion is just the minimum of $\theta$, $\theta^{}_{\MIN}\in [0,\pi/2]$,
which determines the interval in which $\theta$ oscillates, i.e.~$(\theta^{}_{\MIN},\pi-\theta^{}_{\MIN})$ 
and it can be used to introduce the concept of inclination angle, $\theta^{}_{\INC} \in [-\pi/2,\pi/2]$, as
\be
\theta^{}_{\INC} = {\rm sign}(L^{}_{z})\left[\frac{\pi}{2} - \theta^{}_{\MIN}\right]\,. \label{theta-inc}
\ee
Another common definition of orbital inclination angle uses the constants of motion ${\cal I}^{A}$
\be
\cos\iota = \frac{L^{}_{z}}{\sqrt{L^{2}_{z}+C}}\,.  \label{iota-angle}
\ee

Alternatively, the orbits can also be characterized in
terms of three {\em fundamental} frequencies (see e.g.~\cite{Schmidt:2002qk,Drasco:2003ky,Fujita:2009bp}) with
respect to the Boyer-Lindquist coordinate time (they can be also constructed using proper time or any other 
time): 
$\Omega^{}_{r}$, associated with the radial motion (from periapsis to apoapsis and back); 
$\Omega^{}_{\theta}$, associated with polar motion; and 
$\Omega^{}_{\phi}$, associated with azimuthal motion.  These frequencies are important because precessional 
orbital effects are due to mismatches between them and because they can be used to decompose, among other things, 
the GW in a Fourier expansion~\cite{Drasco:2003ky}.  

Expressions for these frequencies in terms of quadratures have been obtained for Kerr in~\cite{Schmidt:2002qk},
and recently also in~\cite{Drasco:2003ky,Fujita:2009bp}. Appendix~\ref{fundamental-frequencies-and-periods} provides
explicit formulas for the fundamental frequencies and periods used in this paper.

A final consideration regarding geodesic motion that is going to be important in this paper
is the $3+1$ splitting of the four-velocity into spatial and time components.  This
splitting is associated with the time variable used to evolve the geodesic equations, in this paper 
Boyer-Lindquist time $t$.  The four-velocity of the SCO, 
${u}^{\alpha} = d{z}^{\mu}/d\tau$, choosing the Boyer-Lindquist time parameterization $z^{\mu}(t) = (t,z^{i}(t))$,
can then be decomposed as 
\be
{u}^{\mu} = \left(\frac{dt}{d\tau},\frac{dz^{i}}{d\tau}\right) \equiv \Gamma\,(1, v^{i} )\,,
\label{sco-4velocity}
\ee
where $\Gamma$ and $v^{i}$ are given by
\be
\Gamma = u^{t} = \frac{dt}{d\tau}\,,\qquad
v^{i} = \Gamma^{-1} u^{i} = \frac{dz^{i}}{dt} \,.
\label{gammafactor-and-spatial-velocity}
\ee
Using the normalization condition $\met^{\KERR}_{\mu \nu} {u}^{\mu} {u}^{\nu} = -1$,
the factor $\Gamma$, the GR generalization of the special-relativistic Lorentz factor, can be 
written in terms of the metric and the components of the
velocity $v^{i}$ as follows:
\be
\Gamma =  \left(-{\met}^{\KERR}_{tt} - 2 {\met}^{\KERR}_{ti}v^{i} - 
{\met}^{\KERR}_{ij} v^{i}v^{j}\right)^{-1/2}\,.
\label{Gamma-def}
\ee
%

\subsection{New kludge Osculating Trajectories}

In this work we consider a version of our scheme in which we only include the dissipative effects 
of the self-force, i.e.~those that only affect to principal orbital elements. 
In Sec.~\ref{discussion-conclusions} we discuss how 
to introduce conservative pieces of the self-force in the new kludge scheme.  
The time scale of change of ${\cal I}^{A}(\tau)/{\cal O}^{A}(\tau)$, 
the radiation-reaction time scale ${T}^{}_{\RR}$, is much bigger than the orbital time scales ${T}^{}_{\ORB}$, 
such that the ratio of these time scales satisfies: ${T}^{}_{\ORB}/{T}^{}_{\RR} \sim q$.

Following the method of osculating orbits described in Sec.~\ref{self-force-emris}, we can describe
the orbital evolution as given locally in time by the geodesic Eqs.~\eqref{tdot-GR}-\eqref{phidot-GR},
where the constants of motion are promoted to time-dependent quantities:
\be
{\cal I}^{A} \; \longrightarrow \; {\cal I}^{A}(\tau)\,,\qquad
{\cal O}^{A} \; \longrightarrow \; {\cal O}^{A}(\tau)\,. \label{timedependentconstants}
\ee
Although here we parameterize the time dependence of these quantities in terms of proper time,
in practice we will use coordinate time, that corresponds to time associated with distant observers. 
For our discussion, we denote the solution of the evolution Eqs.~\eqref{tdot-GR}-\eqref{phidot-GR} 
with the modifications of Eq.~\eqref{timedependentconstants} by ${z}^{\mu}(\tau) = [{t}(\tau),{r}(\tau),{\theta}(\tau),{\phi}(\tau)]$,
but it is clear that it would no longer be a solution of the original geodesic equations.
We do not decompose this solution, which takes into account self-force effects, into a background geodesic 
plus a deviation~\cite{Gralla:2008fg}, as the deviations grow secularly in time and after a certain number of cycles it
cannot be considered a {\em small} deviation of the background geodesic orbit. 
Instead, in the spirit of the osculating orbits method, we treat $z^{\mu}$ as a new, self-consistent trajectory that 
is {\emph{continuously}} corrected away from geodesic motion during evolution. In this sense, we are 
constructing an orbit that is made out of geodesic patches corresponding to different constants of motion.  
The transition from one patch to another is given by the (multipolar, post-Minkowskian) self-force, lacking a more accurate prescription.
The length (duration) of the geodesic patches, or equivalently, the frequency at which the constants of motion
are updated, is a free parameter that we can choose in the new kludge numerical implementation.   

The evolution equations for ${\cal I}^{A}(\tau)/{\cal O}^{A}(\tau)$ can be obtained from the equations of osculating
orbits, Eqs.~\eqref{osculatory1} and~\eqref{osculatory2}.  In our case, the inversion of these equations
to find $d{\cal I}^{A}(\tau)/d\tau$ can be done easily by using the symmetries of the Kerr geometry. 
By applying $\met^{\KERR}_{\alpha\beta}\zeta^{\beta}_{(t)}$, 
$\met^{\KERR}_{\alpha\beta}\zeta^{\beta}_{(\phi)}$, and $\xi^{}_{\alpha\beta}u^{\beta}$ to Eq.~\eqref{osculatory2} 
in combination with Eqs.~\eqref{E-def}-\eqref{c_carter} we obtain the evolution equations for $E$,
$L^{}_{z}$, and $C/Q$ respectively.  
\ba
\frac{dE}{d\tau} & = & - {\zeta}^{(t)}_{\alpha} a^{\alpha}_{\SF}\,, 
\label{evol-cons-motion-E}  \\
\frac{d{L}^{}_{z}}{d\tau} & = & {\zeta}^{(\phi)}_{\alpha} {a}^{\alpha}_{\SF}\,,
\label{evol-cons-motion-L}  \\
\frac{d{Q}}{d\tau} & = &  2\, {\xi}^{}_{\alpha \beta} {u}^{\alpha} a_{\SF}^{\beta}\,,
\label{evol-cons-motion-Q} \\ 
\frac{d{C}}{d\tau} & = & \frac{d{Q}}{d\tau} + 2 \left(a E -  L^{}_{z} \right) 
\left(\frac{d{L}^{}_{z}}{d\tau}  - a \frac{d{E}}{d\tau} \right) \,. 
\label{evol-cons-motion-C}
\ea
The evolution of these quantities with respect to coordinate time $t$ introduces a 
factor $\Gamma^{-1}$ in the above equations, due to the relation: $d/dt = \Gamma^{-1}d/d\tau$.  
The most important quantity here is the self-acceleration $a_{\SF}^{\alpha}$,
which as we shall see scales $\sim \Gamma^{2}$ [see Eq.~\eqref{full-acc}] and is 
given by the metric perturbations $h^{\REG}_{\alpha\beta}$.  This is probably the main
ingredient of self-force descriptions of EMRIs, which is described in more detail in the next section.

For certain special orbits, the rate of change of the orbital elements satisfies
special relations due to the symmetries.  One of these are equatorial
orbits, i.e.~orbits with $\theta(t) = \pi/2$.  We can see that from Eq.~\eqref{thetadot-GR} this implies
$C=0$.  Therefore, equatorial orbits are characterized by $\theta(t) = \pi/2$ and $C=0$.   
Equation~\eqref{evol-cons-motion-C} allows us to write  
$dC/d\tau = 2{\cal C}^{}_{\alpha}a^{\alpha}_{\SF}$, where
\be
{\cal C}^{}_{\alpha} = (\delta_{\alpha}^{\beta}+u^{}_{\alpha}u^{\beta})
\left[\xi^{}_{\beta\lambda}u^{\lambda}+(aE-Lz)\left(\zeta^{(\phi)}_{\beta}+a\zeta^{(t)}_{\beta}\right)\right]\,.
\ee
The vector ${\cal C}^{}_{\alpha}$ vanishes for equatorial geodesics, which implies that the Carter constant
$C(t)$ is always zero and $Q(t)$ can be obtained directly from a combination of $dE/dt$ and $dLz/dt$. 
The self-force then maps equatorial geodesics to equatorial geodesics, i.e.~the equatorial
character of geodesics is preserved upon self-force evolution. 

Another type of orbits with extra symmetries are equatorial and circular orbits.  These orbits have
a helical symmetry described by an approximate Killing vector (which is exact on the geodesic intervals).
This symmetry can be derived from the fact that for these special orbits $d\phi/dt = \Omega^{}_{\phi} =
\mbox{const.}$ [see Eq.~\eqref{phidot-GR}].  The angular velocity $\Omega^{}_{\phi}$ can be written as 
(see~\cite{Bardeen:1972fi}):
\be
\Omega^{}_{\phi} = \pm \frac{v^{3}_{o}}{M^{}_{\MBH}\left(1\pm\frac{a}{M^{}_{\MBH}}v^{3}_{o}\right)}\,,
\ee
where the upper/lower sign corresponds to prograde/retrograde circular orbits (i.e.~that corotate/counterrotate
with the MBH spin and have $L^{}_{z}>0$/$L^{}_{z}<0$), 
$v^{}_{o}\equiv\sqrt{M^{}_{\MBH}/r^{}_{o}}$, and $r^{}_{o}$ is the Boyer-Lindquist radial coordinate
of the circular orbit. Then, the helical Killing vector is $\zeta^{\alpha}_{\HEL} 
= \zeta^{\alpha}_{(t)} + \Omega^{}_{\phi}(r)\zeta^{\alpha}_{(\phi)}
= \partial^{\alpha}_{t} + \Omega^{}_{\phi}(r)\partial^{\alpha}_{\phi}$.  The associated constant of
motion is then $\Lambda\equiv\met^{\KERR}_{\alpha\beta}\zeta^{\alpha}_{\HEL}u^{\beta} = 
-E + \Omega^{}_{\phi}L^{}_{z}$.   The evolution of $\Lambda$ is $d\Lambda/d\tau =
{\zeta}^{\HEL}_{\alpha} {a}^{\alpha}_{\SF}$ and it can be shown that $d\Lambda/d\tau = 0$ for
locally circular-equatorial orbits.  Hence, the evolution of the angular momentum in
the spin direction is related to the evolution of the energy by $dL^{}_{z}/dt = \Omega^{-1}_{\phi}
dE/dt$.

Finally, let us consider circular but nonequatorial orbits, i.e. orbits with $r=r^{}_{o}=\mbox{const}.$ but
$C\neq 0$. There is no helical symmetry for these orbits due to the fact that the MBH spin is not 
aligned with the orbital angular momentum.  Nevertheless, as shown in~\cite{Kennefick:1995za,Ryan:1995xi} 
the radiation-reaction evolution of these orbits preserves their circular character (see also~\cite{Hughes:1999bq}).
These orbits are then characterized by the vanishing of the right-hand side of Eq~\eqref{rdot-GR},
$R(r^{}_{o})\equiv \left[ \left( r^{2}_{o} + a^{2} \right) E - a L^{}_{z} \right]^{2} - 
\left(Q+r^{2}_{o}\right)\Delta(r^{}_{o})= 0$ (the radial coordinate does not change), and its first radial derivative, 
$R'(r^{}_{o})\equiv (dR(r)/dr)^{}_{r^{}_{o}} = 0$ (the orbit is always at a turning point of the radial motion).  In addition,
the condition $R''(r^{}_{o})<0$ has to be satisfied for the circular orbit to be stable. 
Following~\cite{Hughes:1999bq}, the preservation of the circularity of the orbit along the inspiral 
translates into the following two conditions: $\dot{R} =0$ and $\dot{R}' = 0$.  From these two 
conditions we can express the time evolution of the radius of the orbit and the evolution of
the Carter constant in terms of the evolution of the energy and angular momentum of the orbit:
\be
\left(\begin{array}{c} \dot{C} \\ \dot{r}^{}_{o} \end{array}\right) = -\frac{1}{d}
\left(\begin{array}{cc} c^{}_{11} & c^{}_{12} \\ c^{}_{21} & c^{}_{22} \end{array}\right)
\left(\begin{array}{c} \dot{E} \\ \dot{L}^{}_{z}\end{array}\right)\,,
\label{Cdot-rdot}
\ee
where the coefficients $c^{}_{AB}$ ($A,B=1,2$) and $d$ are functions of $(M^{}_{\MBH},a;E,L^{}_{z},C,r^{}_{o})$
and are given in Appendix~\ref{circular-nonequatorial}.  Therefore, the evolution of $C$ and $r^{}_{o}$ can be obtained
by evaluating $dE/dt$ and $dL^{}_{z}/dt$ from the multipolar, post-Minkowskian self-force. 

To finish this section, and as a consistency check, we take the Newtonian limit of the rate of change of the 
orbital elements in Eqs.~\eqref{evol-cons-motion-E}-\eqref{evol-cons-motion-Q}. In this limit, we 
find that to leading order
\ba
\dot{E} &=& a_{\RR}^{t} = v_{i} a_{\RR}^{i}\,, \\
\dot{L}_{z} &=& r^{2} \, \sin^{2}{\theta} \, a_{\RR}^{\phi}  \,, \\
\dot{Q} &=& - 2 \, r^{2} \, \sin^{2}{\theta} \, a \, a_{\RR}^{\phi} \,, \\
\dot{C} &=& - 4 \, M_{\MBH} \, a \,  r \, \cos^{2}{\theta} \sin^{2}{\theta} \, a_{\RR}^{\phi} \,,
\ea
which agrees with the Newtonian expressions of~\cite{Flanagan:2007tv} after transforming to Cartesian 
coordinates. Notice, however, that Eqs.~\eqref{evol-cons-motion-E}-\eqref{evol-cons-motion-Q} contain 
many more terms than in~\cite{Flanagan:2007tv}, due to the fact that in the latter the $\Gamma$ factor is
expanded in the low-speed and weak-field limit.

\subsection{Multipolar Post-Minkowskian Self-Acceleration}
Here we discuss a method to obtain an approximation for the self-force~\eqref{misataquwa2}.
The most rigorous approach would be to compute this force within the framework of BH perturbation theory, 
but as already argued, such a task is still under development and in computational terms would be 
very costly. A different approach is to extract the self-force from the PN equations of motion. 
Such a path was taken by Pound and Poisson~\cite{Pound:2007th} for a Schwarzschild background
and Gair, et al.~for a Kerr background~\cite{Gair:2010iv}. 
This approach, however, is built under a PN approximation, which is not ideal for EMRI modeling as the 
SCO can reach relativistic velocities, with significant relativistic $\Gamma$ factor, 
as it orbits close to the MBH. 

Instead of either of these approaches, we here approximate the self-force via a multipolar, post-Minkowskian 
expansion (see e.g.~\cite{Blanchet:1984wm,Iyer:1993xi,Iyer:1995rn}):
\ba
\met_{00}^{\PM} &=& -1 + 2 V - 2 V^{2} + 2 V^{}_{\RR} + {\cal{O}}(G^{9/2})\,,
\label{metricPM00} \\
\met_{0i}^{\PM} &=& -4 V^{i} - 4 V^{i}_{\RR} + {\cal{O}}(G^{9/2})\,,
\label{metricPM0i} \\
\met_{ij}^{\PM} &=& \delta_{ij} \left[1 + 2 V + 2 V^{}_{\RR} + {\cal{O}}(G^{4}) \right]\,, \label{metricPMij}
\ea
where $V$ and $V^{i}$ are {\emph{time-symmetric}} potentials, in the sense that they are made out of
the half-sum of retarded and advanced integrals of the stress-energy tensor over the source, i.e.~the 
half-sum of the retarded and advanced Green functions associated with the perturbative equations. 
As a consequence, these potentials are invariant under time inversion.
The quantities $V^{}_{\RR}$ and $V_{\RR}^{i}$ are {\emph{time-asymmetric}} radiation-reaction potentials, 
constructed from the half-difference of retarded and advanced waves, and hence, odd under time
inversion.  These potentials are given by~\cite{Blanchet:1984wm,Blanchet:1996vx} 
\ba
V^{}_{\RR}(t^{}_{\HAR},\bm{x}^{}_{\HAR}) &=&  - \frac{1}{5} x^{ij}_{\HAR} M_{ij}^{(5)}(t_{\HAR}) 
+  \frac{1}{189} x^{ijk}_{\HAR} M_{ijk}^{(7)}(t_{\HAR}) \nonumber \\
& - & \frac{1}{70} \bm{x}^{2}_{\HAR} x^{ij}_{\HAR} M_{ij}^{(7)}(t_{\HAR}) + {\cal{O}}(x_{ij}^{\HAR} M_{ij}^{(9)})\,, 
\label{scalar_rr_potential} \\
V^{i}_{\RR}(t^{}_{\HAR},\bm{x}^{}_{\HAR}) &=& \frac{1}{21} {x}^{<ijk>}_{\HAR} M_{jk}^{(6)}(t_{\HAR}) \nonumber \\
& - & \frac{4}{45} \epsilon_{ijk} x^{jl}_{\HAR} S_{kl}^{(5)}(t_{\HAR}) + {\cal{O}}(x^{ijk}_{\HAR} M_{ij}^{(8)})\,,
\label{vector_rr_potential}
\ea
where we recall that $(x^{\alpha}_{\HAR}) = (t^{}_{\HAR},x^i_{\HAR})$ are spacetime harmonic coordinates, 
$\epsilon^{}_{ijk}$ is the antisymmetric Levi-Civita symbol, and 
\be
\hat{x}^{<ijk>}_{\HAR} \equiv x^{ijk}_{\HAR} - \frac{3}{5}\bm{x}^2_{\HAR} \delta^{(ij}_{}x^{k)}_{\HAR}\,.
\ee 
is the symmetric trace-free (STF) projection of the multi-index quantity $x^{ijk} = x^{i} x^{j} x^{k}$. 
The first term in $V^{}_{\RR}$ [Eq.~\eqref{scalar_rr_potential}] 
corresponds to the well-known Burke-Thorne radiation-reaction potential~\cite{Burke:2010wb}:
\be
V^{}_{\rm Burke-Thorne}(t^{}_{\HAR},\bm{x}^{}_{\HAR}) = - \frac{1}{5} x^{ij}_{\HAR} M_{ij}^{(5)}(t_{\HAR}) \,.
\label{burke-thorne-potential}
\ee

The quantities $M_{ij}^{(n)}$, $M_{ijk}^{(n)}$, 
and $S_{ij}^{(n)}$ are the $n$th-time-derivative of the STF mass 
quadrupole, mass octopole and current quadrupole moments. Formally, the radiation-reaction
potentials depend on the integral of certain derivatives of the asymmetric sum (half-difference of retarded and advanced
waves) of multipole moments (see, e.g.~Eqs.~$(2.8)$ in~\cite{Blanchet:1996vx}). Equation~\eqref{vector_rr_potential}, however,
is obtained by expanding these integrals in a slow-velocity approximation, after which the arguments of the
radiation-reaction potentials depend on time only. This is consistent with the fact that the radiation-reaction force
is to be evaluated in the source zone of the SCO, and not in the wave zone.

Let us provide explicit expressions for these multipole moments. 
To lowest order, the mass moments are given by 
\be
M^{}_{ij} = \eta\, m \; z^{}_{<ij>}\,,
\qquad
M^{}_{ijk} = \eta\, \delta m  \; z^{}_{<ijk>}
\label{mass-moments}
\ee
and the current moment is
\be
S_{ij} = \eta\,\delta m\; \epsilon^{}_{kl <i} z_{j>}{}^{k} \dot{z}^{l}\,,
\label{current-moments}
\ee
where angle-brackets are STF projections. To higher order, these moments become more complicated 
as there are now nonlinear contributions from the nonreactive potentials (i.e.~nonlinear 
contributions from the background) as well as tail and memory contributions. These expressions can 
be found for example in Eq.~$(3.1)$--$(3.3)$ of~\cite{Arun:2007sg} and Eq.~$(5.3)$--$(5.5)$ 
of~\cite{Blanchet:2006gy}. In BH perturbation theory language, these higher-order terms would 
contribute conservative and dissipative corrections to the dissipative equations of motion. 
We neglect these contributions in the current version of the new kludge approach, but they can 
be easily incorporated in future improvements of the scheme.

The metric given in Eqs.~\eqref{metricPM00}-\eqref{metricPM0i} is expanded in the far-field limit, a resummation of which 
is necessary in order to use it for self-force calculations. All terms that are independent of 
the radiation-reaction potentials can be identified with MBH background geometry terms, 
corrected perhaps by the presence of the SCO. Neglecting the latter, we can {\emph{exactly resum}} 
the metric in Eqs.~\eqref{metricPM00}-\eqref{metricPMij} so that it can written as follows
\ba
\met^{\PM}_{tt} &=& \met_{tt}^{\KERR,\HAR} + h^{\RR}_{tt} + {\cal{O}}(G^{9/2})\,, \\
\met^{\PM}_{ti} &=& \met_{ti}^{\KERR,\HAR} + h^{\RR}_{ti}  + {\cal{O}}(G^{9/2})\,, \\
\met^{\PM}_{ij} &=& \met_{ij}^{\KERR,\HAR} + h^{\RR}_{ij} + {\cal{O}}(G^{4})\,,
\label{resummed-metric}
\ea
where $\met_{\mu \nu}^{\KERR,\HAR}$ is the Kerr metric in harmonic coordinates (we shall discuss the issue of 
coordinates in more detail in Sec.~\ref{coord-sec}) and where we have introduced the  
metric perturbation, $h^{\RR}_{\mu\nu}$, which is given by
\be
h^{\RR}_{tt} = 2\, V^{}_{\RR} \,,~~
h^{\RR}_{ti} = - 4\, V^{i}_{\RR}\,,~~
h^{\RR}_{ij} = 2\,\delta^{}_{ij} V^{}_{\RR} \,. \label{rrperturbations}
\ee
These are the metric perturbations that are going to describe the \emph{radiation-reaction} effects in our new kludge
scheme, and hence they are our approximation to the regularized metric perturbations $h^{\REG}_{\alpha\beta}$
in Eqs.~\eqref{misataquwa1} and~\eqref{misataquwa2}.

Regarding the MBH background geometry, it is useful for the purposes of this work to introduce 
some scalar, vector, and tensor {\em potentials} in harmonic coordinates, which contain information equivalent to the potentials
$V$ and $V^{i}$ in Eqs.~\eqref{metricPM00}-\eqref{metricPMij}:
\be
K \equiv K^{}_{00} = \met_{00}^{\KERR,\HAR} + 1\,, \quad 
Q \equiv Q^{00} = \met^{00}_{\KERR,\HAR} + 1\,, \label{scalar-local-potentials}
\ee
\be
K^{}_{i} \equiv K_{0i} = \met_{0i}^{\KERR,\HAR}\,, \quad
Q^{i} \equiv Q^{0i} = \met^{0i}_{\KERR,\HAR}\,, \label{vector-local-potentials}
\ee
\be
K^{}_{ij} \equiv \met_{ij}^{\KERR,\HAR} - \delta^{}_{ij}\,, \quad
Q^{ij} \equiv \met^{ij}_{\KERR,\HAR} - \delta^{ij}\,, \label{tensor-local-potentials}
\ee
or more compactly $K^{}_{\mu \nu} \equiv \met_{\mu \nu}^{\KERR,\HAR} - \eta^{}_{\mu \nu}$
and $Q^{\mu \nu} \equiv \met^{\mu \nu}_{\KERR,\HAR} - \eta^{\mu \nu}$.   The expressions from
these local potentials can be derived directly from the expressions of the components of the
Kerr metric in harmonic coordinates given in Appendix~\ref{app:KerrMetricHar}.

We are now in a position to compute the self-acceleration in terms of the radiation-reaction potentials
of Eqs.~\eqref{scalar_rr_potential} and~\eqref{vector_rr_potential} through Eq.~\eqref{rrperturbations}
and Eq.~\eqref{misataquwa2}.  In what follows we develop the expression for the acceleration in order
to describe its structure and the role of the different terms that appear on it.  The first step is to use
the decomposition of the SCO four-velocity given in Eqs.~\eqref{sco-4velocity} and~\eqref{gammafactor-and-spatial-velocity}
in such a way as to rewrite Eq.~\eqref{misataquwa2} in the following form:
\be
a^{\alpha}_{\RR} = - \Gamma^{2} P^{\alpha \beta} \left( A^{(1)}_{\beta} + A^{(2)}_{\beta} \right)\,,
\label{full-acc}
\ee
where we have factored out the dependence on the relativistic $\Gamma$ factor [see Eq.~\eqref{gammafactorwithpotentials} for
the expression of $\Gamma$ in terms of the potentials $K^{}_{\mu\nu}$ and the spatial velocity $v^{i}$]
$P^{\alpha\beta}$ is the projector orthogonal to the SCO four-velocity, 
$P^{\alpha\beta} = \met^{\alpha\beta}_{\KERR}+u^{\alpha}u^{\beta}$ (in terms of the potentials 
$Q^{\mu\nu}$ it is given in Eq.~\eqref{orthogonalprojector-Qs}).  The two pieces of the self-acceleration
in Eq.~\eqref{full-acc} contain different terms.  The first piece, $A^{(1)}_{\alpha}$, only contains gradients
of the radiation-reaction potentials and the spatial velocity $v^{i}$, whereas the second piece,
$A^{(2)}_{\alpha\beta}$, contains explicit couplings between the MBH potentials (actually their derivatives) 
and the radiation-reaction potentials.  The form of these two terms is the following:
\be 
A^{(1)}_{\alpha} = {\cal G}^{\RR}_{\mu\nu\alpha}\,v^{\mu}v^{\nu}\,, \label{a-piece-1}
\ee
where
\be
{\cal{G}}^{\RR}_{\mu \nu \alpha} \equiv \frac{1}{2} \left( \partial^{}_{\mu} h^{\RR}_{\nu \alpha} 
+  \partial^{}_{\nu} h^{\RR}_{\mu \alpha} -  \partial^{}_{\alpha} h^{\RR}_{\mu \nu} \right)\,,  
\label{G-tensor}
\ee
and
\be
A^{(2)}_{\alpha} = -h^{\RR}_{\alpha\beta}\,\Gamma^{\beta}_{\mu\nu}\,v^{\mu}v^{\nu}\,, 
\label{a-piece-2}
\ee
where the connection is to be computed with the Kerr background metric in harmonic coordinates. 
Putting all these different ingredients together and
separating space and time components, we arrive at the following structure for the piece $A^{(1)}_{\alpha}$:
\be
A^{(1)}_{t} = {\cal A}^{\RR}\,, \qquad 
A^{(1)}_{i} = {\cal A}^{\RR}_{i} \,, \label{acc1}
\ee
where the expressions for the quantities ${\cal A}^{\RR}$ and ${\cal A}^{\RR}_{i}$, which depend only on 
the three-velocity $v^{i}$ and  spacetime derivatives of the radiation-reaction potentials $V^{\RR}$ and 
$V^{\RR}_{i}$, are given in Eqs.~\eqref{calaRR} and~\eqref{calaiRR} of Appendix~\ref{app-coeffs}.  
The second piece of the self-acceleration, $A^{(2)}_{\alpha}$, has the following structure:
\ba
A^{(2)}_{t} & = & \left[ {\cal{B}}^{}_{\RR} + {\cal{C}}^{}_{\RR} + {\cal{D}}^{}_{\RR} \right] V^{}_{\RR} \nonumber \\
&+& \left[ {\cal{B}}_{\RR}^{i} + {\cal{C}}_{\RR}^{i} + {\cal{D}}_{\RR}^{i} \right] V^{\RR}_{i} \,,
\label{acc2t}
\ea
\ba
A^{(2)}_{i} &=& -2 \left[ {\cal{B}}_{\RR} + {\cal{C}}_{\RR} + {\cal{D}}_{\RR} \right] V^{\RR}_{i} \nonumber \\
&-& \frac{1}{2}\delta^{}_{ij} \left[ {\cal{B}}_{\RR}^{j} + {\cal{C}}_{\RR}^{j} + {\cal{D}}_{\RR}^{j} \right] V^{}_{\RR} \,,
\label{acc2i}
\ea
where  the quantities ${\cal{B}}^{}_{\RR}$, ${\cal{B}}_{\RR}^{i}$, ${\cal{C}}^{}_{\RR}$, 
${\cal{C}}_{\RR}^{i}$, ${\cal{D}}^{}_{\RR}$, and ${\cal{D}}_{\RR}^{i}$, which depend on the SCO three-velocity $v^{i}$
and the MBH potentials $(K,K^{}_{i},K^{}_{ij})$ and $(Q,Q^{i},Q^{ij})$ and their spatial derivatives, 
are all given explicitly in 
Eqs.~\eqref{B-RR-eq}-\eqref{Di-RR-eq} of Appendix~\ref{app-coeffs}. 

One can check that in the limit ${\met}_{\mu \nu}^{\KERR,\HAR} \to \eta_{\mu \nu}$ (or equivalently, 
$(K,K^{}_{i},K^{}_{ij})\to (2V(1-V),-4V^{}_{i},2V\delta^{}_{ij})$ and $|v| \ll 1$ (in units $c=1$), the spatial components 
of the acceleration in Eq.~\eqref{full-acc} reduce exactly to Eq.~$(3.11)$ in~\cite{Iyer:1993xi,Iyer:1995rn} order by order. 
This can be checked by realizing that the self-acceleration here, Eq.~\eqref{full-acc}, is related to the self-acceleration 
in~\cite{Iyer:1993xi,Iyer:1995rn}, let us call it $a^{i}_{\rm IW}$, via 
\be
a^{i}_{\rm IW} = \frac{d^{2}z^{i}}{dt^{2}} = \frac{dv^{i}}{dt} = \Gamma^{-2}\left( a^{i}_{\RR} - v^{i} a^{t}_{\RR}
\right)\,.
\ee
%
 
\subsection{From Boyer-Lindquist to Harmonic Coordinates}
\label{coord-sec}

The resummed MBH background metric in Eq.~\eqref{resummed-metric}, the background Kerr metric, 
must be written in harmonic coordinates in order for all terms to be in the same coordinate
system and to be consistent with other ingredients of the new kludge scheme, like the
waveform generation procedure described in Sec.~\ref{waveform-generation}.  However, there are parts of this
scheme that are easier to implement in Boyer-Lindquist coordinates, such as the integration of the
geodesic equations.  Therefore, it is crucial to determine out how to transform from Boyer-Lindquist to harmonic 
coordinates. 

Harmonic coordinates refers to any member of the family of coordinate systems, say $\{x^{\alpha}_{\HAR}\}$, 
that satisfy the equation
\be
\square\, x^{\alpha}_{\HAR} = 0\,, 
\label{har-coor-cond}
\ee
where $\square$ is the D'Alembertian operator: $\square\equiv \met^{\alpha\beta}\nabla^{}_{\alpha}\nabla^{}_{\beta}$.  
When we write down the D'Alembertian operator in this coordinate system, this condition is equivalent to the
condition: $\met^{\alpha\beta}_{\HAR}\Gamma^{\mu,\HAR}_{\alpha\beta} = 0$, 
which in turn is equivalent to requiring $\partial_{\beta} {\textgoth{g}}^{\alpha \beta}_{\HAR} = 0$, where 
${\textgoth{g}}^{\alpha \beta}_{\HAR} = \sqrt{-g^{}_{\HAR}} \; 
\met^{\alpha \beta}_{\HAR}$. Notice that the harmonic coordinate condition
and the harmonic gauge condition are different things.  The first expresses a property of a given coordinate
system in a given spacetime, whereas the second refers to the correspondence between the background and
perturbed spacetimes in perturbation theory.  In this sense, the gauge condition 
$\partial^{}_{\mu}h^{\mu\nu} = 0$, where $h^{\mu\nu} ={\textgoth{g}}^{\alpha \beta}-\eta^{\alpha\beta}$,
enforces harmonic coordinates at first order in perturbations around flat spacetime. 
Although the Lorenz gauge  condition~Eq.~\eqref{boxh} is similar to the harmonic gauge condition, 
these conditions are in general different, coinciding only for perturbations around Minkowski spacetime.

The most commonly employed coordinate systems in GR to describe BH geometries are Schwarzschild or Boyer-Lindquist 
coordinates, neither of which is actually harmonic. Although the time and azimuthal Boyer-Lindquist coordinates, 
$t$ and $\phi$, satisfy Eq.~\eqref{har-coor-cond}, the radial and polar coordinates, $r$ and $\theta$, do not. 
One can further check that Eddington-Finkelstein and Kerr-Schild coordinates are also nonharmonic. 
Of course, if one considers a vacuum, Minkowski spacetime, then Eq.~\eqref{har-coor-cond} becomes trivial, 
$\partial^{\alpha} \partial_{\alpha} x^{\mu} = 0$, which is generically satisfied by many
coordinate systems. 

Harmonic coordinates for the Kerr metric have been studied extensively in the literature. 
A harmonic slicing of the Kerr metric was found in~\cite{Cook:1997qc}, where the time-function $T$
is related to Boyer-Lindquist coordinates via
\be
T = t +\frac{r^{2}_{+}+a^{2}}{r^{}_{+}-r^{}_{-}}\ln\left|\frac{r-r^{}_{+}}{r-r^{}_{-}} \right|\,,
\ee
where $r^{}_{\pm}$ are given in Eq.~\eqref{rpm}. Although the slicing is harmonic, the full set of four-dimensional
coordinates is not. Ref.~\cite{Ruiz:1986re} did find a full set of harmonic coordinates for Kerr but no 
explicit expressions for the metric components were given.  In~\cite{1987PThPh..78.1186A}, based on 
work by Ding (see Ref.~\cite{1987PThPh..78.1186A} for references on this), a different transformation 
to harmonic coordinates was found, where the time slicing was not modified. This is very convenient for
the new kludge scheme, which is why we adopted this choice in this paper. 
  
Following~\cite{1987PThPh..78.1186A}, we can map the Kerr metric from Boyer-Lindquist 
coordinates $(t,r,\theta,\phi)$ to harmonic  ones $(t^{}_{\HAR},x^{}_{\HAR},y^{}_{\HAR},z^{}_{\HAR})$ 
via the coordinate transformation
\be
t^{}_{\HAR} = t\,, \label{tHAR-from-BL}
\ee
\be
x^{}_{\HAR} = \sqrt{\left(r - M_{\MBH}\right)^{2} + a^{2}} \; \sin{\theta} \cos[\phi - \Phi(r)]\,,
\label{xHAR-from-BL}
\ee
\be 
y^{}_{\HAR} = \sqrt{\left(r - M_{\MBH}\right)^{2} + a^{2}} \; \sin{\theta} \sin[\phi - \Phi(r)]\,,
\label{yHAR-from-BL}
\ee
\be
z^{}_{\HAR} = \left(r - M_{\MBH}\right) \; \cos{\theta}\,. 
\label{zHAR-from-BL}
\ee
while the inverse transformation is
\be
t = t^{}_{\HAR}\,,   
\qquad
\phi = \Phi(r) + \arctan\left(\frac{y^{}_{\HAR}}{x^{}_{\HAR}}\right)\,, \label{tphiBL-from-HAR}  
\ee
\be
r = M_{\MBH}+ \frac{1}{\sqrt{2}} \left[ r_{\HAR}^2 - a^2 + \sqrt{\left(r_{\HAR}^2 
- a^{2}\right)^{2} + 4 a^2 z_{\HAR}^2 }\right]^{1/2}\,, \label{rBL-from-HAR}  
\ee
\be
\theta = \arccos\left(\frac{z^{}_{\HAR}}{r-M_{\MBH}}\right)\,,
\label{thetaBL-from-HAR}
\ee
We have here introduced the short-hand notation 
\be
r_{\HAR}\equiv(x_{\HAR}^{2} + y_{\HAR}^{2} + z_{\HAR}^{2})^{1/2}\,, \label{rHAR}
\ee
and the angle function $\Phi(r)$, which is given by 
\be
\Phi(r) = \frac{\pi}{2} - \arctan\left\{\frac{\frac{r-M_{\MBH}}{a} + 
\Omega(r)}{1 - \frac{r-M_{\MBH}}{a}\,\Omega(r)}\right\}\,,  \label{Phi-solved}
\ee
with
\ba
\Omega(r)  = \tan\left[\frac{a}{2\sqrt{M_{\MBH}^2-a^2}}\ln
\left(\frac{r-r^{}_{-}}{r-r^{}_{+}}\right)\right]\,. \label{omega-of-r}
\ea
The coordinate transformation between Boyer-Lindquist and 
harmonic coordinates requires expressions for $(\cos\Phi,\sin\Phi)$ and the Jacobian and Hessian of 
the transformation, which we provide in Appendix~\ref{app-blkerrinharmonic}.

We have checked that Eq.~\eqref{har-coor-cond} is satisfied identically for 
$(x^{\alpha}_{\HAR}) = (t^{}_{\HAR},x^{}_{\HAR},y^{}_{\HAR},z^{}_{\HAR})$ in the Kerr background. 
Moreover, one can see that this transformation reduces to the standard one from Schwarzschild to 
harmonic coordinates~\cite{poisson-lecture-notes} in the $a \to 0$ limit.  In this
limit, the transformation~\eqref{xHAR-from-BL}-\eqref{zHAR-from-BL} reduces to the 
Euclidean transformation from spherical to Cartesian coordinates, with $r^{}_{\HAR}$ playing
the role of the spherical coordinate, related to the radial coordinate $r$ by $r^{}_{\HAR}=r-M^{}_{\bullet}$.

This explicit coordinate transformation allows us to compute certain ingredients of the new kludge
scheme in Boyer-Lindquist coordinates and then transform the result to harmonic coordinates when needed. 
For instance, the Kerr metric in harmonic coordinate is then simply 
\be
\met_{\mu \nu}^{\KERR,\HAR} = \met_{\rho \sigma}^{\KERR,\BL} \; \frac{\partial x^{\rho}_{\BL}}{\partial x^{\mu}_{\HAR}} 
\frac{\partial x^{\sigma}_{\BL}}{\partial x^{\nu}_{\HAR}}
\ee
where the Jacobian $\partial x^{\rho}_{\BL}/{\partial x^{\mu}_{\HAR}}$ is given in Appendix~\ref{app:Jacob} 
and the transformation is shown explicitly  in Appendix~\ref{app:KerrMetricHar}. 

In particular, in the current implementation of the new kludge scheme, we choose to perform the quasigeodesic evolution
using variables associated with Boyer-Lindquist coordinates (see Sec.~\ref{num-implementation}). 
Then, from the information of the trajectory in Boyer-Lindquist coordinates we can compute
the velocity and accelerations in harmonic coordinates using the following relations
\ba
\bm{\dot{r}^{}_{\HAR}} &=& \frac{\partial \bm{r^{}_{\HAR}}}{\partial x^{i}_{\BL}}\; \dot{x}^{i}_{\BL} 
= \frac{\partial \bm{r^{}_{\HAR}}}{\partial r}\;\dot{r} + 
\frac{\partial \bm{r^{}_{\HAR}}}{\partial \theta}\;\dot{\theta} +
\frac{\partial \bm{r^{}_{\HAR}}}{\partial \phi}\;\dot{\phi}\,,
\label{BL-vel}
\\
\bm{\ddot{r}^{}_{\HAR}} & = & \frac{\partial \bm{r^{}_{\HAR}}}{\partial x^{i}_{\BL}}\; \ddot{x}^{i}_{\BL}
+  \frac{\partial^2 \bm{r^{}_{\HAR}}}{\partial x^{i}_{\BL}\,\partial x^{j}_{\BL}}\; 
\dot{x}^{i}_{\BL}\, \dot{x}^{j}_{\BL} \nonumber \\
& = & \frac{\partial \bm{r^{}_{\HAR}}}{\partial r}\;\ddot{r} + 
\frac{\partial \bm{r^{}_{\HAR}}}{\partial \theta}\;\ddot{\theta} +
\frac{\partial \bm{r^{}_{\HAR}}}{\partial \phi}\;\ddot{\phi} \nonumber \\
& + & \frac{\partial^2 \bm{r^{}_{\HAR}}}{\partial^2 r}\; \dot{r}^2 + \frac{\partial^2 \bm{r^{}_{\HAR}}}{\partial^2 \theta}\; \dot{\theta}^2 
+ \frac{\partial^2 \bm{r^{}_{\HAR}}}{\partial^2 \phi}\; \dot{\phi}^2 \nonumber \\
& + & 2\left( \frac{\partial^2 \bm{r^{}_{\HAR}}}{\partial r\,\partial \theta}\; \dot{r}\,\dot{\theta}
+ \frac{\partial^2 \bm{r^{}_{\HAR}}}{\partial r\,\partial \phi}\; \dot{r}\,\dot{\phi}
+ \frac{\partial^2 \bm{r^{}_{\HAR}}}{\partial \theta\,\partial \phi}\; \dot{\theta}\,\dot{\phi}\right)\,.
\label{BL-acc}
\ea
where the Hessian ${\partial^2 \bm{r^{}_{\HAR}}}/{\partial x^{i}_{\BL}\,\partial x^{j}_{\BL}}$ is given 
in Appendix~\ref{app:Hessian}. Notice that we only need the spatial components of the self-acceleration in 
Eqs.~\eqref{evol-cons-motion-E}-\eqref{evol-cons-motion-Q}. With this acceleration, we can then compute
the change in the constants of the motion as we osculate from one geodesic to the next. 

On the other hand, thinking about extending the new kludge scheme to systems in which the central
object, the MBH in our case, is not described by the Kerr metric but by a different metric tensor
(either because the central object corresponds to an exotic distribution of mass or because it is governed 
by an alternative theory of gravity, or perhaps both), it may happen that transforming to harmonic coordinates
may be a very difficult task, even when given an explicit coordinate transformation as the one above 
is known (although in this case one may resort to numerical computations). 
Fortunately, one does not need a full 
coordinate transformation to read off the multipole moments of the background. 
Thorne~\cite{Thorne:1980rm} has shown that it suffices to work with so-called 
\emph{asymptotically Cartesian and mass-centered coordinates of order $N$} (ACMC-$N$). 
These coordinates are defined such that the background metric is time-independent and has 
a certain spherical harmonic structure (see, e.g.~Eq.$(11.1a)$ in~\cite{Thorne:1980rm}). 
Thorne has further shown that harmonic coordinates are ACMC-$\infty$, and found an explicit 
map between Boyer-Lindquist and ACMC-$2$ coordinates~\cite{Thorne:1980rm}. In 
Appendix~\ref{app-coord-details} we construct an ACMC-$4$ coordinate system and compare it 
to the exact harmonic coordinates described in this section. The difference between a 
standard ACMC-$4$ and its harmonic generalization appears only near the horizon, and in 
particular, it should affect results when PN corrections to the multipole moments are taken 
into account. We work here directly in harmonic coordinates as given by 
Eqs.~\eqref{tHAR-from-BL}-\eqref{zHAR-from-BL} and explained in detail in 
Appendix~\ref{app-blkerrinharmonic}.

\subsection{Waveform Generation}\label{waveform-generation}

Once the orbital evolution, including self-force back-reaction, has been computed, 
one can construct the resulting GWs. Expressions for these as a function of the 
trajectories can be obtained by solving Eq.~\eqref{evol-eq-hab} either numerically or 
analytically, via some approximation scheme.   Here we use the solution for the waveforms
that comes from the combination of post-Minkowskian and PN expansions and which gives rise
to standard multipolar expressions (see e.g.~\cite{Blanchet:2002av} for a review). In this
way we can write the plus- and cross-polarized solution 
to Eq.~\eqref{evol-eq-hab} as an infinite series expansion in terms of derivatives of mass 
and current multipole moments~\cite{Thorne:1980rm,Kidder:2007rt}:
\be
h_{+} - i\, h_{\times} = \frac{1}{\sqrt{2}} \sum_{\ell,m} \left[U^{\ell m}(t_{r}) - i 
V^{\ell m}(t_{r})\right] {}^{}_{-2}Y^{\ell m},\
\label{formal-PN-exp}
\ee
where $t_{r}$ denotes retarded time and $U^{\ell m}$ and $V^{\ell m}$ are {\emph{radiative}} 
mass and current multipole moments:
\ba
U^{\ell m} &\equiv& \frac{16 \pi}{(2 \ell +1)!!} \sqrt{\frac{(\ell+1) (\ell+2)}
{2 \ell (\ell-1)}}\, {\cal{U}}^{}_{L}\, {\cal{Y}}_{L}^{\ell m*}\,, \\
V^{\ell m} &\equiv& \frac{-32 \pi \ell}{(2 \ell +1)!!} \sqrt{\frac{(\ell+2)}
{2 \ell (\ell+1) (\ell - 1)}}\, {\cal{V}}^{}_{L} \,{\cal{Y}}_{L}^{\ell m*}\,,
\ea
where $({\cal{U}}_{L},{\cal{V}}_{L})$ are symmetric, trace-free, mass-type and current-type 
multipole moment tensors (see, e.g.~Eq.$(15)$ in~\cite{Kidder:2007rt}). Even for the case of 
comparable-mass BH mergers, such an expansion taken only up to quadrupole order ($l=2$) has 
been shown to be sufficient to recover the waveform to excellent accuracy~\cite{Buonanno:2006ui}. 

This prescription is then complete, once expressions for the radiative moments are given in 
terms of derivatives of the orbital trajectories. Such identification, however, is difficult, 
as these moments are defined in the {\emph{far-zone}} (many gravitational wavelengths away 
from the center of mass of the binary), and have no knowledge of the {\emph{source}} multipole 
moments, defined in the {\emph{near-zone}} (less than a gravitational wavelength from the 
center of mass). Asymptotic matching can be used to relate the radiative to the source 
multipole moments~\cite{Blanchet:2002av}, yielding explicit expressions that depend only 
on derivatives of the orbital trajectories in the near-zone. 
Here, in this initial version of the new kludge scheme, we only consider just the leading order contributions 
to the multipoles moments, i.e.~we identify the source and radiative moments: 
${\cal{U}}^{}_{L} = M^{}_{L}$ and ${\cal{V}}^{}_{L} = S^{}_{L}$, ignoring in this way  
subleading corrections that correspond to tail and memory effects.

In a transverse-traceless gauge, one can rewrite the harmonically decomposed metric perturbation 
in the following simpler form:
\ba
h_{ij}^{\TT} &=& \sum_{\ell=2}^{\infty} \left[
\frac{4}{\ell!} \frac{1}{r} M_{L-2}^{(\ell)}(t^{}_{r}) N^{}_{L-2}
\right. \nonumber \\
&+& \left. \frac{8 \ell}{(\ell+1)!} \frac{1}{r} \epsilon^{}_{kl(i} S^{}_{j) k L-1}(t^{}_{r}) 
n^{}_{l} N^{}_{L-2} \right]^{\STF}\,,
\ea
which, when we consider only multipoles up to the mass hexadecapole and the current octopole, 
reduces to 
\ba
h_{ij}^{\TT} &=& \frac{2}{r} \ddot{M}_{ij}^{\STF} + \frac{2}{3 r} \left[ \dddot{M}_{ijk} n^{k} 
+ 4 \epsilon^{kl}{}_{(i} \ddot{S}_{j)k} n_{l}\right]^{\STF}  \nonumber \\
&+& 
\frac{1}{6r} \left[ \ddddot{M}_{ijkl} n^{k} n^{l}
+ 6  \epsilon^{kl}{}_{(i} \dddot{S}_{j)km} n^{l} n^{m}\right]^{\STF}\,.
\label{hTT}
\ea
The expressions for these multipole moments have been given in Eqs.~\eqref{mass-moments} and~\eqref{current-moments},
except for the mass hexadecapole and current octopole multipoles, which are given respectively by
\be
M^{}_{ijkl} = \eta\,m\;z^{}_{<ijkl>}\,, \label{mass-hexadecapole}
\ee
\be
S^{}_{ijk} = \eta\,m\;\epsilon^{}_{lm<i}z^{}_{jk>}{}^{l}\,\dot{z}^{m}\,. \label{current-octopole}
\ee
Again, these moments contain higher-order corrections that can also be easily included in subsequent improvements
of the new kludge scheme. The plus and cross-polarized projections can then be constructed via 
\be
h^{}_{+,\times} = e^{ij}_{+,\times} h_{ij}^{\TT},
\label{h+x}
\ee
where $e^{ij}_{+,\times}$ is the plus- and cross-polarization tensors.

Finally, the observables  that one wishes to compute are the GW response functions, which are 
given by a  projection of the plus- and cross-polarized waveform with the beam-pattern functions 
of the detectors.  For a detector like LISA there are two such functions, $F_{+,I/II}$ and $F_{\times,I/II}$ 
(see, e.g.~\cite{Cutler:1998cc,Barack:2003fp} for expressions of these functions) and the response
is thus
\be
h \equiv \frac{\sqrt{3}}{2\,D^{}_{L}} \sum_{A=I,II}^{}\left(F^{}_{+,A}\, h^{}_{+} 
+ F^{}_{\times,A} h^{}_{\times}\right) \,,
\label{response}
\ee
where $D^{}_{L}$ is the luminosity distance from the source to the observer and the prefactor of 
$\sqrt{3}/2$ is due to the triangular arrangement of the LISA detector.

\subsection{Summary of the New Kludge Approach}

The physical quantity one is interested in is the response function, which is
given by Eq.~\eqref{response} in terms of the beam-pattern functions 
(see, e.g.~\cite{Barack:2003fp}) and the plus- and cross-polarized waveform. The latter
is given in Eq.~\eqref{h+x} in terms of the transverse-traceless metric perturbation. 
The first approximation we make is to expand $h_{ij}^{\TT}$ in post-Minkowskian
multipole moments $M_{ij}$, $M_{ijk}$ and $S_{ij}$ 
through Eq.~\eqref{hTT}, where we neglect tail and memory corrections. 
The second approximation we make is to treat these moments
in a Newtonian-like fashion through Eq.~\eqref{mass-moments} and~\eqref{current-moments} 
in terms of the trajectory of the bodies, neglecting post-Newtonian corrections
due to nonlinearities. These two approximations provide the response function as a function of
the trajectory of the bodies in harmonic coordinates. 

The orbital trajectories are obtained by solving the geodesic equations enhanced by time-varying orbital 
elements in Boyer-Lindquist coordinates. The time variation of the orbital
elements is prescribed by the radiation-reaction acceleration in Eqs.~\eqref{acc1}, \eqref{acc2t} and~\eqref{acc2i} 
in harmonic coordinates and in terms of quantities that depend on the harmonic Kerr background  
[with the map given in Eqs.~\eqref{ACMC-H-6-t}-\eqref{ACMC-H-6-phi}] and reactive potentials, in turn given in 
Eq.~\eqref{vector_rr_potential}. To evolve osculating orbits, one must therefore map the Boyer-Lindquist acceleration and
velocity to harmonic coordinates via Eqs.~\eqref{BL-vel}-\eqref{BL-acc}, so as to compute the rates of change of the orbital
elements, which then in turn allows us to map between osculating geodesics. Once the SCO's worldline
has been completely obtained, we can use it in harmonic coordinates in the waveform prescription.

The kludge nature of the approach then becomes clear. We employ a combination of
approximation schemes that include a multipolar, post-Minkowskian expansion (for the far-zone 
metric perturbation and for the local prescription of the self-force) a post-Newtonian expansion (for the multipole moments
in terms of the trajectories) and a BH perturbation theory expansion (when treating
the trajectories as self-adjusting geodesics).  All of this is tied together via a nontrivial
numerical implementation that is described next. 

\section{Numerical Implementation of the New Kludge Scheme}
\label{num-implementation}

In this section we provide details of how we have implemented each of the ingredients of the
new kludge scheme described in the previous section.  The numerical code that we have developed
is written in C language~\cite{Kernighan:1988kr} and uses different parts of the GNU scientific library~\cite{Galasi:2006mg} 
as we describe below.

\subsection{Integration of the Equations of Motion}

We need to integrate numerically the set of ODEs consisting of Eqs.~\eqref{tdot-GR}-\eqref{phidot-GR},
and at the same time we need to update the value of the {\em constants} of motion $(E,L^{}_{z},C/Q)$.
The separation of the geodesic equations has produced equations for the radial and polar 
Boyer-Lindquist coordinates, Eqs.~\eqref{rdot-GR} and~\eqref{thetadot-GR} respectively, that 
have turning points (at pericenter and apocenter in the case of the radial coordinate, and at
the location of the orbital inclination angle in the case of the polar coordinate), and at these
points we have either $\dot{r} = 0$ or $\dot{\theta} = 0$.
This means that these are not the best variables for the numerical integration, as ODE solvers
present convergence problems at turning points. 
 
To avoid this problem, we introduce new
variables in the place of the radial and polar Boyer-Lindquist coordinates, $r$ and $\theta$.
These new coordinates are angle variables defined by the following expressions: 
\be
r = \frac{p M_{\MBH}}{1 + e \cos{\psi}}\,, 
\qquad 
\cos^{2}{\theta} = \cos^{2}{\theta}_{\rm min} \cos^{2}{\chi}\,. \label{ODEanglestoBL}
\ee
We can write the equations for $r$ and $\theta$ in terms of their turning points and 
other extrema (points at which the time derivatives vanish but are not accessible to the
motion).  In the case of the radial motion, we can write the right-hand side of Eq.~\eqref{rdot-GR}
as
\be
(1-E^{2})(r^{}_{\APO}-r)(r-r^{}_{\PERI})(r-r^{}_{3})(r-r^{}_{4})\,, \label{new-rhs-rdot}
\ee
where $r^{}_{3}$ and $r^{}_{4}$ satisfy $r^{}_{\APO}>r^{}_{\PERI}>r^{}_{3}>r^{}_{4}$.  In
the same way, we can write the right-hand side of the equation for the polar motion
[Eq.~\eqref{thetadot-GR}] in the following form
\be
\frac{a^{2}(1-E^{2})}{1-z}\left(z^{}_{+}-z\right)\left(z-z^{}_{-}\right) \,, \label{new-rhs-thetadot}
\ee
where 
\be
z = \cos^{2}\theta\,,\qquad
z^{}_{-} = \cos^{2}\theta^{}_{\MIN} \label {z-def} \,,
\ee
and $z^{}_{+} > z^{}_{-}$.  We describe in Appendix~\ref{app-const-of-motion} the relations between these
extrema, i.e.~$(r^{}_{\APO},r^{}_{\PERI},r^{}_{3},r^{}_{4})$ and $(z^{}_{-}, z^{}_{+})$, and
how to find them.

For convenience, we parametrize the trajectory in terms of the Boyer-Lindquist coordinate time $t$, which is
also a time harmonic coordinate, instead of the proper time $\tau$.  This is done using Eq.~\eqref{tdot-GR}
to rewrite the evolution equations with respect to $t$. In this way, the resulting equations for 
$(\psi(t),\chi(t),\phi(t))$ are (see also, e.g.~\cite{Drasco:2003ky}): 
\begin{widetext}
\be
\frac{d\psi}{dt} = \frac{M^{}_{\MBH}\sqrt{1-E^{2}}\sqrt{\left[p(1-e)-p^{}_{3}(1+e\cos\psi)\right] 
\left[p(1+e)-p^{}_{4}(1+e\cos\psi)\right]}}{(1-e^{2})\left[\Psi(\psi) +a^{2}E\,z(\chi)\right]}\,,
\label{ode-dpsidt}
\ee
\be
\frac{d\chi}{dt} = \frac{\sqrt{a^{2}(1-E^{2})\left(z^{}_{+}-z^{}_{-}\cos^{2}\chi\right)}}
{\Psi(\psi) +a^{2}E\,z(\chi)} \,, \label{ode-dchidt}
\ee
\be
\frac{d\phi}{dt} = \frac{1}{\Psi(\psi) +a^{2}E\,z(\chi)}\left\{ 
\frac{2M^{}_{\MBH}ar(\psi)E}{\Delta(r(\psi))} + \left(\frac{1}{1-z(\chi)} - \frac{a^{2}}{\Delta(r(\psi))}
\right)L^{}_{z} \right\}\,, \label{ode-dphidt}
\ee
\end{widetext}
where we have introduced the following definitions: $p^{}_{3}\equiv r^{}_{3}(1-e)/M^{}_{\MBH}$,
$p^{}_{4}\equiv r^{}_{4}(1+e)/M^{}_{\MBH}$, 
\be
\Psi(\psi) \equiv \left[\frac{(r^{2}(\psi)+a^{2})^{2}}{\Delta(r(\psi))} - a^{2}\right]E 
- \frac{2M^{}_{\MBH}ar(\psi)L^{}_{z}}{\Delta(r(\psi))}\,,
\ee
and $r(\psi)$ and $z(\chi)$ are given through Eq.~\eqref{ODEanglestoBL}.
Therefore, the actual outcome of the numerical integration of the ODEs of Eqs.~\eqref{ode-dpsidt}-\eqref{ode-dphidt} 
is a time series of the three angles $(\psi(t),\chi(t),\phi(t))$, which grows monotonically in time.
The numerical method we use to integrate these ODEs is the Bulirsch-Stoer extrapolation method 
(\cite{Bulirsch:1966bs,*Stoer:1993sb}) as described by~\cite{Ito:1997aj,*Fukushima:1996aj} (see
also~\cite{Press:1992nr}).

\subsection{Estimation of Time Derivatives}

Probably the main challenge in the numerical implementation of the new kludge scheme is the
evaluation of the time derivatives of the different quantities involved.  To understand the
nature of this problem let us focus on the computation of our multipolar, post-Minkowskian self-force.
If we look at the expression for the radiation-reaction potentials $V^{}_{\RR}$ 
and $V^{i}_{\RR}$, Eqs.~\eqref{scalar_rr_potential} and~\eqref{vector_rr_potential},
we realize that we need to compute up to the seventh time derivative of the mass quadrupole
and octopole moments and up to the fifth time derivative of the current quadrupole moment.
But since we also need to compute time derivatives of these potentials [see Eqs.~\eqref{full-acc}
and~\eqref{acc1} and also Eqs.~\eqref{calaRR} and~\eqref{calaiRR}] we also need the
eighth time derivative of the mass quadrupole and octopole moments and the sixth time derivative
of the current quadrupole moment.  

In principle, one could think about computing these derivatives
analytically by using the equations of motion, Eqs.~\eqref{tdot-GR}-\eqref{phidot-GR}.
The problem is that we need the time derivatives of the trajectory in harmonic coordinates, and to 
pass from the ODE angles $(\psi(t),\chi(t),\phi(t))$ to harmonic coordinates we need to first use 
Eq.~\eqref{ODEanglestoBL} to go from these angles to Boyer-Lindquist coordinates, and then 
Eqs.~\eqref{tHAR-from-BL}-\eqref{zHAR-from-BL} to go from Boyer-Lindquist to harmonic coordinates. 
Therefore, we would need to obtain analytically higher time-derivatives of the ODE angles (up to eighth order), 
which involves using the Christoffel symbols of the Kerr metric and several of their derivatives, and also
to differentiate several times Eqs.~\eqref{ODEanglestoBL} and~\eqref{tHAR-from-BL}-\eqref{zHAR-from-BL}.
In practice, this makes the analytical computations unfeasible, even using modern computer 
algebra systems, and even if they were not, one would worry about the reliability of the numerical 
evaluation of the huge expressions that one would result.  

For these reasons, we resort to a numerical evaluation of these derivatives. The
starting point is the fact that we can compute the trajectory, the velocity, and the acceleration
almost directly from the integration of the ODE Eqs.~\eqref{tdot-GR}-\eqref{phidot-GR} and with
a high accuracy.  From there, and using purely analytical expressions, we can directly obtain the
second time derivatives of the mass quadrupole and octopole moments and the first time derivative of
the current quadrupole moment.  Thus, we just need to compute up to the six additional time derivative for the
mass moments (i.e., starting from their second time derivative) and up to the five additional time derivative
for the current quadrupole moment (i.e., starting from its first time derivative).

Computing numerical derivatives, in contrast to numerical integration, is a subtle task 
(see, e.g.~\cite{Press:1992nr}).  For instance, if we consider finite difference formulas for 
the different derivatives, the computation requires the particular combinations of the function we
want to differentiate at points close to the evaluation point, and these combinations are divided by
a power of the offset between the different evaluation points. If one chooses the offset to be too small,
high-order cancellations in the combinations of the function evaluations can occur beyond machine precision, 
yielding a meaningless final result. If instead one chooses the offset to be too big, the function might be evaluated
at points where its behavior is very different from the one near the evaluation points, which in turn can also lead 
to large errors in the numerical derivatives. In many situations one can find an interval of offset values in which the 
high-order derivatives are sufficiently accurate, but such an interval depends on the orbit characteristics and it is not
easy to predict. Although we tried many different finite difference rules (from rules involving a few points
to rules involving more than 20 evaluation points), as well as other generic numerical differentiation techniques
(such as numerical interpolation or Chebyshev differentiation), 
a large amount of fine-tuning that was difficult to predict seemed essential in all cases.

The key point to improve the differentiation algorithm is to realize that the methods we have just discussed are
more general than needed for the EMRI problem. In the latter, one is always dealing with functions with certain
properties that can be exploited to construct a better numerical differentiation method. 
The key feature is that multipole moments are functionals of the
trajectories, which are piecewise timelike bounded Kerr geodesics, and in turn 
can be characterized by three fundamental frequencies (in the generic case, see 
the discussion in Sec.~\ref{MBHgeometryandgeodesics}).  Following~\cite{Drasco:2003ky}, we know
that a general functional of Kerr orbits, let us call it $f[\psi,\chi,\phi](t)$, can
be expanded in a multiple Fourier series of these frequencies, that is
\be
f[\psi,\chi,\phi](t) = \sum^{}_{k,m,n} f^{}_{k,m,n} e^{-i\,
(k\Omega^{}_{r}+m\Omega^{}_{\theta}+n\Omega^{}_{\phi})\,t}\,, \label{multifourier}
\ee
where $(k,m,n)$ are integers running from $-\infty$ to $+\infty$ and $f^{}_{k,m,n}$
are complex coefficients such that $f^{}_{-k,-m,-n} = \bar{f}^{}_{k,m,n}$.  There
are three special cases in which this expansion is simplified: (i) Circular-equatorial
orbits;  (ii) Equatorial noncircular orbits; (iii) Circular nonequatorial orbits. In case (i), 
the Fourier series contains only a single frequency, the azimuthal one $\Omega^{}_{\phi}$. 
In case (ii), there are two independent frequencies, $\Omega^{}_{r}$ and 
$\Omega^{}_{\phi}$. In case (iii), there are also two independent frequencies, but they are 
$\Omega^{}_{\theta}$ and $\Omega^{}_{\phi}$.

Our procedure to estimate time derivatives is then to first fit an expansion like that of Eq.~\eqref{multifourier} 
to the multipole moments that are required using a standard least-square
fitting algorithm, and then to estimate time derivatives via
\be
f^{(N)}[\psi,\chi,\phi](t) =  \sum^{}_{k,m,n}{f}^{N}_{k,m,n} 
 e^{-i\,(k\Omega^{}_{r}+m\Omega^{}_{\theta}+n\Omega^{}_{\phi})\,t}\,, \label{multifourierderivative}
\ee
where
\be
{f}^{N}_{k,m,n} = (-i)^{N}
(k\Omega^{}_{r}+m\Omega^{}_{\theta}+n\Omega^{}_{\phi})^{N}\,f^{}_{k,m,n} \,.
\ee

The Fourier fits to the multipole moments, therefore, play a crucial role in the accuracy of
the high-order derivatives. We carry these fits out by evaluating the function to be fitted 
on a certain number of points along a geodesic piece of the orbit.  As we have already mention, we can compute
analytically the first time derivatives of the multipoles, so the time-dependent functions that
we actually fit are: $M^{(2)}_{ij}(t)$, $M^{(2)}_{ijk}(t)$, and $S^{(1)}_{ij}(t)$ for the
radiation-reaction potentials, and also $M^{}_{ijkl}(t)$ and $S^{}_{ijk}(t)$ for the waveforms.
The parameters that we need to choose for the least-squares fit are: (i) the size in time of
the interval where we fit the function; (ii) the number of points in this interval where the
function is going to be evaluated (i.e.~the number of points to be fitted); and (iii) the
number of harmonics/frequencies that we include in the finite expansion of Eq.~\eqref{multifourier}. 
For choice (i), we take a fixed fraction of the shortest orbital period (i.e.~of the minimum of
$T^{}_{r} = 2\pi/\Omega^{}_{r}$, $T^{}_{\theta} = 2\pi/\Omega^{}_{\theta}$, and 
$T^{}_{\phi} = 2\pi/\Omega^{}_{\phi}$; see Appendix~\ref{fundamental-frequencies-and-periods}). 
For choice (ii), we use between $50-500$ points, 
depending on the case we are dealing with (generic or very particular) and the precision
we want to achieve. For choice (iii), we use between 2 and 5 harmonics of the fundamental
frequencies. Adding more harmonics would increase the accuracy of the derivatives, but we
have empirically found that $5$ harmonics is usually sufficient.

For the practical implementation of the least-squares fit we use the GNU Scientific 
Library~\cite{Galasi:2006mg}.  We have performed a number of experiments for different
types of functions (and also for multipole moments of the orbital trajectory) and we have
found that this technique is very robust and provides very high accuracy even for the highest
derivatives.  For instance, for the sixth time derivative we find typical accuracies of 
one part in $10^{5}$.  Taking into account that the magnitude of the time derivatives of 
the multipole moments decreases significantly with the order of the derivative, this 
accuracy is more than enough for our purposes.  Another important feature of this technique
is that is has appeared to be quite robust with respect to the three choices of parameters
we have discussed above.

\section{Numerical Results}
\label{num-results}

In this section, and in order to illustrate the new kludge scheme, we present some numerical results 
from a numerical code that we have developed to implement new kludges, as well as some
comparison with other results in the literature. The examples
of new kludge evolutions shown here are all for prograde orbits but there are no
obstacle to produce similar results for retrograde orbits.

Let us first consider our proper use of harmonic coordinates, an ingredient of new kludges that is very different
from traditional kludge implementations. In the latter, (see, e.g.~\cite{Babak:2006uv}) harmonic coordinates
are approximated via (Euclidean) Cartesian Boyer-Lindquist coordinates, i.e.~$(x^{}_{\BL},y^{}_{\BL},
z^{}_{\BL}) = (r\sin\theta\cos\phi\,,r\sin\theta\sin\phi,r\cos\theta)$. Such coordinates will differ from 
true harmonic coordinates greatly in the strong-field regime. In turn, this will modify the resulting trajectories and waveforms, as
the proper choice of coordinates is crucial in the calculation of multipole moments, both in radiation-reaction computations 
and in waveform production.  

\begin{figure*}[htb]
\centering
\includegraphics[width=0.45\textwidth]{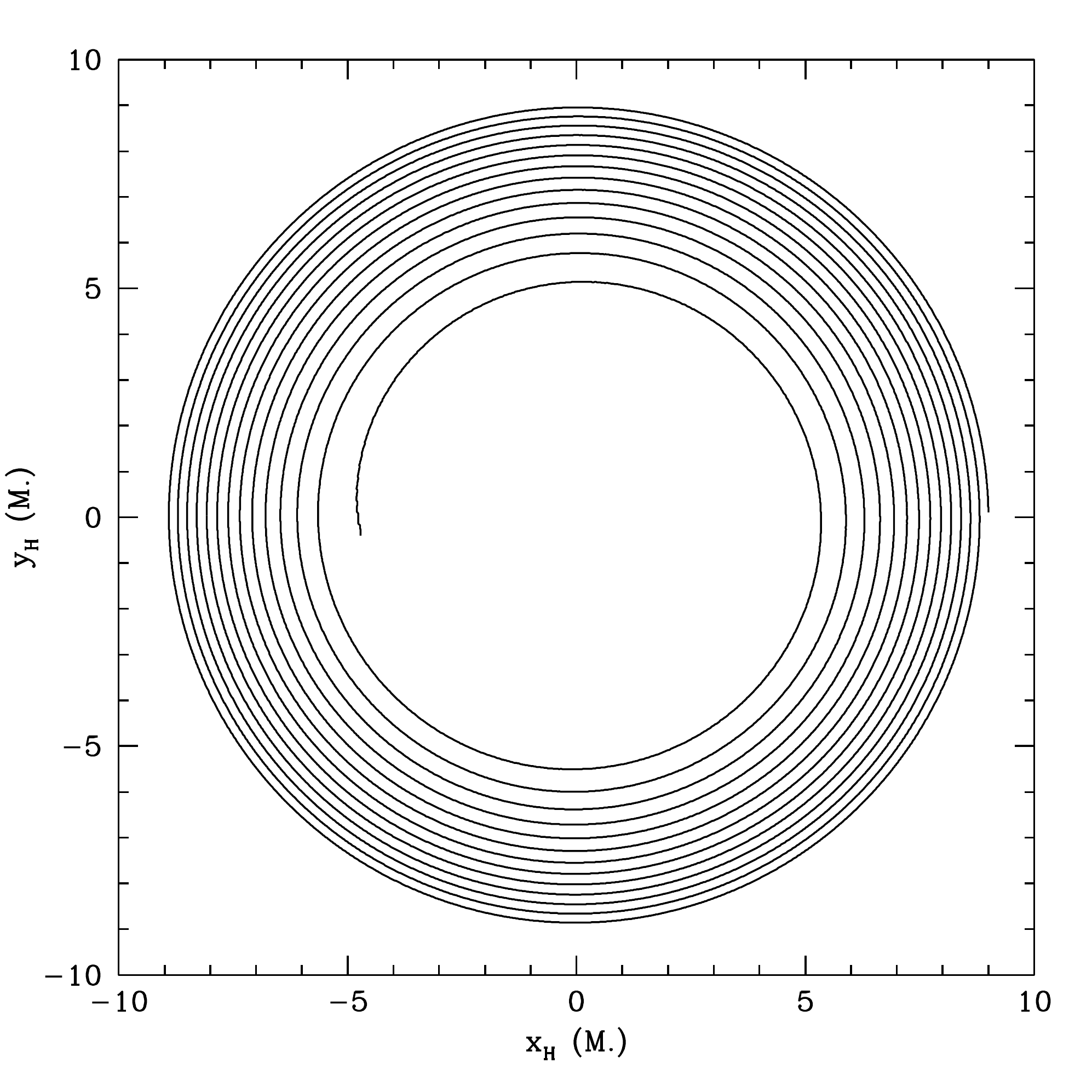}
\includegraphics[width=0.45\textwidth]{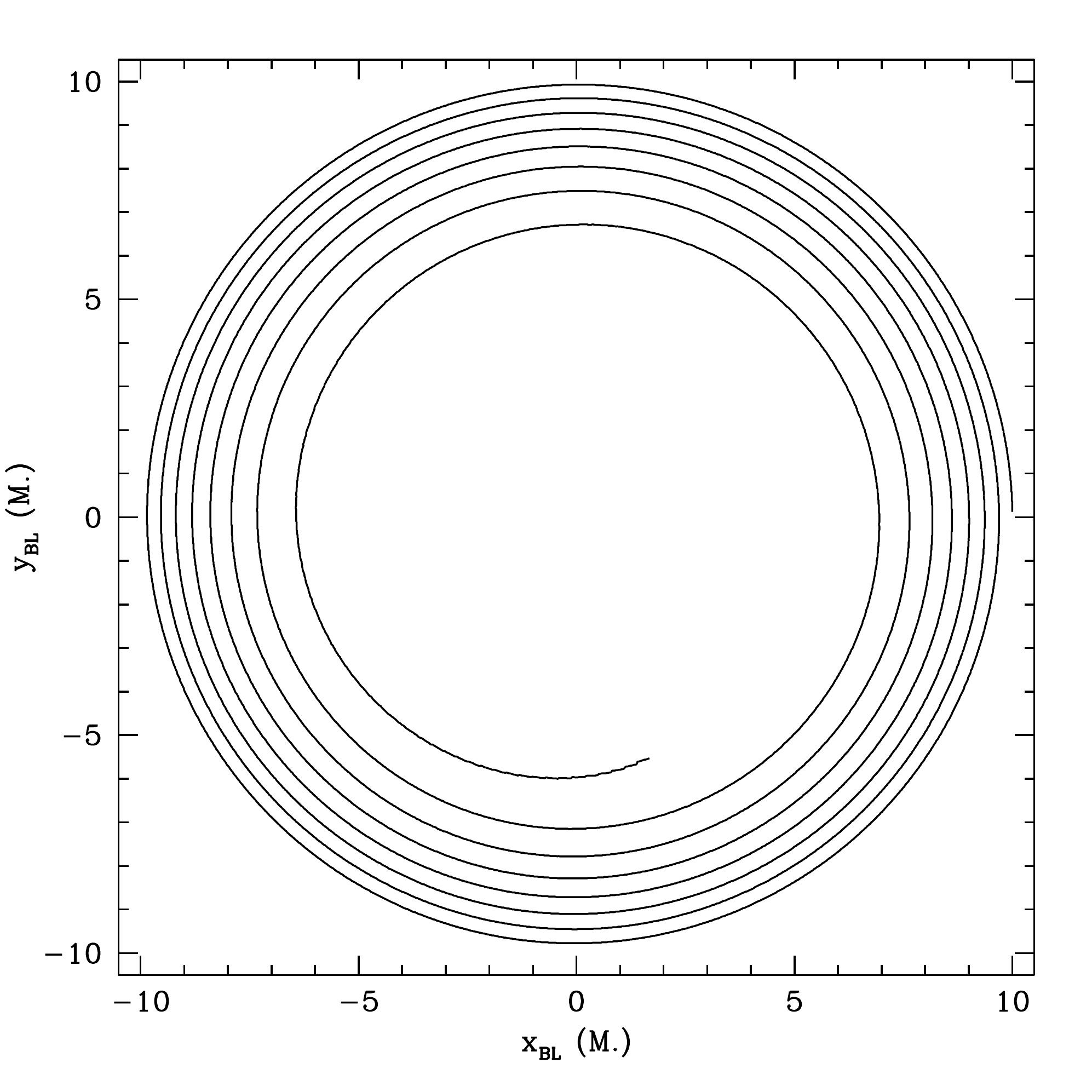}
\caption{Circular and equatorial inspiral of a binary system characterized by $a/M^{}_{\MBH}=0.1$ 
and $q=1/10$.  The inspiral starts at $(x,y)=(10,0)M^{}_{\MBH}$.  The figure on the left shows 
the inspiral using harmonic coordinates and the figure on the right using Cartesian Boyer-Lindquist
coordinates. \label{circular-equatorial-orbit-harmonic}}
\end{figure*}

Let us then compare how much error is introduced by the use of the wrong coordinate system in EMRI waveform construction. 
For this, we employ circular equatorial orbits as they are simpler when comparing the waveform phase. 
Fig.~\ref{circular-equatorial-orbit-harmonic} shows a circular-equatorial inspiral orbit for a system with:
$a/M^{}_{\MBH} = 0.1$, $q = 1/10$ and $p^{}_{o} = 10$, where $p^{}_{o}$ is the initial value of the semilatus rectum.
That is, both orbits have the same initial semilatus rectum, and hence the same initial energy and orbital frequency $\Omega^{}_{\phi}$.
We have chosen such a large mass ratio in this case, so that one could see the trajectory tracks in the figure. 
The orbit on the left panel has been built and represented using the harmonic coordinates $x^{\alpha}_{\HAR}$
whereas the orbit on the right has been built and represented using the Cartesian Boyer-Lindquist
coordinates $x^{}_{\BL}$ (these coordinates are used both for the integration 
of the geodesic equations of motion and for the estimation of the self-force). 

%
%

The trajectories in this figure are stopped at the radius of the last stable circular equatorial orbit (LSO), 
given by~\cite{Bardeen:1972fi}:
\be
r^{}_{\LSO} = M^{}_{\MBH} \left\{ 3 + Z^{}_{2} \mp \sqrt{ \left(3-Z^{}_{1}\right)
\left(3+Z^{}_{1} + 2Z^{}_{2}\right)}\right\} \,, 
\ee
where the upper (lower) sign is for prograde (retrograde) orbits, and
\ba
Z^{}_{1} = 1 + \left(1-\frac{a^{2}}{M^{2}_{\MBH}}\right)^{\frac{1}{3}}\left[ 
\left(1+\frac{a}{M^{}_{\MBH}}\right)^{\frac{1}{3}} +
\left(1-\frac{a}{M^{}_{\MBH}}\right)^{\frac{1}{3}} \right]\,, \nonumber 
\ea
\ba
Z^{}_{2} = \sqrt{ 3\frac{a^{2}}{M^{2}_{\MBH}} + Z^{2}_{1} }\,. \nonumber
\ea
For the evolutions shown in Fig.~\ref{circular-equatorial-orbit-harmonic}, the LSO $r\approx
5.67M^{}_{\MBH}$ and $r^{}_{\HAR}\approx 4.67M^{}_{\MBH}$, consistent with $r_{\HAR} \sim r - M_{\MBH}$.

One must be careful when comparing trajectory tracks, since these are completely coordinate dependent (by definition). 
Instead, we can compare the number of GW cycles, which are directly observable. The number of 
orbital cycles in the case of harmonic coordinates is essentially double that of the Boyer-Lindquist Cartesian coordinate case, 
and consequently, the same is true for the associated waveforms (not shown here).  
This extremely large distance is perhaps a bit of an overestimate, since we considered very 
strong-field EMRIs and used the two different systems of coordinates in both the waveform 
generation and the calculation of the multipolar, post-Minkowskian self-force. This last point is important because in many kludge
schemes the coordinates only enter in the trajectory and in the waveform construction, whereas
the radiation-reaction part is based on PN results or on BH perturbation theory. In any case, 
this example shows that the proper and consistent choice of coordinates can play a major role
in the final waveforms produced.

Let us now consider how other choices in the new kludge scheme modify the final waveforms produced. 
One approximation one can make to simplify new kludges is to ignore the local potentials
$K^{}_{\mu\nu}$ and $Q^{\mu\nu}$ introduced in Eqs.~\eqref{scalar-local-potentials}-\eqref{tensor-local-potentials},
i.e.~to pick $K^{}_{\mu\nu} = Q^{\mu\nu} = 0$, or in the case that we use the approximate harmonic
coordinates of Appendix~\ref{app-coord-details} we can use the expansions of Appendix~\ref{app-Far-Field-Qs-and-Ks}
for these potentials to any order.  Obviously this can make a big difference and in our simulations
we have always used the Kerr local potentials in exact harmonic coordinates 
[Eqs.~\eqref{scalar-local-potentials}-\eqref{tensor-local-potentials}].

\begin{figure*}[htb]
\centering
\includegraphics[width=0.47\textwidth]{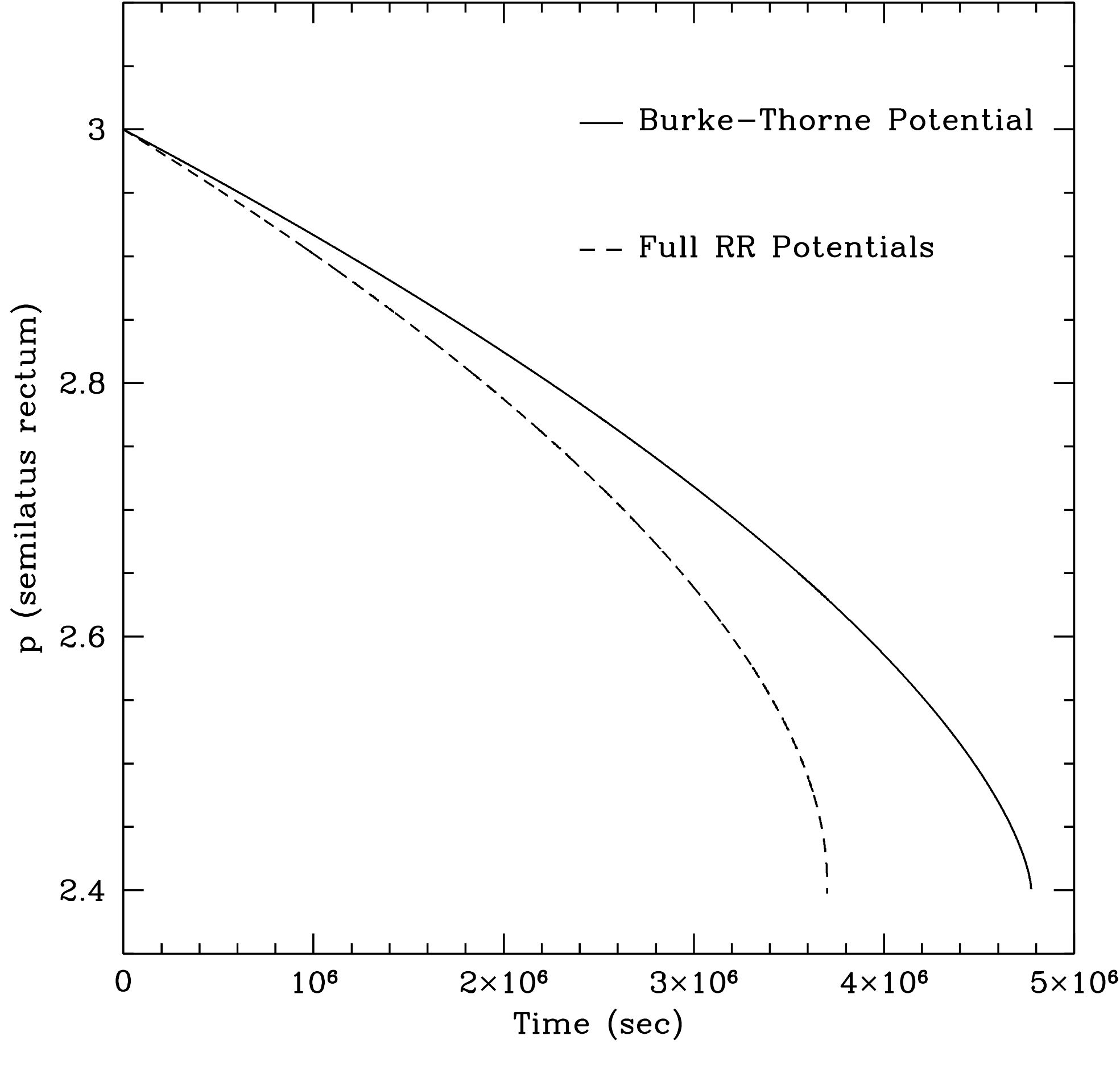}
\includegraphics[width=0.45\textwidth]{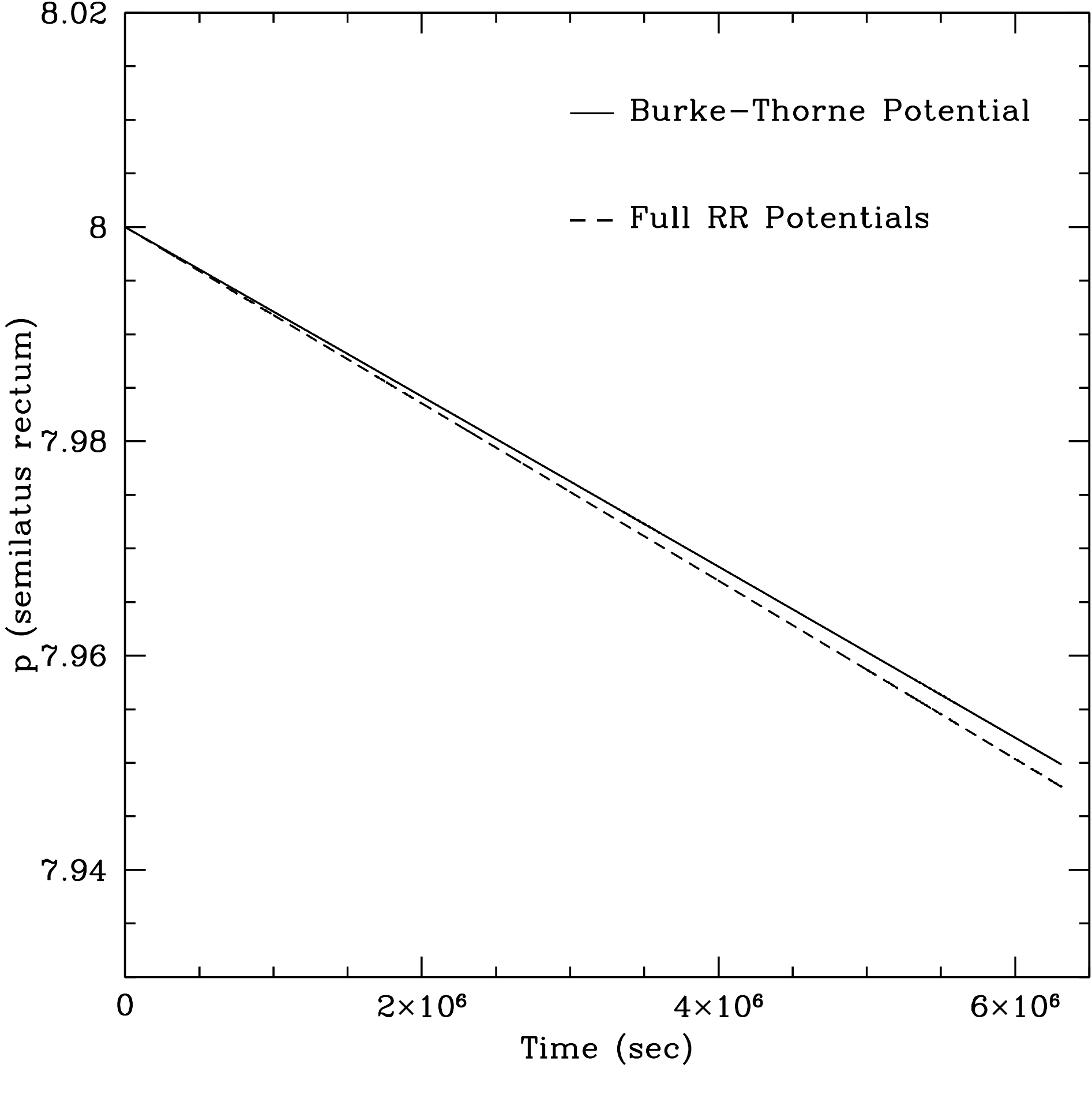}
\caption{Evolution of the semilatus rectum in circular and equatorial inspirals of a binary system characterized by:
$M^{}_{\MBH} = 4.5\times10^{6}M^{}_{\odot}$, $a/M^{}_{\MBH}=0.98$, and $q=10^{-5}$.  
The figure on the left shows an inspiral that has started with $p^{}_{o} = 3$ and the figure on the right
an inspiral that has started with $p^{}_{o} = 8$.   While the one on the left stopped near the last stable
circular orbit, the one on the right was stopped after $0.2$ yr of evolution.
\label{semilatusrectumevolution}}
\end{figure*}

Another approximation one can make is to use just the radiation-reaction potential of Burke and Thorne 
[Eq.~\eqref{burke-thorne-potential}] (with $V^{i}_{\RR}=0$), 
instead of the full potentials of Eqs.~\eqref{scalar_rr_potential} and~\eqref{vector_rr_potential} 
(see~\cite{Iyer:1993xi,Iyer:1995rn}). In order to illustrate the waveform difference in this case we have 
studied the inspiral of a system with $M^{}_{\MBH} = 4.5\times10^{6}M^{}_{\odot}$, $a/M^{}_{\MBH}=0.98$ and $q=10^{-5}$.
In Fig.~\ref{semilatusrectumevolution} we show the evolution of the semilatus rectum for two cases,
one with $p(t=t^{}_{o}) \equiv p^{}_{o} = 3$ (the plot on the left), being $t^{}_{o}$ the evolution initial time, 
and the other one with $p^{}_{o} = 8$.  
%
%
As we can see from the figure, the higher-derivative corrections to the Burke-Thorne radiation-reaction 
potential [Eqs.~\eqref{scalar_rr_potential} and~\eqref{vector_rr_potential}] increase the radiation-reaction
effects, in the sense that $p$ and the other orbital elements change more rapidly when these corrections are included.
Moreover, these corrections are more significant in the strong-field region, near the last stable orbit, and less
so as the distance to the MBH increases.  

\begin{figure}[htb]
\centering
\includegraphics[width=0.47\textwidth]{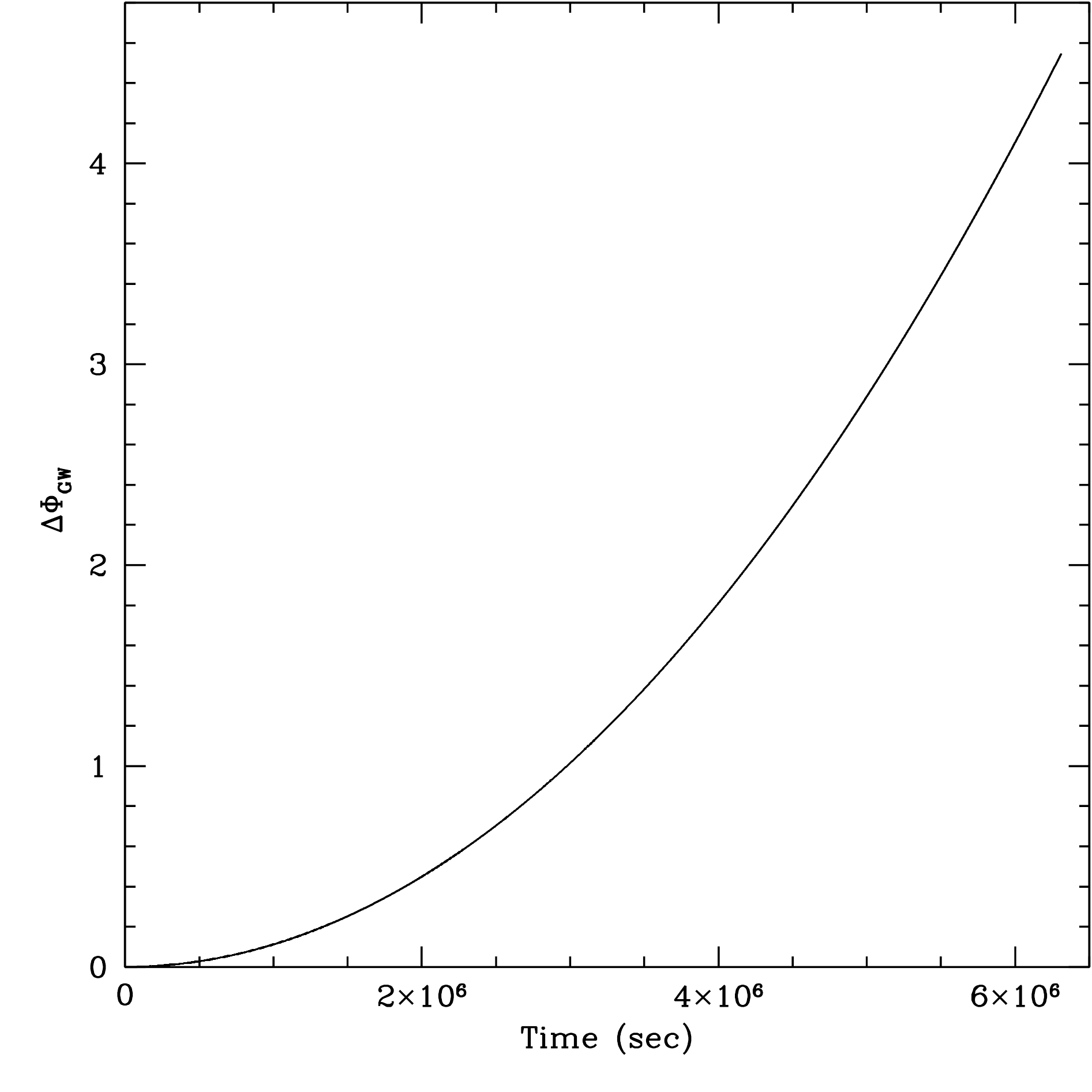}
\caption{Evolution of the GW phase difference between two inspirals characterized by:
$M^{}_{\MBH} = 4.5\times10^{6}M^{}_{\odot}$, $a/M^{}_{\MBH}=0.98$, $q=10^{-5}$
and $p^{}_{o} = 8$.  The GW phase difference is computed using the formula:
$\Delta\Phi^{}_{\GW}(t) = \Phi^{\rm Burke-Thorne}_{\GW}(t) - \Phi^{\rm Full~RR}_{\GW}(t)$, where
$\Phi^{\rm Burke-Thorne}_{\GW}$ is the GW phase for an evolution that uses only
the Burke-Thorne potential [Eq.~\eqref{burke-thorne-potential}] and $\Phi^{\rm Full~RR}_{\GW}$ 
is the GW phase for an evolution that uses the full radiation-reaction potentials 
[Eqs.~\eqref{scalar-local-potentials}-\eqref{tensor-local-potentials}].
\label{phasedifference}}
\end{figure}

We can assess quantitatively the difference due to introducing corrections to the Burke-Thorne potential 
by looking at the GWs emitted.
To that end, we introduce the following definition: 
\be
h^{}_{+} - i h^{}_{\times}  = {\cal A}^{}_{\GW}\; e^{i\Phi^{}_{\GW}}\,,
\ee
where ${\cal A}^{}_{\GW} = \sqrt{h^{2}_{+} + h^{2}_{\times}}$ is the GW amplitude and 
$\Phi^{}_{\GW} = \arctan{(h^{}_{\times}/h^{}_{+})}$ is the accumulated GW phase.
The GW phase difference induced by the presence of the corrections to the Burke-Thorne potential
is shown in Fig.~\ref{phasedifference} for the evolution with $p^{}_{o}=8$ that corresponds to
the right panel of Fig.~\ref{semilatusrectumevolution}. Observe that the GW phase
difference increases with time to accumulate up to $4.54$ rad for a total evolution time of $0.2$ yr.
This means we can expect a dephasing of more than $3$ cycles for a $1$ yr evolution, i.e.~the 
radiation-reaction corrections to the Burke-Thorne potential are of relevance for precise
and long EMRI evolutions.  The situation is even more dramatic for the strong-field evolution
that starts with $p^{}_{o} = 3$ (left panel of Fig.~\ref{semilatusrectumevolution}), where
the evolution with only the Burke-Thorne potentials takes $55.2$ days while the one with the full
radiation-reaction potentials takes $42.8$ days before reaching an unstable orbit, which translates in a difference
of $2862.5$ GW cycles.
%
%

Finally, let us look at the choices associated with the waveform construction model. 
In this part of the new kludge scheme, we can choose the multipolar order of the expansion of the gravitational radiation field 
as described in Sec.~\ref{waveform-generation}. Three possibilities present themselves here: (i) {\em Quadrupolar waveforms}; 
(ii) {\em Octopole-Quadrupolar waveforms}; and (iii) {\em Hexadecapole-Octopolar waveforms}. Case (i)
is the lowest-order approximation and consists of considering only the mass quadrupole term
in the waveform expansion.  This is equivalent to considering up to second time derivatives of the trajectory,
i.e.~up to accelerations. Case (ii) accounts for the next order multipole, that is, the mass octopole and the current quadrupole. 
This is equivalent to considering up to third time derivatives of the trajectory, i.e.~up to the {\em jerk}: $j^{i} \equiv d^{3}x^{i}/dt^{3}$.
Case (iii) adds one more multipole, that is, the mass hexadecapole and the current octopole.  This is equivalent to considering 
up to the fourth time derivatives of the trajectory, i.e.~up to the {\em snap}: $s^{i} \equiv d^{4}x^{i}/dt^{4}$.

In general, adding more multipoles does not affect the GW phase but it introduces amplitude 
corrections that depend on the inspiral character. For instance, comparing the quadrupole waveforms
with the hexadecapole-octopole waveforms for the evolutions corresponding to Fig.~\ref{semilatusrectumevolution}, 
we find no difference in the GW phase, but a $3.3\%$ (for the case with $p^{}_{o}=8$) and $4.8\%$ 
(for the case with $p^{}_{o}=3$) difference in the averaged GW amplitude.  In this comparison we have
used the full radiation-reaction potentials for all evolutions.

Let us now present some results for orbits more generic than circular equatorial, since our kludges can easily handle these as well.
Consider first circular nonequatorial orbits, since these are the simplest kind
of orbits that allow us to study the evolution of the orbital inclination,
either as described by $\theta^{}_{\INC}$ [see Eq.~\eqref{thetamin}] or by $\iota$ 
[see Eq.~\eqref{iota-angle}].  In this sense, it is interesting to test the conclusions
of~\cite{Hughes:1999bq} (see also~\cite{Glampedakis:2002cb}) that inclination remains almost constant 
for most such EMRIs and also to compare some quantitative results in order to assess 
the present new kludge implementation.  

The approximation $d\iota/dt=0$ has been considered by different authors in order to obtain an evolution equation for
the Carter constant:
\be
\frac{d{\iota}}{dt}= 0 \quad \Rightarrow \quad
\dot{C} = \frac{2C}{L^{}_{z}}\dot{L}^{}_{z}\,.  \label{cdotequationaprox}
\ee
Given that it is not simple to estimate the evolution of $C$ within the framework of perturbation
theory (see~\cite{Hughes:2005qb,Ganz:2007rf}), this approximation provides a clear path to the construction of generic
EMRI waveforms. The PN leading-order prediction for the evolution of the inclination given by Ryan~\cite{Ryan:1995wh}
predicts a grow in the inclination.  Although this prediction is known to overestimate the inclination growth 
(see~\cite{Glampedakis:2002cb} for a discussion), it agrees with BH perturbation
computations valid in the strong field. 

Table~\ref{res-circular-nonequatorial} 
presents some results of the evolution of the radius $r^{}_{o}$ and Carter constant $C$ for 
circular nonequatorial orbits characterized by:
\begin{itemize}
\item[(i)] $a/M^{}_{\MBH} = 0.05$, $r^{}_{o}/M^{}_{\MBH} = 100$, and $\iota = 60.0$ deg;
\item[(ii)] $a/M^{}_{\MBH} = 0.95$, $r^{}_{o}/M^{}_{\MBH} = 100$, and $\iota = 60.05$ deg;
\item[(iii)] $a/M^{}_{\MBH} = 0.05$, $r^{}_{o}/M^{}_{\MBH} = 7$, and $\iota = 60.17$ deg;
\item[(iv)] $a/M^{}_{\MBH} = 0.95$, $r^{}_{o}/M^{}_{\MBH} = 7$, and $\iota = 60.43$ deg$\,$.  
\end{itemize}
\begin{table*}
\begin{tabular}{ccc|c|cccc}
\hline\hline
$a/M^{}_{\MBH}$ & $r^{}_{o}/M^{}_{\MBH}$ & $\iota$ (deg) & Quantity  &  PN               & Teukolsky & New Kludge & New Kludge  \\
                &                        &               &           & (Ref.~\cite{Ryan:1995xi})   & (Ref.~\cite{Hughes:2001jr}) & Burke-Thorne & Full RR \\
\hline
$0.05$ & $100$ & $60.00$ & $q^{-1}\,(d{r}^{}_{o}/dt)$    &  $-1.2797\times 10^{-5}$ & $-1.2676\times 10^{-5}$ & $-1.1700\times 10^{-5}$ & $-1.1708\times 10^{-5}$\\
       &       &         & $(M^{}_{\MBH}/q)\,(d{\iota}/dt)$ & $7.0439\times 10^{-12}$ & $6.6936\times 10^{-12}$ & $6.1597\times 10^{-12}$ & $6.5089\times 10^{-12}$ \\ 
\hline
$0.95$ & $100$ & $60.05$ & $q^{-1}\,(d{r}^{}_{o}/dt)$    &  $-1.2733\times 10^{-5}$ & $-1.2610\times 10^{-5}$ & $-1.1622\times 10^{-5}$ & $-1.1634\times 10^{-5}$ \\
       &       &         & $(M^{}_{\MBH}/q)\,(d{\iota}/dt)$ & $1.3389\times 10^{-10}$ & $1.2040\times 10^{-10}$ & $1.1628 \times 10^{-10}$ & $1.2273\times 10^{-10}$ \\
\hline
$0.05$ & $7$  &  $60.17$ & $q^{-1}\,(d{r}^{}_{o}/dt)$    &  $-3.6762\times 10^{-2}$  &  $-1.0964\times 10^{-1}$  &  $-8.3338\times 10^{-2}$  & $-8.9633\times 10^{-2}$ \\
       &      &          & $(M^{}_{\MBH}/q)\,(d{\iota}/dt)$ &     $1.5867 \times 10^{-5}$  &  $1.0875\times 10^{-5}$  &  $9.6088\times 10^{-6}$  &  $1.6233\times 10^{-5}$ \\
\hline
$0.95$ & $7$ &  $60.43$ & $q^{-1}\,(d{r}^{}_{o}/dt)$    &   $-2.7499\times 10^{-2}$  &  $-4.6574\times 10^{-2}$  &  $-3.4547\times 10^{-2}$  &  $-3.6825\times 10^{-2}$  \\
    &        &      & $(M^{}_{\MBH}/q)\,(d{\iota}/dt)$ &    $3.0806\times 10^{-4}$  &  $1.2073\times 10^{-4}$  &  $1.6023\times 10^{-4}$  &  $2.5962\times 10^{-4}$  \\
\hline\hline
\end{tabular}
\caption{\label{res-circular-nonequatorial} Evolution of the radius, $r^{}_{o}$, and inclination angle, $\iota$,
of circular nonequatorial orbits characterized by (columns 1st to 3rd from top to bottom): 
(i) $a/M^{}_{\MBH} = 0.05$, $r^{}_{o}/M^{}_{\MBH} = 100$, and $\iota = 60.0$ deg;
(ii) $a/M^{}_{\MBH} = 0.95$, $r^{}_{o}/M^{}_{\MBH} = 100$, and $\iota = 60.05$ deg;
(iii) $a/M^{}_{\MBH} = 0.05$, $r^{}_{o}/M^{}_{\MBH} = 7$, and $\iota = 60.17$ deg;
(iv) $a/M^{}_{\MBH} = 0.95$, $r^{}_{o}/M^{}_{\MBH} = 7$, and $\iota = 60.43$ deg$\,$. 
Column 5th gives the value of $q^{-1}\,(d{r}^{}_{o}/dt)$ and $(M^{}_{\MBH}/q)\,(d{\iota}/dt)$ obtained from the PN calculations
of Ryan~\cite{Ryan:1995xi}; column 6th gives the value obtained by Hughes~\cite{Hughes:2001jr} solving the Teukolsky
equation; and columns 7th and 8th are the values obtained with our current new kludge implementation 
using the Burke-Thorne [Eq.~\eqref{burke-thorne-potential}] and the full [Eqs.~\eqref{scalar_rr_potential} and~\eqref{vector_rr_potential}] 
radiation-reaction potentials respectively.}
\end{table*}
These evolutions use Eq.~\eqref{Cdot-rdot} and the formulas of Appendix~\ref{circular-nonequatorial}.
Observe from the table that the new kludge results are in very good agreement with Teukolsky like evolutions, 
even though the former has not really been refined or optimized. 

\begin{figure}[htb]
\centering
\includegraphics[width=0.47\textwidth]{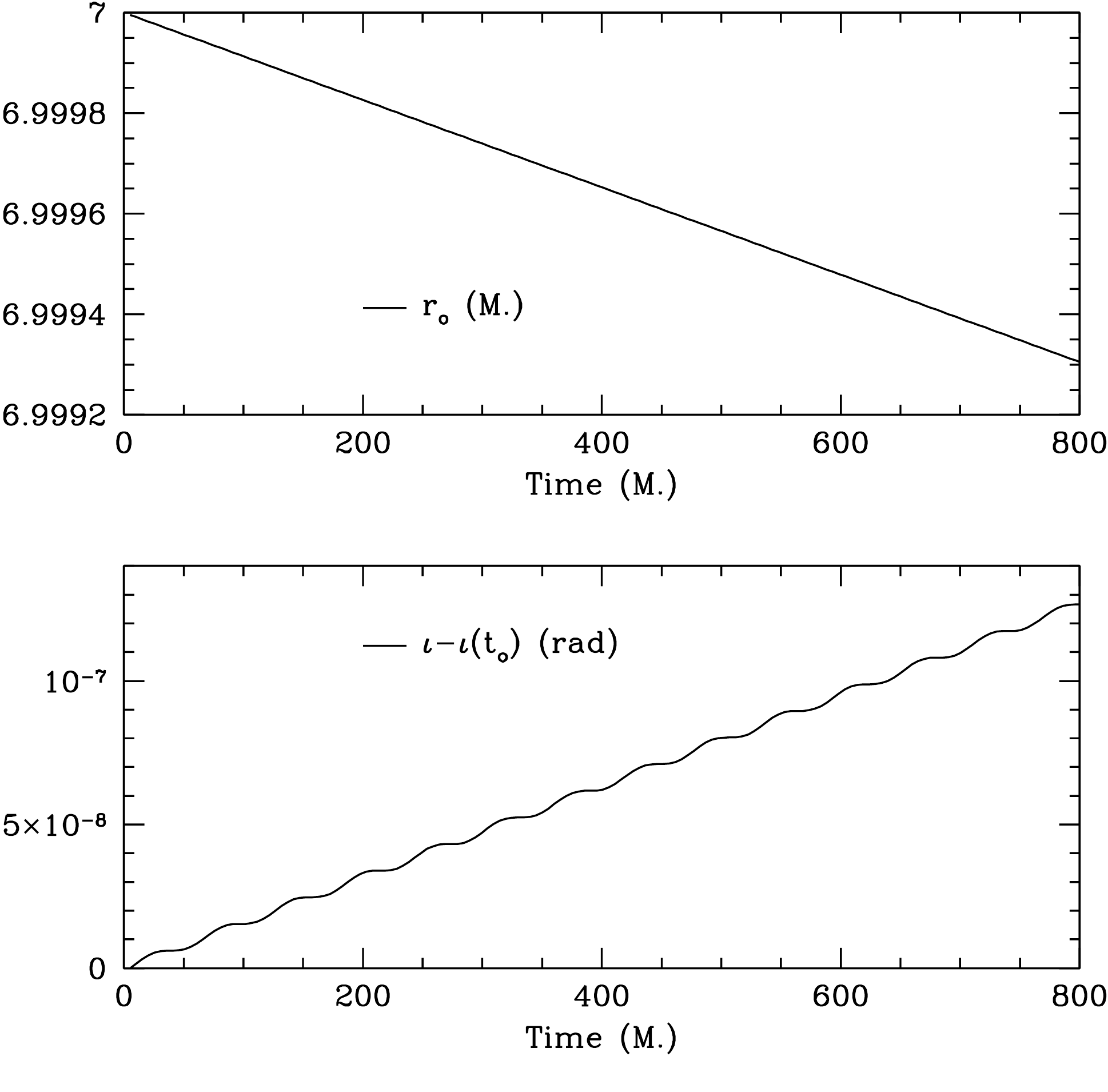}
\caption{Evolution of the radius $r^{}_{o}$ (upper plot) and inclination $\iota$ (lower plot) for the 
circular nonequatorial orbit characterized by $a/M^{}_{\MBH} = 0.05$, $r^{}_{o}/M^{}_{\MBH} = 7$, and $\iota = 60.17$ deg.
This is the third case in Table~\ref{res-circular-nonequatorial}.  The total time of the evolution in units of
the MBH mass is $800 M^{}_{\MBH}$.
\label{evolution-of-ro-and-iota}}
\end{figure}

The local nature of new kludges can also be appreciated in the local, nonuniform temporal changes of the inclination angle, 
i.e.~$d\iota/dt$ is not constant in time but oscillates with the orbital period $T^{}_{\theta}$ 
(see Table~\ref{freqs-circular-nonequatorial} for the value of the fundamental 
frequencies $\Omega^{}_{\theta}$ and $\Omega^{}_{\phi}$ for the cases of Table~\ref{res-circular-nonequatorial}).  
In contrast, we find that $dr^{}_{o}/dt$ is approximately constant within orbital time scales.  
We illustrate these facts in Fig.~\ref{evolution-of-ro-and-iota} where we compare the evolution of both quantities for a total time of
$\Delta t = 800\,M^{}_{\MBH}$ and for the third case of Table~\ref{res-circular-nonequatorial}, characterized by
$a/M^{}_{\MBH} = 0.05$, $r^{}_{o}/M^{}_{\MBH} = 7$, and $\iota = 60.17$ deg. The results
of Table~\ref{res-circular-nonequatorial} can be obtained via a simple linear regression of the evolution of $\iota$ over
a number of orbital periods (we quote the slope as the value of $d\iota/dt$ in Table~\ref{res-circular-nonequatorial}).
\begin{table}
\begin{tabular}{ccc|cc}
\hline\hline
$a/M^{}_{\MBH}$ & $r^{}_{o}/M^{}_{\MBH}$ & $\iota$ (deg) & $M^{}_{\MBH}\Omega^{}_{\theta}$ & $M^{}_{\MBH}\Omega^{}_{\phi}$  \\
\hline
$0.05$          & $100$                  & $60.00$       & $9.9992\times 10^{-4}$ & $1.0000\times 10^{-3}$ \\ 
$0.95$          & $100$                  & $60.05$       & $9.9856\times 10^{-4}$ & $1.0004\times 10^{-3}$ \\
$0.05$          & $7$                    & $60.17$       & $5.3776\times 10^{-2}$ & $5.4065\times 10^{-2}$ \\
$0.95$          & $7$                    & $60.43$       & $4.9813\times 10^{-2}$ & $5.4537\times 10^{-2}$ \\
\hline\hline
\end{tabular}
\caption{\label{freqs-circular-nonequatorial} Values of the fundamental frequencies $\Omega^{}_{\theta}$
and $\Omega^{}_{\phi}$ (see Appendix~\ref{fundamental-frequencies-and-periods} for details) for the four 
cases shown in Table~\ref{res-circular-nonequatorial}.
The values of the frequencies are normalized with respect to the MBH mass $M^{}_{\MBH}$.}
\end{table}
%

%
%

Let us comment further on the results of Table~\ref{res-circular-nonequatorial}, classifying them into
weak-field calculations [cases (i) and (ii)] and strong-field ones [cases (iii) and (iv)].  First, as it
was already discussed, the new kludge results in the weak field do not differ much using the Burke-Thorne or the
full radiation-reaction potentials, whereas in the strong field the differences are significative. Second,
considering the absolute value of the numbers quoted in the table, new kludges
tend to always underestimate the rate of decay of the radius of the circular orbit, $r^{}_{o}$, 
with respect to the Teukolsky results of~\cite{Hughes:2001jr}. This is in contrast with
the PN results of~\cite{Ryan:1995xi}, which overestimate $dr^{}_{o}/dt$ in the weak field and underestimate it in
the strong field.  

Regarding the evolution of the inclination angle $\iota$ the situation is a bit different, although 
we can still see similar differences between using the Burke-Thorne and the full radiation-reaction potentials
in the strong-field computations. For the weak-field computations, the differences are significantly bigger
for the evolution of $\iota$ than for the evolution of $r^{}_{o}$.  To put these numbers in context, 
the errors in the new kludge estimations of the rates of $r^{}_{o}$ and $\iota$, as compared with the Teukolsky estimations,  
are: for $dr^{}_{o}/dt$, $7.63-7.83\%$ for the weak-field computations
and $18.24-25.82\%$ for the strong-field ones; for $d\iota/dt$, $1.93-7.97\%$ for the weak-field computations
and $11.64-115.04\%$ for the strong-field ones.  In the weak field, the new kludge computations, again assuming
that the Teukolsky computations are the correct ones, in general do worse than the PN ones for $dr^{}_{o}/dt$
but they do better for $d\iota/dt$, whereas in the strong field the new kludge computations seem to be better
than the PN ones in general. Of course, this assumes that the Teukolsky waveforms are exactly correct, which
is not the case either. A more detailed comparison between Teukolsky and new kludge results will be carried out elsewhere.

\begin{figure*}[htb]
\centering
\includegraphics[width=0.328\textwidth]{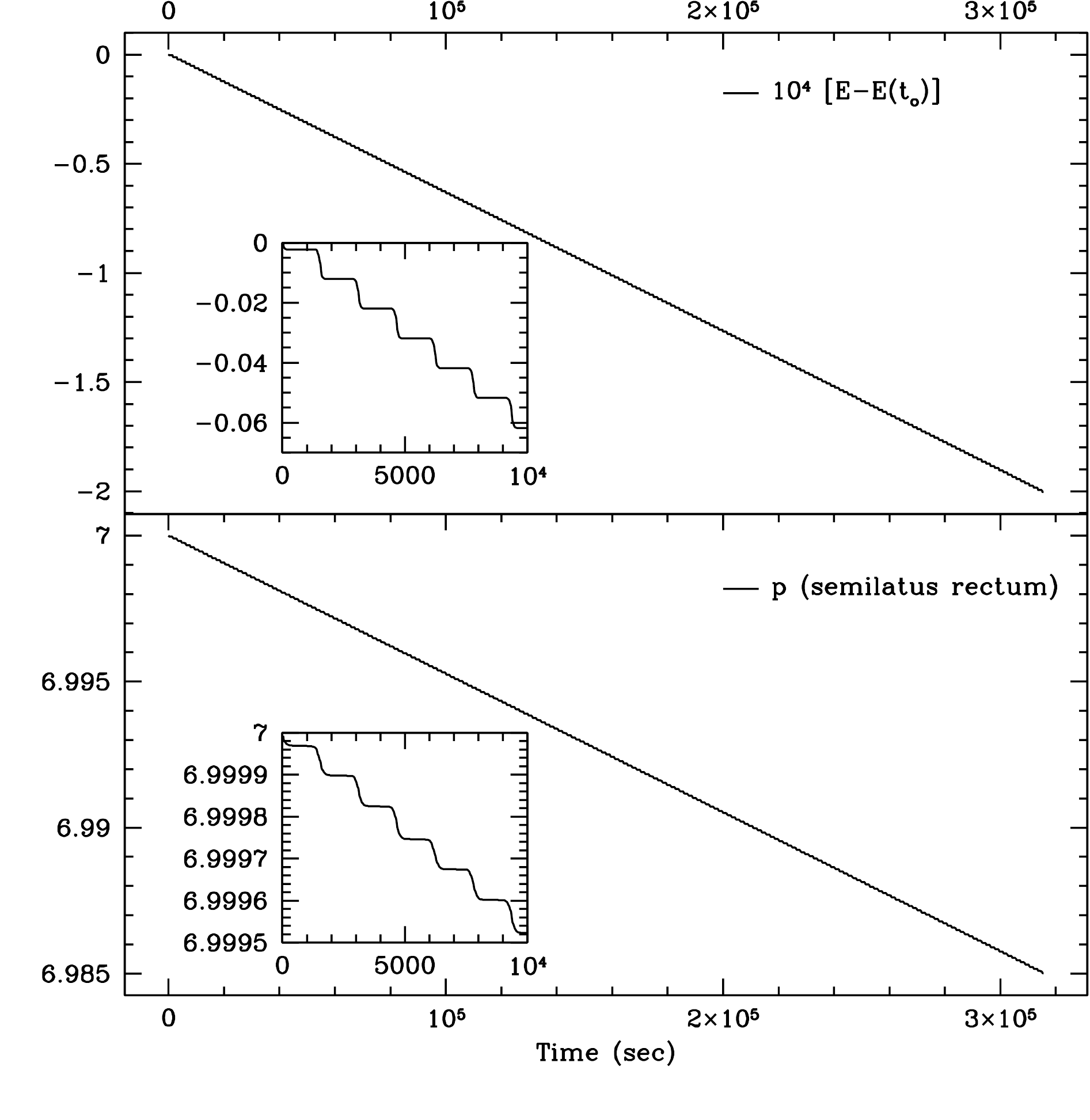}
\includegraphics[width=0.328\textwidth]{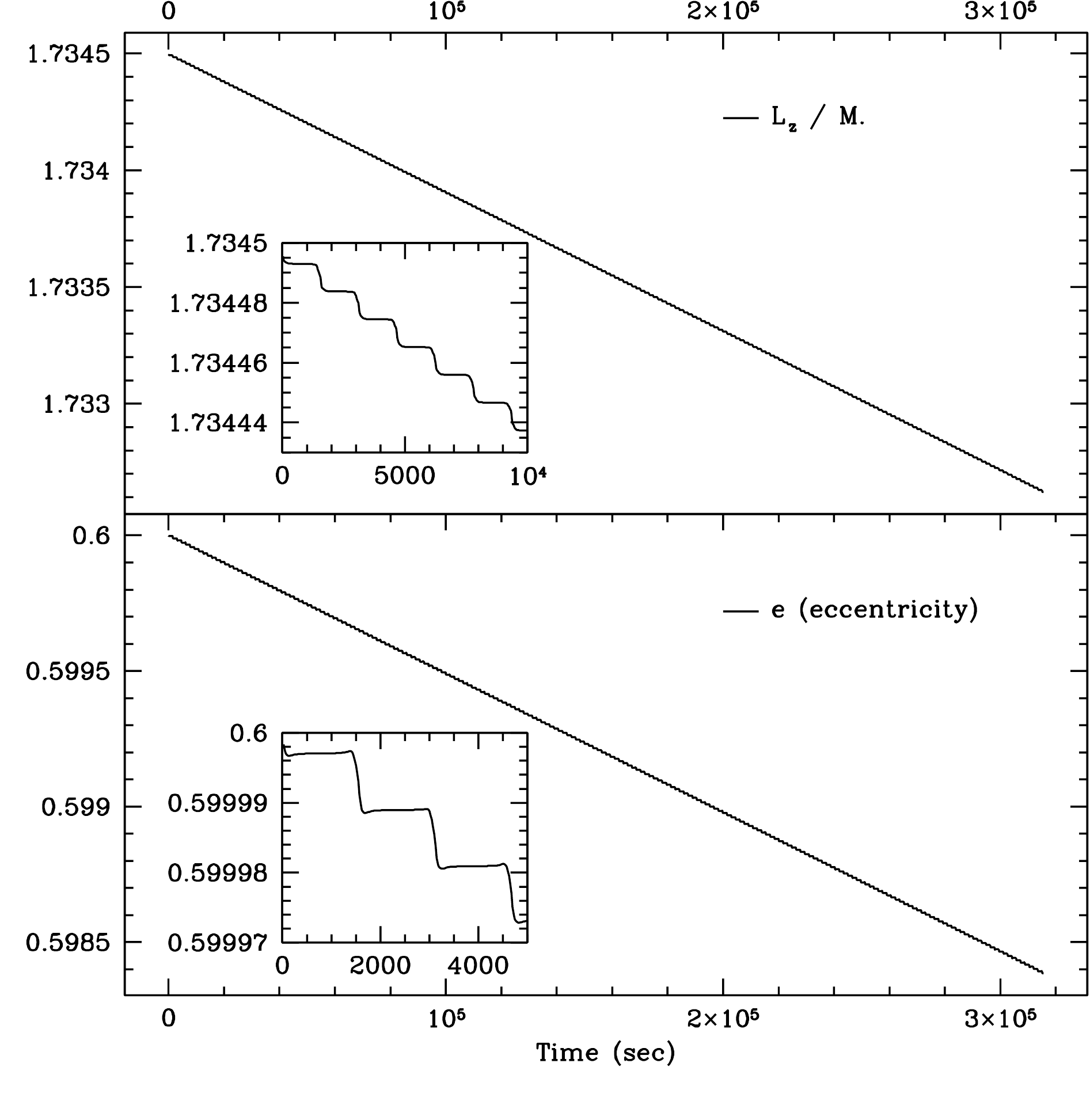}
\includegraphics[width=0.328\textwidth]{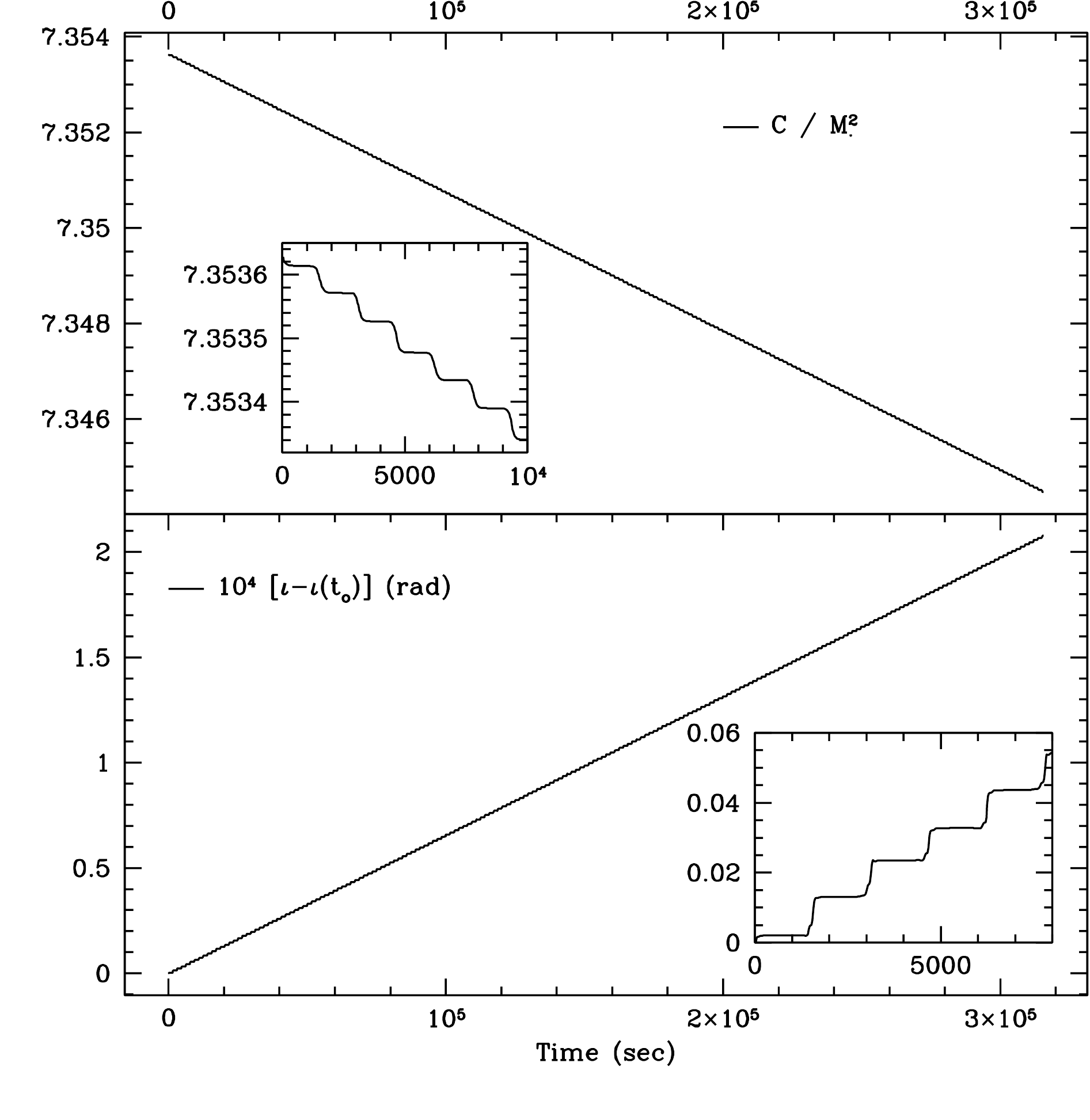}
\caption{Evolution (for a total time of $10^{-2}$ yrs) of an eccentric and inclined inspiral 
of a system characterized by: $M^{}_{\MBH} = 10^{6} M^{}_{\odot}$, $a/M^{}_{\MBH}=0.98$, 
and $q=10^{-5}$.  The plots show the evolution of the following quantities: Energy $E$ 
(top left), actually $10^{4}[E-E(t^{}_{o})]$, where $E(t^{}_{o}) = 0.9575513$; angular
momentum along the spin axis, $L^{}_{z}$ (top center); Carter constant $C$ (top right);
semilatus rectum, $p$ (bottom left), with $p(t^{}_{o}) = 7$; eccentricity, $e$ (bottom
center), with $e(t^{}_{o}) = 0.6$; and inclination angle $\iota$ (bottom right), actually
$10^{4}[\iota-\iota(t^{}_{o})]$, where $\iota(t^{}_{o})=57.39$ deg.  All plots contain
subplots where the detailed evolution during a few orbital periods is
shown. \label{evolution-parameters-generic-orbit}}
\end{figure*}

Let us finish by considering an example of generic, eccentric and inclined orbital evolutions with 
the new kludge scheme.~For such orbits, all the constants of motion/orbital elements 
change in time, plotted in Fig.~\ref{evolution-parameters-generic-orbit} for a sufficiently 
long time, including many orbital periods. The main figure hides the changes in the
orbital time scales, so we have included subplots where we can see that the evolution details
scale with the orbital periods.  As we have already discussed above, this is a consequence of using a 
local in time self-force. These effects are not present in evolutionary schemes based on flux averaging over a certain number of
orbital periods (essentially all other models currently used). 

As we can see in Fig.~\ref{evolution-parameters-generic-orbit}, the
local evolution of the different quantities presents slightly different patterns.  The
difference in these patterns are enhanced if one chooses a more extreme orbit (more
eccentric and more strong field).  We can also appreciate that looking at the
global evolution (over the long time scale) all quantities decay in time except for
the inclination, which grows in time. However, if one looks at the details of the evolution over
the orbital periods, one can see (eg. for the eccentricity) that it can
locally grow in time although the global tendency is to decay.  Therefore, the new kludge
scheme leads to quite rich evolutionary patterns due to its local in time character, which
also makes it a very valuable tool to investigate questions like the appearance of 
transient resonances in EMRIs~\cite{Flanagan:2010cd}.
%
%

\begin{figure}[htb]
\centering
\includegraphics[width=0.47\textwidth]{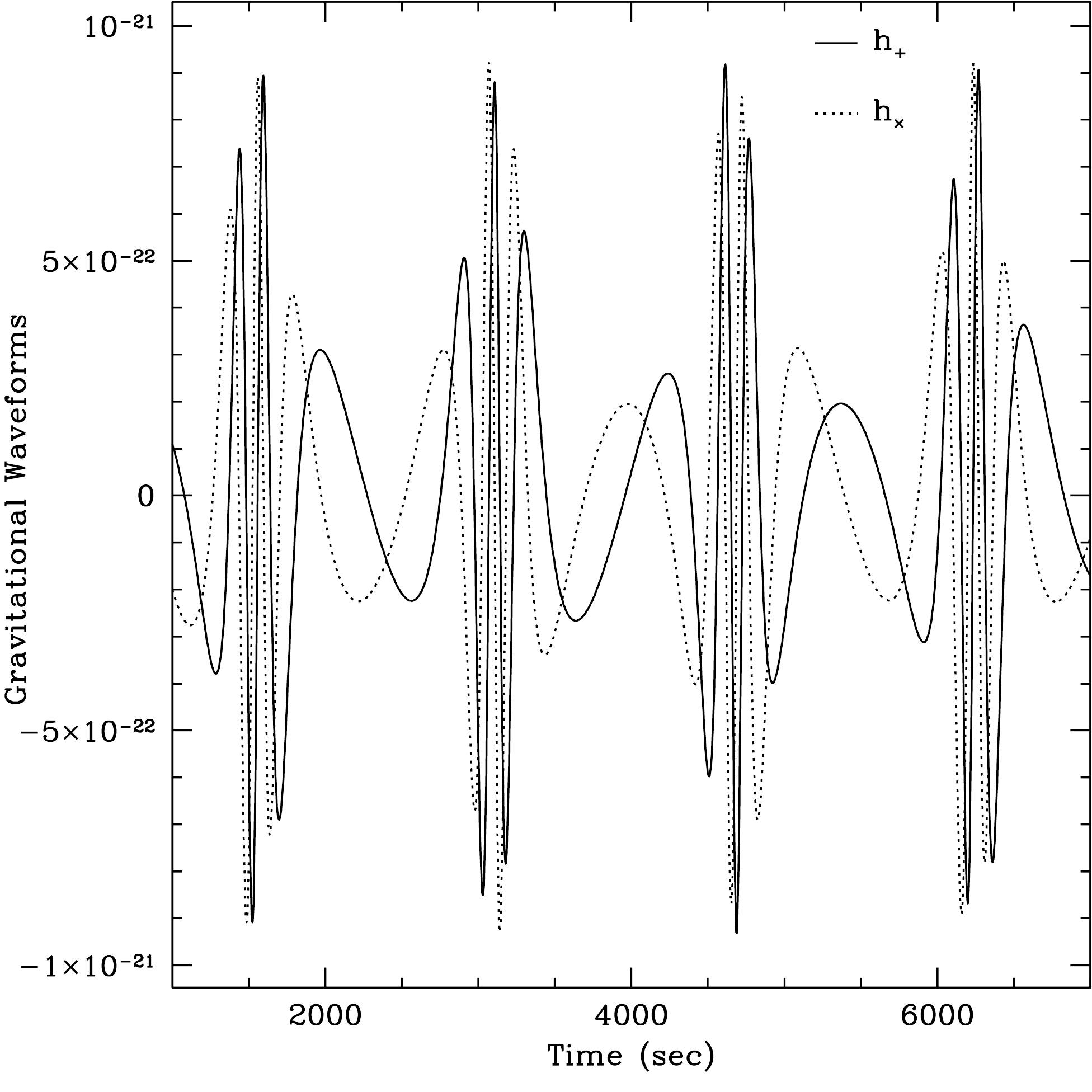}
\caption{Fragments of the GWs emitted from an eccentric and
inclined system [with initial orbital elements $(p,e,\iota)(t^{}_{o}) = (7,0.6,57.39$ deg$)$]
characterized by: $M^{}_{\MBH} = 10^{6} M^{}_{\odot}$, $a/M^{}_{\MBH}=0.98$, 
and $q=10^{-5}$.  The evolution of the constants of motion and orbital elements
of this system are shown in Fig.~\ref{evolution-parameters-generic-orbit}.
The solid line represents the $+$ polarization, $h^{}_{+}(t)$, and the dotted line
represents the $\times$ polarization, $h^{}_{\times}(t)$, as seen by an observer
located along the spin axis.\label{generic-gws}}
\end{figure}

To illustrate the multipolar waveforms produced by the new kludge scheme for general orbits, 
Fig.~\ref{generic-gws} shows short fragments (for the sake of clarity) of the GWs associated
with the last evolution of Fig.~\ref{evolution-parameters-generic-orbit}, i.e. for the inspiral of a system 
with $M^{}_{\MBH} = 10^{6} M^{}_{\odot}$, $a/M^{}_{\MBH}=0.98$, and $q=10^{-5}$, 
and with initial orbital elements $(p,e,\iota) = (7,0.6,57.39$ deg$)$.
The waveforms polarizations shown in Fig.~\ref{generic-gws}, which correspond to an
observer along the spin axis, present the richness of EMRI GWs for eccentric and
inclined orbits.
%
%

\section{Conclusions and Discussion} 
\label{discussion-conclusions}
We have introduced the new kludge scheme to model the dynamics and the
GW emission of EMRIs, which in principle could also be used for 
intermediate-mass ratio systems.   This scheme combines ingredients from 
the multipolar, post-Minkowskian formalism and black hole perturbation theory to evolve a nongeodesic word line (with
respect to the geometry of the binary's large component, assumed here to be a MBH)
and construct waveforms. The orbits are built as a sequence of local geodesics 
whose orbital elements evolve according to a local self-force that we approximate via a multipolar, post-Minkowskian expansion. 
The leading order term of this self-force corresponds to the well-known Burke-Thorne
radiation-reaction potential.  We have seen that a crucial ingredient in this construction
is the mapping from Boyer-Lindquist coordinates, in which we integrate locally in
time the orbits to the harmonic coordinates required both by the multipolar, post-Minkowskian 
self-force and for the GW multipolar expansion. Once we have trajectories in harmonic coordinates, 
it is straightforward to build waveforms.

In practice, the implementation of the new kludge schemes requires a number of numerical/analytical 
ingredients.  First, one must numerically integrate (locally in time) geodesic equations, for
which we use appropriate angle variables that avoid turning points.  From the integration
of these equations, we can use analytical formulas to map the trajectories (and associated velocities and accelerations) 
to harmonic coordinates. Then, information from small fragments of the geodesic orbit (smaller than the fundamental orbital periods) 
is used to build the different multipole moments and compute their time derivatives.  We have found
this to be the most challenging point of the implementation as we need up to eight-order
time derivatives of the multipoles, in particular of the mass quadrupole. 
Numerical differentiation is much more complicated than numerical integration, since
the choice of the offset in finite difference formulas is crucial to obtain correct results.
By experimenting with different numerical techniques we have concluded that general
differentiation techniques (i.e.~valid for any differentiable function) are quite difficult 
to implement successfully (specially for the high-order derivatives required), since the derivatives 
require precise fine-tuning that depends on the fundamental frequencies of
motion. The way out has been to use a numerical method that takes into account what
we know analytically about local motion: the fundamental frequencies.  More specifically, we
find that fitting a truncated multiperiod (the fundamental periods, which we also obtain
numerically in terms of elliptic integrals) Fourier series
to fragments of the local evolution using standard least-squares methods provides the
accuracy that we require for a precise estimation of the radiation-reaction potentials
and the self-force.  With this, we can then evolve the constants of motion, thus mapping
from the initial geodesic to new ones. Such a transition requires the mapping of the new constants of 
motion $(E,L^{}_{z},C/Q)$ to the new orbital elements $(e,p,\iota/\theta^{}_{\INC})$, which can be done using analytic
formulas.  Repeating this procedure iteratively in time, we build the inspiral trajectory or worldline, from
which we then finally construct the waveforms.

Apart from the exact harmonic coordinates of Ding (see Ref.~\cite{1987PThPh..78.1186A})
that we use in our implementation, we have also provided approximate harmonic coordinates
based on the construction of ACMC coordinates (see Appendices~\ref{app-coord-details} 
and~\ref{app-Far-Field-Qs-and-Ks}).  The construction of these coordinates is a useful
exercise that can be used in scenarios where the big component of the binary is not
described by the Kerr geometry but by something else either related with other theories
of gravity or with the idea that MBH at galactic centers may be exotic matter configurations
rather than black holes.

The results presented here serve as a first introduction and proof-of-principle 
of the new kludge scheme. Much work remains to be done to improve and to validate the 
method in order to obtain GWs to the level of accuracy required for 
LISA data analysis.  One clear way of improving the scheme is to use better expressions
for the different multipole moments that go beyond the present leading-order approximation.
One caveat of doing this is that the PN corrections required are in general not known for
generic spinning binaries. In any case, the currently known PN corrections will certainly
improve the accuracy of the new kludge evolutions. Another way to improve the scheme would
be to introduce conservative corrections to the background, for instance as it is done in the EOB
formalism~\cite{Buonanno:1998gg,*Buonanno:2000ef,*Damour:2009sm}.

A more detailed and exhaustive validation of the new kludge scheme would include, as a first step, a comparison of the
evolution of the constants of motion with the fluxes associated with them that are
calculated by solving the Teukolsky equation~\cite{Hughes:1999bq,Hughes:2001jr}. 
As we have already mentioned above, one has to be careful in doing this comparison given 
that the latter employs averages over several cycles, while the new kludge fluxes are 
computed locally at the SCO's location. Once the fluxes have been validated, 
one should compare the waveforms themselves. An overlap study would determine 
the level of agreement between them.   

Another interesting aspect that we can test is whether the new kludge scheme can be used
for IMRIs and even for systems with moderate mass ratios.
Recently, comparisons between self-force, PN, and numerical relativity computations
have arrived to the conclusion that by replacing the mass ratio $q$ by the symmetric
mass ratio $\eta$, the self-force predictions compare quite well with numerical 
relativity and PN predictions in the comparable-mass range~\cite{Barack:2010ny,Tiec:2011bk}.
These results may allow for a simpler description of IMRIs, which otherwise would require 
either long full numerical computations or higher-order perturbative computations, 
or a combination of both. In the new kludge scheme, $q\rightarrow \eta$ is actually
mostly built in already, since the mass ratio information enters through the definition
of the multipole moments [see Eqs.~\eqref{mass-moments}, \eqref{current-moments}, \eqref{mass-hexadecapole},
and~\eqref{current-octopole}]. For these, we already use general binary expressions, 
instead of effective one-body ones with mass $q M^{}_{\MBH}$.
Therefore, it would be interesting to see whether future improvements of the new kludge
scheme can produce reasonably good results for systems other than EMRIs.

The ability to compute approximate local quantities at the location of the SCO,
in particular the multipolar, post-Minkowskian self-force,  
makes the new kludge scheme an interesting tool to study certain local behavior believed
to exist in EMRIs. For example, Flanagan and Hinderer~\cite{Flanagan:2010cd} have recently 
reported that certain rapid changes in orbital elements can arise for generic EMRIs when the 
orbital frequencies become commensurate. During these rapid changes, it has been 
postulated that the EMRI waveforms might suffer a ``glitch,'' not unlike those 
observed in pulsar astronomy. Questions remain though as to the exact nature of 
this effect. One could compare the effect of these glitches in waveforms as 
computed from an averaged-scheme (such as the Teukolsky one) and a local one 
(such as the new kludge approach). These, and other studies, would provide one more 
piece to the EMRI puzzle. 

\acknowledgments We would like to thank Leor Barack, Jon Gair and Frans Pretorius for useful comments 
and suggestions.  We also thank the anonymous referee for criticisms and suggestions that have improved the
paper. CFS acknowledges support from the Ram\'on y Cajal Programme of 
the Ministry of Education and Science of Spain, by a Marie Curie International 
Reintegration Grant No. MIRG-CT-2007-205005/PHY within the 7th European Community 
Framework Programme, and from the Contract No. ESP2007-61712 of the Spanish Ministry 
of Education and Science. CFS also acknowledges the hospitality of Princeton 
University and Massachusetts Institute of Technology where part of this work was completed.  NY acknowledges support 
from National Science Foundation Grant Nos. PHY-0745779 and PHY-1114374, as well as support provided by the National 
Aeronautics and Space Administration through Einstein Postdoctoral Fellowship 
Award No. PF0-110080, issued by the Chandra X-ray Observatory Center, which 
is operated by the Smithsonian Astrophysical Observatory for and on behalf of the 
National Aeronautics Space Administration under Contract No. NAS8-03060. NY also 
acknowledges support from NASA Grant No. NNX11AI49G, under Grant No. 00001944,
and the hospitality of the Institut de Ci\`encies de l'Espai (CSIC-IEEC) where part of this work was completed. 
We also acknowledge the Centro de Supercomputaci\'on de Galicia (CESGA) 
Project No. ICTS-CESGA-200.

\appendix

\section{Useful formulas for the Computation of the Self-Acceleration} \label{app-coeffs}
In this appendix we define the quantities relevant to the multipolar, post-Minkowskian radiation-reaction 
acceleration of Eq.~\eqref{full-acc}, where the pieces $A^{(1)}_{\alpha}$ and $A^{(2)}_{\alpha}$
are given in Eqs.~\eqref{a-piece-1} and~\eqref{a-piece-2} respectively.  We also need the
expression of the projector orthogonal to the SCO four-velocity.  This quantity can be written 
in terms of the potentials $K^{}_{\mu\nu}$ and $Q^{\mu\nu}$ as follows:
\ba
P^{\alpha\beta} & = &\eta^{\alpha\beta}+Q^{\alpha\beta}+\Gamma^{2}v^{\alpha}v^{\beta}\,, 
\label{orthogonalprojector-Ks} \\
P^{}_{\alpha\beta} & = & \eta^{}_{\alpha\beta}+K^{}_{\alpha\beta}+\Gamma^{2}v^{}_{\alpha}v^{}_{\beta}\,.
\label{orthogonalprojector-Qs}
\ea
where the $\Gamma$ factor was already given in Eq.~\eqref{Gamma-def}, but it can also be rewritten,
in terms of the $K^{}_{\mu\nu}$ potentials, as
\be
\Gamma = \frac{1}{\sqrt{1 - v^{2} - K - 2 K^{}_{i} v^{i} - K^{}_{ij} v^{ij}}} \,,
\label{gammafactorwithpotentials}
\ee
where we can see how the relativistic $\Gamma$ factor is modified, with respect to the
Special Relativity form $\Gamma = (1-v^{2})^{-1/2}$, due to Kerr {\em local} potentials.

The first piece of the self-acceleration, $A^{(1)}_{\alpha}$, is determined by the object
${\cal{G}}^{\RR}_{\mu \nu \alpha}$ [Eq.~\eqref{G-tensor}].  The time and spatial components
of $A^{(1)}_{\alpha}$ [see~\eqref{acc1}] are given by
\ba
{\cal A}^{\RR} & = & \left(1 - v^{2} \right) \partial_{t} V_{\RR} + 2 v^{i} \partial_{i} V_{\RR} - 4 v^{ij} 
\partial_{i} V_{j}^{\RR}\,,   \label{calaRR} \\  
{\cal A}^{\RR}_{i} & = & - \left(1 + v^{2}\right) \partial_{i} V_{\RR} + 2 v_{i} \partial_{t} V_{\RR} 
+ 2 v_{i} v^{j} \partial_{j} V^{\RR}  \nonumber \\
&-& 8 v^{j} \partial_{[j} V_{i]}^{\RR} - 4 \partial_{t} V_{i}^{\RR}\,.  \label{calaiRR}
\ea
The second piece of the self-acceleration, $A^{(2)}_{\alpha}$, is linear in the radiation-reaction
potentials $V^{}_{\RR}$ and $V^{\RR}_{i}$ [see Eq.~\eqref{a-piece-2}].  The coefficients of the
linear combinations of these potentials in the time and spatial components of $A^{(2)}_{\alpha}$
[Eqs.~\eqref{acc2t} and~\eqref{acc2i} respectively] have the following form:  
\ba
{\cal{B}}_{\RR} &=& Q^{i} \partial_{i} K\,, \label{B-RR-eq} \\
{\cal{C}}_{\RR} &=& 2 \left(1 - Q \right) v^{i} \partial_{i} K + 4 Q^{i} v^{j}  \partial_{[i} K_{j]}\,, 
\label{C-RR-eq} \\
{\cal{D}}_{\RR} &=& 2 \left(1 - Q\right) v^{ij} \partial_{i} K_{j}   \nonumber \\
& - &  Q^{i} v^{jk}\left(2 \partial_{(j} K_{k)i} - \partial_{i} K_{jk} \right)\,, \label{D-RR-eq} \\
{\cal{B}}^{i}_{\RR} &=& - 2 \left( \delta^{ij} + Q^{ij}\right) \partial_{j} K\,, \label{Bi-RR-eq} \\
{\cal{C}}^{i}_{\RR} &=& 4 Q^{i} v^{j} \partial_{j} K + 8 \left( \delta^{ik} + Q^{ik} \right) v^{j}  
\partial_{[j} K_{k]}\,, \label{Ci-RR-eq} \\
{\cal{D}}^{i}_{\RR} &=& 4 Q^{i} v^{jk} \partial_{j} K_{k}  
\nonumber \\
& + & 2 \left( \delta^{il} + Q^{il} \right) v^{jk} \left( 2 \partial_{(j} K_{k)l} 
- \partial_{l} K_{jk} \right)\,. \label{Di-RR-eq}
\ea
All contractions in these expressions are to be performed with the Kronecker delta, $\delta^{}_{ij}
=\delta^{ij} = {\rm diag}(1,1,1)$, 
and we recall that Latin indices in the middle of the alphabet stand for spatial coordinates only 
(thus, all tensors are purely spatial). 

For the purposes of the new kludge scheme we need to compute these local potentials associated with
the Kerr metric ($K_{\HAR}$, $K^{\HAR}_{i}$, and $K^{\HAR}_{ij}$) and its inverse ($Q_{\HAR}$, $Q^{i}_{\HAR}$, 
and $Q^{ij}_{\HAR}$) in harmonic  coordinates.  They are need for the orthogonal projector 
[Eqs.~\eqref{orthogonalprojector-Ks} and~\eqref{orthogonalprojector-Qs}],
for the $\Gamma$ factor [Eq.~\eqref{Gamma-def}], and for the coefficients of the second piece of the acceleration,
$A^{(2)}_{\alpha}$ [Eqs.~\eqref{B-RR-eq}-\eqref{Di-RR-eq}].  Moreover, for these coefficients we also
need to compute the first spatial derivatives of the metric potentials 
($\partial^{}_{k} K^{}_{\HAR}$, $\partial^{}_{k} K^{\HAR}_{i}$, and $\partial^{}_{k} K^{\HAR}_{ij}$).  
This can be done in an efficient way by using the following relations:
\be
\partial^{}_{k} K^{}_{\HAR} = \met^{\KERR,\BL}_{t\mu}\; {}^{\BL}\Gamma^\mu_{ti}\,
\frac{\partial x^i_{\BL}}{\partial x^k_{\HAR}}\,,
\ee
\ba
\partial^{}_{k} K^{\HAR}_{i} & = & \met^{\KERR,\BL}_{\mu(t}\; {}^{\BL}\Gamma^\mu_{m)n}\,
\frac{\partial x^n_{\BL}}{\partial x^k_{\HAR}}\,
\frac{\partial x^m_{\BL}}{\partial x^i_{\HAR}} \nonumber \\
& + & 2\,\met^{\KERR,\BL}_{tm}\frac{\partial^2 x^m_{\BL}}{\partial x^i_{\HAR} \partial x^k_{\HAR}}\,,
\ea
\ba
\partial^{}_{k} K^{\HAR}_{ij} & = & \left\{ \met^{\KERR,\BL}_{\mu(\ell}\; {}^{\BL}\Gamma^\mu_{m)n}\,
\frac{\partial x^n_{\BL}}{\partial x^k_{\HAR}}\,
\frac{\partial x^m_{\BL}}{\partial x^i_{\HAR}} \right. \nonumber \\
& + & \left. 2\,\met^{\KERR,\BL}_{\ell m}\frac{\partial^2 x^m_{\BL}}{\partial x^i_{\HAR} 
\partial x^k_{\HAR}} \right\} \frac{\partial x^\ell_{\BL}}{\partial x^i_{\HAR}} \,.
\ea
The advantage of these expressions is that they only involve the Jacobian and Hessian of the coordinate
transformation between Boyer-Lindquist and harmonic coordinates, and the Christoffel symbols in Boyer-Lindquist
coordinates. The latter are relatively simple functions and have to be computed anyway for other purposes 
(e.g.~to compute the accelerations of Boyer-Lindquist coordinates).

\section{Harmonic Coordinates for Rotating Black Holes}
\label{app-blkerrinharmonic}

In this appendix we describe more properties of the set of harmonic coordinates for 
rotating BHs employed in this paper. 
Eqs.~\eqref{tHAR-from-BL}-\eqref{zHAR-from-BL} and Eqs.~\eqref{tphiBL-from-HAR}-\eqref{thetaBL-from-HAR}
provide the explicit expressions for the transformation from Boyer-Lindquist to
harmonic coordinates and the inverse transformation. 
From these transformations, we find that the relation between the Boyer-Lindquist radial coordinate $r$
and the spatial harmonic coordinates $(x^{}_{\HAR},y^{}_{\HAR},z^{}_{\HAR})$ is:
\be
\frac{x_{\HAR}^2+y_{\HAR}^2}{(r-M_{\MBH})^2+a^2} + \frac{z_{\HAR}^2}{(r-M_{\MBH})^2} = 1\,,
\ee
which resembles the relation between $r$ and the spatial Kerr-Schild Cartesian coordinates 
(see, e.g.~\cite{Hawking:1973uf}).  This relation allows us to write $r$ as a function of $r_{\HAR}$ 
as given in Eq.~\eqref{rBL-from-HAR}, which we find convenient to rewrite as $r \equiv M_{\MBH} + \varrho$, 
where 
\be
\varrho = \sqrt{\frac{1}{2}\left[ r^2_{\HAR}-a^2 + \sqrt{ (r^2_{\HAR}-a^2)^2 
+ 4(\bm{s}\cdot\bm{r^{}_{\HAR}})^2}\right]}\,.  \label{varrhodef}
\ee
and where the dot here denotes the flat-space three-dimensional scalar product. Moreover,
we have here introduced the following notation in the spirit of the usual three-dimensional flat-space
vector algebra: a \emph{position} vector $\bm{r}^{}_{\HAR} = (x^{}_{\HAR},y^{}_{\HAR},z^{}_{\HAR})$, whose 
spatial norm (under the flat three-dimensional metric) was defined in Eq.~\eqref{rHAR},
and the reduced spin vector $\bm{s}$ which, for consistency with the choices made in the paper
we take it to be aligned with the z-axis: $\bm{s}=(0,0,a)$.  Nevertheless, one does not need to assume 
this specific form for $\bm{s}$, as the expressions we present in this Appendix are invariant under 
the action of the rotation group. We can then use the above expressions for arbitrary directions of the 
spin angular momentum of the BH. Note from 
Eq.~\eqref{varrhodef} that we always have $r^2_{\HAR} \geq \varrho^2\geq r^{2}_{\HAR}-a^{2}$, and clearly
the equality holds for the Schwarzschild case ($a=0$).  These inequalities translate in the following
inequalities for $r$:
\be
r^{}_{\HAR}+M^{}_{\MBH} \geq r \geq \sqrt{r^{2}_{\HAR}-a^{2}} + M^{}_{\MBH}\,.
\ee

An important ingredient of the construction of the harmonic coordinates is the angle function $\Phi(r)$,
which was already given in Eq.~\eqref{Phi-solved}.  However, its proper definition is 
actually~\cite{1987PThPh..78.1186A}
\be
\Phi(r) \equiv - \int_{r}^{\infty} dr' \frac{a M_{\MBH}^{2}}{\Delta(r') [\Delta(r') + M_{\MBH}^{2}]}\,,
\ee
where we recall that $\Delta(r) \equiv r^{2} - 2 M_{\MBH} r + a^{2}$. An alternative expression for $\Phi(r)$
can be obtained by simplifying Eq.~\eqref{Phi-solved} into
\ba
\Phi(r) & = & \frac{\pi}{2} - \arctan\left(\frac{r-M_{\MBH}}{a}\right) \nonumber \\
& - & \frac{a}{2\sqrt{M_{\MBH}^2-a^2}}\ln\left(\frac{r-r^{}_{-}}{r-r^{}_{+}}\right)\,.
\ea
Two important relations to evaluate the coordinate transformation of 
Eqs.~\eqref{tHAR-from-BL}-\eqref{zHAR-from-BL} are:
\be
\cos\Phi = \frac{\frac{r-M_{\MBH}}{a} + \Omega(r)}
{\sqrt{\left[1+\Omega^2(r)\right]\left\{1+\left(\frac{r-M_{\MBH}}{a}\right)^2\right\}}} \,,
\label{cosPHI}
\ee
\be
\sin\Phi = \frac{1- \frac{r-M_{\MBH}}{a}\,\Omega(r)}
{\sqrt{\left[1+\Omega^2(r)\right]\left\{1+\left(\frac{r-M_{\MBH}}{a}\right)^2\right\}}} \,.
\label{sinPHI}
\ee
where $\Omega(r)$ is given in Eq.~\eqref{omega-of-r}. 
The explicit form of the components of the Kerr metric and its inverse in harmonic coordinates is given in 
Appendix A of~\cite{1987PThPh..78.1186A}.  These expressions are quite lengthy and, in that form, not very
efficient for numerical computations.  In what follows, we present an alternative method to compute the components
of the transformed metric in an efficient way. At the same time, we introduce expressions that are invariant under the 
rotation group, in the sense that they do no longer assume the BH spin axis is aligned with the $\hat{z}_{\HAR}$-axis.  

\subsection{Jacobian} \label{app:Jacob}

We start by presenting the expressions of the components of the Jacobian associated with the coordinate 
transformation between Boyer-Lindquist and harmonic coordinates.
In other words, we compute the matrices 
${\bf{J}}^{\BL}_{\HAR} = D(t,r,\theta,\phi)/D(t^{}_{\HAR},x^{}_{\HAR},y^{}_{\HAR},z^{}_{\HAR})$ and
${\bf{J}}^{\HAR}_{\BL} = D(t^{}_{\HAR},x^{}_{\HAR},y^{}_{\HAR},z^{}_{\HAR})/D(t,r,\theta,\phi)$, 
which obviously are inverses of each other: ${\bf{J}}^{\HAR}_{\BL}\cdot {\bf{J}}^{\BL}_{\HAR} = 
{\bf{J}}^{\BL}_{\HAR}\cdot {\bf{J}}^{\HAR}_{\BL}  = {\rm {\bf{Id}}}^{}_{3}$, 
where ${\rm {\bf{Id}}}^{}_{3}$ is the identity matrix in three-dimensions. 
The components of the Jacobian ${\bf{J}}^{\BL}_{\HAR}$ are
\ba
\frac{\partial t}{\partial t^{}_{\HAR}} = 1\,, \qquad 
\frac{\partial t}{\partial \bm{r}^{}_{\HAR}} = \bm{0}\,, \\
\frac{\partial r}{\partial t^{}_{\HAR}} = \frac{\partial \theta}{\partial t^{}_{\HAR}}
= \frac{\partial \phi}{\partial t^{}_{\HAR}} = 0\,, \\
\frac{\partial r}{\partial \bm{r}^{}_{\HAR}} =  \frac{\varrho\,
\left\{ \varrho^2 \bm{r}^{}_{\HAR} - (\bm{s}\cdot\bm{r}^{}_{\HAR}) \bm{s}\right\}}
{\sqrt{r^2_{\HAR}-\varrho^2}\left[\varrho^4 + (\bm{s}\cdot\bm{r}^{}_{\HAR})^2\right]} \,, \\
\frac{\partial \theta}{\partial \bm{r}^{}_{\HAR}} = \frac{\varrho\,
\left\{ (\bm{s}\cdot\bm{r}^{}_{\HAR})\bm{r}^{}_{\HAR} - \varrho^2 \bm{s}\right\}}
{\sqrt{r^2_{\HAR}-\varrho^2}\left[\varrho^4 + (\bm{s}\cdot\bm{r}^{}_{\HAR})^2\right]}   \,, 
\ea
\ba
&&\frac{\partial \phi}{\partial \bm{r}^{}_{\HAR}} =  - \frac{a}{(r^2_{\HAR}-\varrho^2)
(\varrho^2+a^2)}\bm{s}\times\bm{r}^{}_{\HAR} \nonumber\\
&&+ \frac{aM_{\MBH}^2\varrho\,\left\{ \varrho^2 \bm{r}^{}_{\HAR} - (\bm{s}\cdot\bm{r}^{}_{\HAR}) 
\bm{s}\right\}}{(\varrho^2+a^2)(\varrho^2+a^2-M_{\MBH}^2)\left[\varrho^4 + (\bm{s}\cdot\bm{r}^{}_{\HAR})^2\right]} \,,
\ea
where the $\times$ denotes the usual flat-space three-dimensional vector product.  The 
components of the inverse Jacobian, i.e.~${\bf{J}}^{\HAR}_{\BL}$, are
\ba
\frac{\partial t^{}_{\HAR}}{\partial t} = 1\,, \qquad 
\frac{\partial t^{}_{\HAR}}{\partial r} = 
\frac{\partial t^{}_{\HAR}}{\partial \theta} = 
\frac{\partial t^{}_{\HAR}}{\partial \phi} = 0\,, 
\ea
\ba
\frac{\partial \bm{r}^{}_{\HAR}}{\partial r} & = &\frac{\varrho^2 \bm{r}^{}_{\HAR} 
- (\bm{s}\cdot\bm{r}^{}_{\HAR}) \bm{s}}{\varrho(\varrho^2+a^2)}
\nonumber \\
& - & \frac{M_{\MBH}^2}{(\varrho^2+a^2)(\varrho^2+a^2-M_{\MBH}^2)}\bm{s}\times\bm{r}^{}_{\HAR}\,,
\ea
\ba
\frac{\partial \bm{r}^{}_{\HAR}}{\partial \theta} & = &
\frac{\bm{s}\cdot\bm{r}^{}_{\HAR}}{\varrho\,\sqrt{r^2_{\HAR}-\varrho^2}}\bm{r}^{}_{\HAR} 
-\frac{\varrho\,(\varrho^2+a^2)}{a^2\,\sqrt{r^2_{\HAR}-\varrho^2}} \bm{s} \,,
\ea
\ba
\frac{\partial \bm{r}^{}_{\HAR}}{\partial \phi} & = & \frac{1}{a}\bm{s}\times\bm{r}^{}_{\HAR}\,.
\ea
%

\subsection{Kerr Metric in Harmonic Coordinates}\label{app:KerrMetricHar}

The expressions above can be used to compute the transformed Kerr metric and its inverse in 
harmonic coordinates systematically and efficiently.  The covariant components of the 
metric in harmonic coordinates are
\ba
\met^{\KERR,\HAR}_{\,t t} & = & \met^{\KERR,\BL}_{\,t t} \,, \qquad
\met^{\KERR,\HAR}_{\,t \mbox{\small $\bm{r}$}} = \met^{\KERR,\BL}_{\,t\phi}\, \frac{\partial \phi}{\partial \bm{r}^{}_{\HAR}}\,,\\
\met^{\KERR,\HAR}_{\,\mbox{\small $\bm{r}$}\mbox{\small $\bm{r}$}} & = & \met^{\KERR,\BL}_{\,rr}\,
\frac{\partial r}{\partial \bm{r}^{}_{\HAR}}\frac{\partial r}{\partial \bm{r}^{}_{\HAR}} +
\met^{\KERR,\BL}_{\,\theta\theta}\,\frac{\partial \theta}{\partial \bm{r}^{}_{\HAR}}\frac{\partial \theta}{\partial \bm{r}^{}_{\HAR}}
\nonumber \\
& + & \met^{\KERR,\BL}_{\,\phi\phi}\,\frac{\partial \phi}{\partial \bm{r}^{}_{\HAR}}\frac{\partial \phi}{\partial \bm{r}^{}_{\HAR}}\,,
\ea
where unlabeled coordinates on the right-hand sides stand for Boyer-Lindquist coordinates as usual. Similarly, 
the contravariant components of the metric in harmonic coordinates are
\ba
\met_{\KERR,\HAR}^{\,t t } & = & \met^{\,t t}_{\KERR,\BL} \,, \qquad
\met_{\KERR,\HAR}^{\,t \mbox{\small $\bm{r}$}} = \met^{\,t\phi}_{\KERR,\BL}\, \frac{\partial \bm{r}^{}_{\HAR}}{\partial \phi}\,,\\
\met_{\KERR,\HAR}^{\,\mbox{\small $\bm{r}$}\mbox{\small $\bm{r}$}} & = & \met_{\KERR,\BL}^{\,rr}\,
\frac{\partial \bm{r}^{}_{\HAR}}{\partial r}\frac{\partial \bm{r}^{}_{\HAR}}{\partial r} +
\met_{\KERR,\BL}^{\,\theta\theta}\,\frac{\partial \bm{r}^{}_{\HAR}}{\partial \theta}\frac{\partial \bm{r}^{}_{\HAR}}{\partial \theta}
\nonumber \\
& + & \met_{\KERR,\BL}^{\,\phi\phi}\,\frac{\partial \bm{r}^{}_{\HAR}}{\partial \phi}\frac{\partial \bm{r}^{}_{\HAR}}{\partial \phi}\,.
\ea
It is easy to see that, under transformation of the three-dimensional rotation group, the covariant and contravariant 
$(t^{}_{\HAR},t^{}_{\HAR})$ components transform as scalars, 
the covariant and contravariant $(t^{}_{\HAR},r^{i}_{\HAR})$ components transform as vectors,
and the covariant and contravariant $(r^{i}_{\HAR},r^{j}_{\HAR})$ components 
transform as $2$-rank tensors. The expressions of all these components are invariant.

The computation of the components of the Kerr metric and its inverse in harmonic coordinates 
requires expressions for the Boyer-Lindquist components of the 
Kerr metric in harmonic coordinates.   The covariant components are
\ba
\met^{\KERR,\BL}_{\,tt} &=& - \frac{\varrho^2(\varrho^2-M_{\MBH}^2) + (\bm{s}\cdot\bm{r}^{}_{\HAR})^2}
{\varrho^2(\varrho+M_{\MBH})^2+(\bm{s}\cdot\bm{r}^{}_{\HAR})^2}\,,\\
\met^{\KERR,\BL}_{\,t\phi} &=& - \frac{2M_{\MBH}}{a}\frac{\varrho^2(\varrho+M_{\MBH})^2(r^{2}_{\HAR}-\varrho^2)}
{\varrho^2(\varrho+M_{\MBH})^2+(\bm{s}\cdot\bm{r}^{}_{\HAR})^2}\,,\\
\met^{\KERR,\BL}_{\,rr} &=& \frac{\varrho^2(\varrho+M_{\MBH})^2+(\bm{s}\cdot\bm{r}^{}_{\HAR})^2}{\varrho^2(\varrho^2+a^2-M_{\MBH}^2)}\,,\\
\met^{\KERR,\BL}_{\,\theta\theta} &=& \frac{\varrho^2(\varrho+M_{\MBH})^2+(\bm{s}\cdot\bm{r}^{}_{\HAR})^2}{\varrho^2}\,,\\
\met^{\KERR,\BL}_{\,\phi\phi} &=& \frac{r^{2}_{\HAR}-\varrho^2}{a^2}\left[ (\varrho+M_{\MBH})^2+a^2 \right. \nonumber \\
&+& \left.  \frac{2M_{\MBH}\varrho^2(\varrho+M_{\MBH})(r^{2}_{\HAR}-\varrho^2)}{\varrho^2(\varrho+M_{\MBH})^2+(\bm{s}\cdot\bm{r}^{}_{\HAR})^2}\right]\,,
\ea
while the contravariant components are
\begin{widetext}
\ba
\met_{\KERR,\BL}^{\,tt} &=& -\varrho^2\frac{\left[(\varrho+M_{\MBH})^2+a^2\right]^2 - (\varrho^2+a^2-M_{\MBH}^2)(r^{2}_{\HAR}-\varrho^2)}
{\left[\varrho^2(\varrho+M_{\MBH})^2+(\bm{s}\cdot\bm{r}^{}_{\HAR})^2\right] (\varrho^2+a^2-M_{\MBH}^2)}\,, \\
\met_{\KERR,\BL}^{\,t\phi} &=& - \frac{2M_{\MBH}a\varrho^2(\varrho+M_{\MBH})}{\left[\varrho^2(\varrho+M_{\MBH})^2
+(\bm{s}\cdot\bm{r}^{}_{\HAR})^2\right] (\varrho^2+a^2-M_{\MBH}^2)}\,, \\
\met_{\KERR,\BL}^{\,rr} &=& \frac{\varrho^2(\varrho^2+a^2-M_{\MBH}^2)}{\varrho^2(\varrho+M_{\MBH})^2+(\bm{s}\cdot\bm{r}^{}_{\HAR})^2}\,,\\
\met_{\KERR,\BL}^{\,\theta\theta} &=& \frac{\varrho^2}{\varrho^2(\varrho+M_{\MBH})^2+(\bm{s}\cdot\bm{r}^{}_{\HAR})^2}\,,\\
\met_{\KERR,\BL}^{\,\phi\phi} &=& \frac{a^2\left[\varrho^2(\varrho^2-M_{\MBH}^2)+(\bm{s}\cdot\bm{r}^{}_{\HAR})^2\right]}
{\left[\varrho^2(\varrho+M_{\MBH})^2+(\bm{s}\cdot\bm{r}^{}_{\HAR})^2\right] (\varrho^2+a^2-M_{\MBH}^2)(r^{2}_{\HAR}-\varrho^2)}\,.
\ea
\end{widetext}
%

\subsection{Hessian}\label{app:Hessian}

Let us now present explicit expressions for the Hessian of the transformation
between harmonic and Boyer-Lindquist coordinates, needed for the computation of the accelerations 
in harmonic coordinates. Also, for the computation of the derivatives of the metric in
harmonic coordinates we need the Hessians of the Boyer-Lindquist coordinates
with respect to the harmonic ones. More specifically, for the computation of the
coefficients of the second piece of the acceleration, $A^{(2)}_{\alpha}$ 
[Eqs.~\eqref{B-RR-eq}-\eqref{Di-RR-eq}]. We give here these expressions using the 
three-dimensional vector notation introduced above.  The expressions for the Hessian matrix 
$\partial^{2} \bm{r}^{k}_{\HAR}/(\partial x^{i}_{\BL}\,\partial x^{j}_{\BL})$ are:
\begin{widetext}
\ba
\frac{\partial^2 \bm{r}^{}_{\HAR}}{\partial r^2} &=& -\frac{1}{(\varrho^2+a^2)^2}
\left[ 1-\frac{M_{\MBH}^4}{(\varrho^2+a^2-M_{\MBH}^2)^2}\right]\bm{s}\times\bm{s}\times\bm{r}^{}_{\HAR}
+ \frac{2M_{\MBH}^2\,\varrho}{(\varrho^2+a^2)(\varrho^2+a^2-M_{\MBH}^2)^2}\bm{s}\times\bm{r}^{}_{\HAR}\,,
\ea
\ba
\frac{\partial^2 \bm{r}^{}_{\HAR}}{\partial r\,\partial\theta} &=& \frac{1}{(\varrho^2+a^2)
\sqrt{r^2_{\HAR}-\varrho^2}}\left\{ \bm{s}\times\bm{r}^{}_{\HAR}\times\bm{r}^{}_{\HAR} 
-\frac{M_{\MBH}^2(\bm{s}\cdot\bm{r}^{}_{\HAR})}{\varrho(\varrho^2+a^2-M_{\MBH}^2)}
\bm{s}\times\bm{r}^{}_{\HAR} \right\}\,,
\ea
\ba
\frac{\partial^2 \bm{r}^{}_{\HAR}}{\partial r\,\partial\phi} &=& \frac{1}{a(\varrho^2+a^2)}
\left\{ \varrho (\bm{s}\times\bm{r}^{}_{\HAR}) 
-\frac{M_{\MBH}^2}{\varrho^2+a^2-M_{\MBH}^2}\bm{s}\times\bm{s}\times\bm{r}^{}_{\HAR}
\right\}\,,
\ea
\be
\frac{\partial^2 \bm{r}^{}_{\HAR}}{\partial \theta^2} = - \bm{r}^{}_{\HAR}\,,
\qquad
\frac{\partial^2 \bm{r}^{}_{\HAR}}{\partial \theta\,\partial\phi} = 
\frac{(\bm{s}\cdot\bm{r}^{}_{\HAR})}{a\varrho\sqrt{r^2_{\HAR}-\varrho^2}} 
\bm{s}\times\bm{r}^{}_{\HAR}\,,
\qquad
\frac{\partial^2 \bm{r}^{}_{\HAR}}{\partial \phi^2} = \frac{1}{a^2}
\bm{s}\times\bm{s}\times\bm{r}^{}_{\HAR}\,.
\ee
and the expressions for its inverse 
$\partial^{2} x^{k}_{\BL}/(\partial \bm{r}^{i}_{\HAR}\,\partial \bm{r}^{j}_{\HAR})$ are:
\ba
\frac{\partial^2 r}{\partial \bm{r}^{}_{\HAR} \partial\bm{r}^{}_{\HAR}} &  = &
\frac{\varrho}{\varrho^{4}+(\bm{s}\cdot\bm{r}^{}_{\HAR})^{2}}\left\{
\bm{Id}^{}_{3} + \bm{s}\odot\bm{s} 
+  \frac{2}{\varrho^{4}+(\bm{s}\cdot\bm{r}^{}_{\HAR})^{2}}
\left[\varrho^{2}\bm{r}^{}_{\HAR}-(\bm{s}\cdot\bm{r}^{}_{\HAR})\bm{s}\right]
\odot\left[\varrho^{2}\bm{r}^{}_{\HAR}+(\bm{s}\cdot\bm{r}^{}_{\HAR})\bm{s}\right] \right.
\nonumber \\
 & + & \left. \frac{(\bm{s}\cdot\bm{r}^{}_{\HAR})^{2}-3\varrho^{4}}{\left[\varrho^{4}
+(\bm{s}\cdot\bm{r}^{}_{\HAR})^{2}\right]^{2}}
\left[\varrho^{2}\bm{r}^{}_{\HAR}+(\bm{s}\cdot\bm{r}^{}_{\HAR})\bm{s}\right]
\odot\left[\varrho^{2}\bm{r}^{}_{\HAR}+(\bm{s}\cdot\bm{r}^{}_{\HAR})\bm{s}\right]\right\} \,,
\ea
\ba
\frac{\partial^2 \theta}{\partial \bm{r}^{}_{\HAR} \partial\bm{r}^{}_{\HAR}} &  = &
\frac{1}{\sqrt{r^2_{\HAR}-\varrho^2}\left[\varrho^{4}
+(\bm{s}\cdot\bm{r}^{}_{\HAR})^{2}\right]}\left\{  \left[ \left( \frac{r^2_{\HAR}}{r^2_{\HAR}-\varrho^2}
-\frac{4\,\varrho^4}{\varrho^{4}+(\bm{s}\cdot\bm{r}^{}_{\HAR})^{2}}\right)\frac{\partial r}{\partial \bm{r}^{}_{\HAR}}
-\frac{\varrho}{r^2_{\HAR}-\varrho^2}\,\bm{r}^{}_{\HAR}\right] 
\odot\left[(\bm{s}\cdot\bm{r}^{}_{\HAR})\bm{r}^{}_{\HAR}-\varrho^{2}\bm{s}\right]
\right. \nonumber \\
 & + & \left. \varrho(\bm{s}\cdot\bm{r}^{}_{\HAR})\left[ \bm{Id}^{}_{3} - \bm{s}\odot\bm{r}^{}_{\HAR} \right] \right\}\,,
\ea
\ba
\frac{\partial^2 \phi}{\partial \bm{r}^{}_{\HAR} \partial\bm{r}^{}_{\HAR}} &  = &
\bm{\phi}^{1}_{\bm{r}^{}_{\HAR}\bm{r}^{}_{\HAR}} + 
\frac{\partial^2 \Phi}{\partial \bm{r}^{}_{\HAR} \partial\bm{r}^{}_{\HAR}} 
= \bm{\phi}^{1}_{\bm{r}^{}_{\HAR}\bm{r}^{}_{\HAR}} + 
\Phi'' \frac{\partial r}{\partial \bm{r}^{}_{\HAR}}\odot\frac{\partial r}{\partial \bm{r}^{}_{\HAR}}
+\Phi'\frac{\partial^2 r}{\partial \bm{r}^{}_{\HAR} \partial\bm{r}^{}_{\HAR}}\,,
\ea
\end{widetext}
where $\bm{\phi}^{1}_{\bm{r}^{}_{\HAR}\bm{r}^{}_{\HAR}}$ is the following symmetric matrix
\ba
\bm{\phi}^{1}_{\bm{r}^{}_{\HAR}\bm{r}^{}_{\HAR}} = \left( \begin{array}{ccc}
\frac{2x^{}_{\HAR}y^{}_{\HAR}}{(x^{2}_{\HAR}+y^{2}_{\HAR})^2} & \frac{y^{2}_{\HAR}-x^{2}_{\HAR}}{(x^{2}_{\HAR}+y^{2}_{\HAR})^2} & 0 \\
\ast & -\frac{2x^{}_{\HAR}y^{}_{\HAR}}{(x^{2}_{\HAR}+y^{2}_{\HAR})^2} & 0 \\
\ast & \ast & 0
\end{array}
\right)\,,
\ea
and where $\odot$ denotes symmetric tensor product: $\bm{a}\odot\bm{b} = (\bm{a}\otimes\bm{b}
+\bm{b}\otimes\bm{a})/2$.

\section{ACMC and approximate Harmonic Coordinate Systems}
\label{app-coord-details}

In this Appendix, we construct an ACMC-$6$ coordinate system and a system of approximate harmonic
coordinates associated with these ACMC-$6$ coordinates that we compare with the exact 
harmonic coordinates presented in Sec.~\ref{coord-sec}.   Thorne~\cite{Thorne:1980rm} constructed 
an explicit map between Boyer-Lindquist coordinates $(x^{\alpha}_{\BL})=(t,r,\theta,\phi)$ 
and ACMC-$2$ for a Kerr BH with mass $M^{}_{\MBH}$ and Kerr spin parameter $a$. The extension 
of this map from Boyer-Lindquist coordinates to ACMC-$6$ coordinates, $(x^{\alpha}_{\ACMC})=(T,X,Y,Z)$, 
is given below for the first time:
\be
t = T\,, \label{ACMC-6-t}
\ee
\be
r = R + \frac{a^{2}}{2 R} \cos^{2}{\bar{\theta}} - \frac{5a^{4}}{8 R^{3}} \cos^{4}{\bar{\theta}}
+ \frac{21a^{6}}{16 R^{5}}\cos^{6}{\bar{\theta}}\,, 
\label{ACMC-6-r}
\ee
\be
\theta = \bar\theta - \frac{a^{2}}{4 R^{2}} \sin(2{\bar{\theta}})\left\{ 1  
- \frac{3 a^{2}}{2 R^{2}} \cos^{2}{\bar{\theta}} + \frac{10 a^{4}}{3R^{4}}\cos^{4}{\bar{\theta}} \right\}\,, 
\label{ACMC-6-theta}
\ee
\be
\phi = \arctan\left(\frac{Y}{X}\right)\,,  \label{ACMC-6-phi}
\ee
where we have introduced the shorthands $R = (X^{2} + Y^{2} + Z^{2})^{1/2}$, and $\bar{\theta} = \arccos(Z/R)$.

Harmonic coordinates are guaranteed to be ACMC-$\infty$ (see~\cite{Thorne:1980rm}), but the 
converse is not necessarily true: ACMC coordinates of order $N$ are not necessarily harmonic. 
Nonetheless, one can enhance the ACMC-$6$ coordinate transformations of Eqs.~\eqref{ACMC-6-t}-\eqref{ACMC-6-phi} 
to construct a map from Boyer-Lindquist coordinates to approximate harmonic (AH) coordinates 
$(x^{\alpha}_{\AHAR})=({t}^{}_{\AHAR},{x}^{}_{\AHAR},{y}^{}_{\AHAR},{z}^{}_{\AHAR})$:
\be
t ={t}^{}_{\AHAR}\,, \label{ACMC-H-6-t}
\ee
\ba
r & = & {r}^{}_{\AHAR} + M_{\MBH} - \frac{a^{2}[1-(\bm{\hat{s}}\cdot\bm{n}^{}_{\AHAR})^{2}]}{2\,{r}^{}_{\AHAR}}  
- \frac{5a^{4}(\bm{\hat{s}}\cdot\bm{n}^{}_{\AHAR})^{4}}{8\,{r}_{\AHAR}^{3}}  \nonumber \\
& + & \frac{21a^{6}(\bm{\hat{s}}\cdot\bm{n}^{}_{\AHAR})^{6}}{16\,{r}_{\AHAR}^{5}} \,, \label{ACMC-H-6-r}
\ea
\ba
\theta & = & \theta^{}_{\AHAR} -\frac{a^{2}(\bm{\hat{s}}\cdot\bm{n}^{}_{\AHAR})
\sqrt{1-(\bm{\hat{s}}\cdot\bm{n}^{}_{\AHAR})^{2}}}{2\,r^{2}_{\AHAR}}
\left\{ 1 - \frac{3a^{2}(\bm{\hat{s}}\cdot\bm{n}^{}_{\AHAR})^{2}}{2\,r^{2}_{\AHAR}} \right. \nonumber \\
& +  & \left. \frac{10a^{4}(\bm{\hat{s}}\cdot\bm{n}^{}_{\AHAR})^{4}}{3\,r^{4}_{\AHAR}} \right\} \,,
\label{ACMC-H-6-theta}
\ea
\be
\phi = \arctan\left(\frac{\bm{\hat{y}}^{}_{\AHAR}\cdot\bm{n}^{}_{\AHAR}}
{\bm{\hat{x}}^{}_{\AHAR}\cdot\bm{n}^{}_{\AHAR}}\right)\,, 
\label{ACMC-H-6-phi}
\ee
where $\bm{\hat{s}}$ is a unit vector (in the Euclidean sense) along the MBH spin axis which, 
according to the conventions of this paper, has components $\bm{s}=(0,0,1)$.
Moreover we have introduced the following notation in the Euclidean vector calculus style: 
$\bm{r}^{}_{\AHAR} = r_{\AHAR}^{i} = ({x}^{}_{\AHAR}, {y}^{}_{\AHAR}, {z}^{}_{\AHAR})$, 
${r}^{}_{\AHAR} = ({x}_{\AHAR}^{2} + {y}_{\AHAR}^{2} + {z}_{\AHAR}^{2})^{1/2}$,
$\bm{n}^{}_{\AHAR} = n_{\AHAR}^{i}\ = bm{r}^{}_{\AHAR}/r^{}_{\AHAR}$, 
$\theta^{}_{\AHAR} = \arccos(\bm{\hat{s}}\cdot\bm{n}^{}_{\AHAR})$, 
$\bm{\hat{y}}^{}_{\AHAR}$ and $\bm{\hat{x}}^{}_{\AHAR}$ are three-dimensional (spatial), orthogonal vectors in the plane
orthogonal to the spin axis.

These approximate harmonic coordinates are not only ACMC-$6$ but also harmonic up to terms of 
order ${\cal{O}}(M_{\MBH}^{6}/{r}_{\AHAR}^{6})$, as one can check by verifying that 
$\partial^{\AHAR}_{\beta} (\sqrt{-\met^{}_{\KERR,\AHAR}}\; \met^{\alpha \beta}_{\KERR,\AHAR}) = 0$, 
where $\met_{\alpha \beta}^{\KERR,\AHAR}$ is the transformed Kerr 
metric~\footnote{This transformation is slightly different from that found in~\cite{Aguirregabiria:2001vk}, 
for zero integration constants. One can show by direct evaluation, however, that both 
transformations lead to harmonic coordinates that are also ACMC.} and $\partial^{\AHAR}_{\alpha}$ 
denotes partial differentiation with respect to these coordinates.  

One should note that this coordinates are only {\emph{pseudo}}-harmonic, in that they do not respect the 
harmonic coordinate condition to all orders in $M_{\MBH}/{r}^{}_{\AHAR}$. Therefore, these coordinates differ 
somewhat from the transformation in Eqs.~\eqref{tHAR-from-BL}-\eqref{zHAR-from-BL}.
We can show this by expanding Eqs.~\eqref{tphiBL-from-HAR}-\eqref{thetaBL-from-HAR}
in $M_{\MBH}/r^{}_{\HAR} \ll 1$ and $a/r^{}_{\HAR} \ll 1$:
\be
t = t^{}_{\HAR}\,, 
\ee
\ba
r &=& r^{}_{\HAR} + M_{\MBH} - \frac{a^{2}}{2\,{r}^{}_{\HAR}}[1-(\bm{\hat{s}}\cdot\bm{n}^{}_{\HAR})^{2}] \nonumber \\
& - & \frac{a^{4}}{8\,r_{\HAR}^3} \left[1-6(\bm{\hat{s}}\cdot\bm{n}^{}_{\HAR})^{2} 
+ 5(\bm{\hat{s}}\cdot\bm{n}^{}_{\HAR})^{4}\right]  \\
& - & \frac{a^{6}}{16\,r_{\HAR}^{5}}\left[1 -15(\bm{\hat{s}}\cdot\bm{n}^{}_{\HAR})^{2} 
+ 35(\bm{\hat{s}}\cdot\bm{n}^{}_{\HAR})^{4}
-21(\bm{\hat{s}}\cdot\bm{n}^{}_{\HAR})^{6}\right] \,, \nonumber 
\ea
\ba
\theta &=& \theta^{}_{\HAR} -\frac{a^{2}(\bm{\hat{s}}\cdot\bm{n}^{}_{\HAR})
\sqrt{1-(\bm{\hat{s}}\cdot\bm{n}^{}_{\HAR})^{2}}}{2\,r^{2}_{\HAR}} \left\{ 1 \nonumber \right. \\
& + & \frac{3 a^{2}}{4 r^{2}_{\HAR}}\left[1-2(\bm{\hat{s}}\cdot\bm{n}^{}_{\HAR})^{2}\right] 
 \nonumber \\
& + & \left. \frac{5 a^{4}}{24r_{\HAR}^{4}} \left[1-4(\bm{\hat{s}}\cdot\bm{n}^{}_{\HAR})^{2}\right]
\left[3-4(\bm{\hat{s}}\cdot\bm{n}^{}_{\HAR})^{2}\right] \right\} \,,
\ea
\ba
\phi &=& \phi^{}_{\HAR} 
- \frac{a M_{\MBH}^{2}}{3\, r_{\HAR}^{3}} \nonumber \\ 
& - & \frac{a M_{\MBH}^{2}}{10\,r_{\HAR}^{5}}\left[
a^{2} - 5a^{2}(\bm{\hat{s}}\cdot\bm{n}^{}_{\HAR})^{2} + 2M^{2}_{\MBH}  \right]\,,
\ea
where we have used a similar vector notation as in Eqs.~\eqref{ACMC-H-6-t}-\eqref{ACMC-H-6-phi} with
$\theta^{}_{\HAR} = \arccos(\bm{\hat{s}}\cdot\bm{n}^{}_{\HAR})$ and 
$\phi^{}_{\HAR} = \arctan({\bm{\hat{y}}^{}_{\HAR}\cdot\bm{n}^{}_{\HAR}}/{\bm{\hat{x}}^{}_{\HAR}\cdot\bm{n}^{}_{\HAR}})$.
As we can see, disagreements arise in the radial transformation at order ${\cal{O}}(a^{4})$, but this 
difference is proportional to a monopole (without angular dependence)
and a quadrupole term (quadratic in angular dependence), which do not modify the ACMC condition at 
octopole order.  Similar disagreements arise also in the angular sector of the transformation. 

We have thus shown that the ACMC coordinate map we found in Eqs.~\eqref{ACMC-H-6-t}-\eqref{ACMC-H-6-phi}
reproduces the main ingredients of the full harmonic coordinate transformation. In fact, since we have 
shown  that Eqs.~\eqref{ACMC-H-6-t}-\eqref{ACMC-H-6-phi} lead to a metric that satisfies the harmonic 
coordinate condition, we can infer that the difference between this equation and 
Eqs.~\eqref{tphiBL-from-HAR}-\eqref{thetaBL-from-HAR} must amount to a refinement of the coordinate system.

\section{Far-Field Expansion of the Kerr Metric in Approximate Harmonic Coordinates}
\label{app-Far-Field-Qs-and-Ks}

In this appendix we expand the Kerr metric in the far-field using the approximate harmonic 
coordinates $x^{\alpha}_{\AHAR}$ of Appendix~\ref{app-coord-details} [the coordinate 
transformations from Boyer-Lindquist coordinates are given in 
Eqs.~\eqref{ACMC-H-6-t}-\eqref{ACMC-H-6-phi}].  Then, in transforming the Kerr metric from
Boyer-Lindquist coordinates to these approximate harmonic coordinates we assume 
$M_{\MBH}/r^{}_{\AHAR} \ll 1$ and $a/r^{}_{\AHAR} \ll 1$.  We give the expressions of the
Kerr metric and its inverse in terms of the {\em local} potentials $K^{}_{\alpha\beta}$
and $Q^{\alpha\beta}$. 
The far-field expansions of the $K^{}_{\mu \nu}$ potentials in the approximate harmonic coordinates
of Eqs.~\eqref{ACMC-H-6-t}-\eqref{ACMC-H-6-phi}, $K_{\alpha\beta}^{\AHAR}$, is given by 
\ba
K^{\AHAR} & = & K_{00}^{\AHAR} =  \frac{2 M_{\MBH}}{r^{}_{\AHAR}}\left\{ 1 
- \frac{M^{}_{\MBH}}{r^{}_{\AHAR}} \right. \nonumber \\
& + & \frac{1}{2\,r_{\AHAR}^{2}} \left[2M^{2}_{\MBH} + a^{2}\left(1 - 3(\bm{\hat{s}}\cdot\bm{n}^{}_{\AHAR})^{2}\right)\right]
\nonumber \\ 
& - & \left. \frac{M_{\MBH}}{r_{\AHAR}^{3}}\left[ M^{2}_{\MBH} 
+ a^{2}\left(1-4(\bm{\hat{s}}\cdot\bm{n}^{}_{\AHAR})^{2}\right)\right] \right\}  \,,  \label{kerrK} 
\ea
\ba
K_{i}^{\AHAR} & = & K_{0i}^{\AHAR} = -\frac{2 M^{}_{\MBH}}{r_{\AHAR}^{2}} \left\{ 1 - \frac{M^{}_{\MBH}}{r^{}_{\AHAR}} 
\right.  \\
& + & \left. \frac{1}{2\,r_{\AHAR}^{2}} \left[2M^{2}_{\MBH} + a^{2}\left(1 - 5(\bm{\hat{s}}\cdot\bm{n}^{}_{\AHAR})^{2}\right)\right]
\right\} \left(\bm{\hat{s}}\times\bm{n}^{}_{\AHAR}\right)^{}_{i}\,, \label{kerrKi} \nonumber
\ea
\ba
&& K_{ij}^{\AHAR} =  \frac{2 M^{}_{\MBH}}{r^{}_{\AHAR}}\left\{ \delta^{}_{ij} 
+ \frac{M^{}_{\MBH}}{2\,r^{}_{\AHAR}} \left( \delta^{}_{ij} + n^{\AHAR}_{i} n^{\AHAR}_{j} \right) 
\right. \nonumber \\
& + & \frac{1}{2\,r_{\AHAR}^{2}}\left[ 2M^{2}_{\MBH} n^{\AHAR}_{i} n^{\AHAR}_{j} 
+ a^{2}\left(1-3(\bm{\hat{s}}\cdot\bm{n}^{}_{\AHAR})^{2} \right)\delta^{}_{ij} \right] \nonumber \\
& + & \frac{1}{2M^{}_{\MBH}r_{\AHAR}^3} \left[2M^{4}_{\MBH}n^{\AHAR}_{i} n^{\AHAR}_{j}
+ M^{2}_{\MBH}a^{2}\left\{ \left( (\bm{\hat{s}}\cdot\bm{n}^{}_{\AHAR})^{2} -2 \right) \delta^{}_{ij}
\right. \right. \nonumber \\
& + & \left. 3\left(1 - (\bm{\hat{s}}\cdot\bm{n}^{}_{\AHAR})^{2} \right)n^{\AHAR}_{i} n^{\AHAR}_{j}
\right\} + \frac{a^{4}}{4}\left\{ \left(1 - 3(\bm{\hat{s}}\cdot\bm{n}^{}_{\AHAR})^{2} \right) \delta^{}_{ij} 
\right. \nonumber \\
& - & \left. 2\left(2 - 9(\bm{\hat{s}}\cdot\bm{n}^{}_{\AHAR})^{2} \right)n^{\AHAR}_{i} n^{\AHAR}_{j} \right\}
\nonumber \\
& - & \frac{a^{2}\left(8M^{2}_{\MBH}-a^{2}\right)}{4}(\bm{\hat{s}}\cdot\bm{n}^{}_{\AHAR})\left(n^{\AHAR}_{i}\hat{s}^{}_{j}
+n^{\AHAR}_{j}\hat{s}^{}_{i}\right) \nonumber \\
& + & \left. \left. \frac{3a^{2}\left(4M^{2}_{\MBH}-a^{2}\right)}{4}\hat{s}^{}_{i}\hat{s}^{}_{j} 
\right] \right\} \,. \label{kerrKij} 
\ea
The far-field expansions of the $Q^{\mu \nu}$ potentials in the approximate harmonic coordinates,
of Eqs.~\eqref{ACMC-H-6-t}-\eqref{ACMC-H-6-phi}, $Q^{\alpha\beta}_{\AHAR}$, is given by 
\ba
Q^{}_{\AHAR} & = & Q^{00}_{\AHAR} = -  \frac{2 M^{}_{\MBH}}{r^{}_{\AHAR}}\left\{ 1 
+ \frac{M^{}_{\MBH}}{r^{}_{\AHAR}} \right. \nonumber \\
& + & \frac{1}{2\,r_{\AHAR}^{2}} \left[2M^{2}_{\MBH} + a^{2}\left(1 
- 3(\bm{\hat{s}}\cdot\bm{n}^{}_{\AHAR})^{2}\right)\right] \nonumber \\
& + & \left. \frac{M^{}_{\MBH}}{r_{\AHAR}^{3}} \left( M^{2}_{\MBH} - a^{2} \right) \right\} \,, \label{kerrQ} 
\ea
\ba
Q^{i}_{\HAR} & = & Q^{0i}_{\HAR} = -\frac{2 M^{}_{\MBH}}{r_{\AHAR}^{2}} \left\{ 1 - \frac{M^{}_{\MBH}}{r^{}_{\AHAR}} 
\right.  \label{kerrQi}  \\
& + & \left. \frac{1}{2\,r_{\AHAR}^{2}} \left[4M^{2}_{\MBH} + a^{2}\left(1 
- 5(\bm{\hat{s}}\cdot\bm{n}^{}_{\AHAR})^{2}\right)\right]
\right\} \left(\bm{\hat{s}}\times\bm{n}^{}_{\AHAR}\right)^{i}\,, \nonumber
\ea
\ba
&&Q^{ij}_{\AHAR} =  -\frac{2 M^{}_{\MBH}}{r^{}_{\AHAR}}\left\{ \delta^{ij} 
- \frac{M^{}_{\MBH}}{2\,r^{}_{\AHAR}} \left( 3\delta^{ij} - n_{\AHAR}^{i} n_{\AHAR}^{j} \right) \right. \nonumber \\
& - & \frac{1}{2\,r_{\AHAR}^{2}}\left[ 2M^{2}_{\MBH} n_{\AHAR}^{i} n_{\AHAR}^{j} 
- \left( 4M^{2}_{\MBH} + a^{2}\left(1-3(\bm{\hat{s}}\cdot\bm{n}^{}_{\AHAR})^{2} \right)\right)\delta^{ij} \right] \nonumber \\
& + & \frac{1}{2M^{}_{\MBH}r_{\AHAR}^3} \left[3M^{4}_{\MBH}n_{\AHAR}^{i} n_{\AHAR}^{j}
- M^{2}_{\MBH}\left\{   a^{2}\left( 1 + 3(\bm{\hat{s}}\cdot\bm{n}^{}_{\AHAR})^{2}\right) n_{\AHAR}^{i} n_{\AHAR}^{j}
\right. \right. \nonumber \\
& + & \left. \left( 5M^{2}_{\MBH} + a^{2}\left(2 - 9(\bm{\hat{s}}\cdot\bm{n}^{}_{\AHAR})^{2} \right)\right)\delta^{ij}
\right\} \nonumber \\
& + & \frac{a^{4}}{4}\left\{ \left(1 - 3(\bm{\hat{s}}\cdot\bm{n}^{}_{\AHAR})^{2} \right)\delta^{ij}
- 2\left(2-9(\bm{\hat{s}}\cdot\bm{n}^{}_{\AHAR})^{2}\right) n_{\AHAR}^{i} n_{\AHAR}^{j} \right\}
\nonumber \\
& + & \frac{a^{2}\left(8M^{2}_{\MBH}+3a^{2}\right)}{4}(\bm{\hat{s}}\cdot\bm{n}^{}_{\AHAR})\left(n_{\AHAR}^{i}\hat{s}^{j}
+n_{\AHAR}^{j}\hat{s}^{i}\right) \nonumber \\
& - & \left. \left. \frac{a^{2}\left(4M^{2}_{\MBH}+3a^{2}\right)}{4}\hat{s}^{i}\hat{s}^{j} \right] \right\} \,.
\label{kerrQij}
\ea
where we recall that the symbol $\times$ refers to the Euclidean vector product and the dot product to 
the Euclidean scalar product.  Moreover, $n^{i}_{\AHAR} \equiv x_{\AHAR}^{i}/r^{}_{\AHAR}$ and 
indices are raised and lowered with the flat metric. 
We have checked that the above metric satisfies the differential harmonic coordinate condition 
to the order of approximation. That is, we have checked that 
$\partial^{\AHAR}_{\beta} (\sqrt{-\met^{}_{\KERR,\AHAR}}\; \met^{\alpha \beta}_{\KERR,\AHAR}) = 0$,
where here $\met_{\alpha \beta}^{\KERR,\AHAR}$ is the expanded Kerr metric in the
approximate harmonic coordinates as determined by the expansions of the potentials above, 
and $\partial^{\AHAR}_{\alpha}$ denotes partial differentiation with respect to the
approximate harmonic coordinates $(x^{\alpha}_{\AHAR})$.

\section{Relations between different orbital parameterizations}
\label{app-const-of-motion}

Geodesic orbits in Kerr spacetime are fully determined by the three constants of motion 
${\cal I}^{A} = (E,L^{}_{z},C \; \rm{or} \; Q)$.  In this work, we restrict
our attention to bounded motion ($E^{2}<1$, see~\cite{Wilkins:1972rs}).  Then,
it is also very useful to characterize the orbit in terms of the orbital parameters
${\cal O}^{A} = (p,e,\iota \; \rm{or} \; \theta^{}_{\INC})$, which provide more transparent 
geometrical information about the properties of the orbit.    Both sets of
parameters are important, the set ${\cal I}^{A}$ is more adapted to the separation
of the geodesic equations and to the radiation-reaction computations, whereas
the set ${\cal O}^{A}$ is better in terms of orbit characterization. 
Therefore, it is important to know the relation between
these two sets of constants of motion and how to map them.  This is specially
important when radiation-reaction changes these parameters and we need to know
the new values of ${\cal O}^{A}$ once the new values of ${\cal I}^{A}$ have been
computed using Eqs.~\eqref{evol-cons-motion-E}-\eqref{evol-cons-motion-C}.
The relations we present below follow from developments 
in~\cite{Schmidt:2002qk,Drasco:2003ky,Fujita:2009bp}.

Let us first consider the case in which the set $(p,e,\theta^{}_{\INC})$ is known,
i.e.~we parametrize the orbit in terms of these parameters and we need to find
the parameters ${\cal I}^{A}$ for evolving the equations of motion.  
We mainly use $\theta^{}_{\INC}$
instead of $\iota$ for convenience, but we shall also present formulas related to the 
inclination angle $\iota$. The goal is to find the mapping from
$(p,e,\theta^{}_{\INC})$ to $(E,L^{}_{z},C \; \rm{or} \; Q)$.
First of all, given $\theta^{}_{\INC}$, the minimum value that $\theta$ can take,
$\theta^{}_{\MIN}$, is [see Eq.~\eqref{theta-inc}]
\ba
\theta^{}_{\MIN} = \frac{\pi}{2}- \mbox{sign} (L^{}_{z})\, \theta^{}_{\INC}\,,
\label{thetamin}
\ea
where the sign of $L^{}_{z}$ determines whether the orbit is prograde (positive) or
retrograde (negative).  The sign of $L^{}_{z}$ is encoded in $\theta^{}_{\INC}$ as 
follows: If $0<\theta^{}_{\INC}<\pi/2$, then $\mbox{sign}(L^{}_{z}) = 1$; if~ 
$-\pi/2<\theta^{}_{\INC}<0$, then $\mbox{sign}(L^{}_{z}) = -1$.  The particular
case $\theta^{}_{\INC}=0$ is singular in the sense that both signs are possible
for $L^{}_{z}$.

Since $\theta = \theta^{}_{\MIN}$ is a minimum, we have $\dot{\theta}({\theta^{}_{\MIN}}) = 0$, 
which means that the right-hand side of Eq.~\eqref{thetadot-GR} has to vanish at 
$\theta^{}_{\MIN}$, from which we obtain an expression for $C$
\ba
C = z^{}_{-} \left[ \frac{L^{2}_{z}}{1-z^{}_{-}} + a^{2}(1-E^{2}) \right]\,.
\label{Cexpression}
\ea
where $z^{}_{-}$ is given in Eq.~\eqref{z-def}.
Now that we have an expression for $C$ in terms of $E$ and $L^{}_{z}$, let us find
expressions for $(E,L^{}_{z})$ in terms of $p$ and $e$. This can be done from
the analysis of the radial motion, and using the expressions of the apocenter 
and pericenter radii, $r^{}_{\APO}$ and $r^{}_{\PERI}$ [Eq.~\eqref{peri-and-apo}], 
in terms of $(p,e)$.
These values of $r$ are extrema and hence, the right-hand side of
Eq.~\eqref{rdot-GR} vanishes at $r=r^{}_{\APO}$ and at $r=r^{}_{\PERI}$, leading to two equations for the
three unknowns $(E,L^{}_{z},C)$.  The Carter constant, however, is given in terms of $(E,L^{}_{z})$ in
Eq.~\eqref{Cexpression}, which then leads to two equations for two unknowns, $(E,L^{}_{z})$. These
equations take the following polynomial structure
\ba
\alpha^{}_{I} E^{2} + 2\beta^{}_{I} E L^{}_{z} + \gamma^{}_{I} L^{2}_{z} + 
\lambda^{}_{I} = 0 \,,
\label{equationsforEandL}
\ea
where the subindex $I$ stands for apocenter or pericenter and where
the coefficients $\alpha^{}_{I}$, $\beta^{}_{I}$, $\gamma^{}_{I}$, and
$\lambda^{}_{I}$, in the case of noncircular orbits ($r^{}_{\PERI} \neq
r^{}_{\APO}$), are given by
\ba
\alpha^{}_{I} & = & \left(r^{2}_{I}+a^{2}\right)\left(r^{2}_{I}+a^{2}z^{}_{-}\right)
+2M_{\MBH}r^{}_{I}a^{2}\left(1-z^{}_{-}\right)\,, \label{alpha-eq} \\
\beta^{}_{I} & = & - 2M_{\MBH}r^{}_{I}a\,, \label{beta-eq} \\
\gamma^{}_{I} & = & - \frac{1}{1-z^{}_{-}}\left[ r^{2}_{I} + a^{2}z^{}_{-} 
-2M_{\MBH}r^{}_{I}\right]\,, \label{gamma-eq} \\
\lambda^{}_{I} & = & -\left(r^{2}_{I}+a^{2}z^{}_{-}\right)\Delta(r^{}_{I}) \nonumber \\
& = & -\left(r^{2}_{I}+a^{2}z^{}_{-}\right)\left(r^{2}_{I} -2M_{\MBH}r^{}_{I}+a^{2}\right)
\label{lambda-eq} \,.
\ea
In the case of circular orbits these two relations are exactly the same and we
need an extra equation.  This equation comes from the fact that in the circular
case $dr/d\tau$ must always vanish and hence also the radial derivative of 
Eq.~\eqref{rdot-GR} must vanish.  Then, the first equation in the circular case is
given by Eqs.~\eqref{alpha-eq}-\eqref{lambda-eq} with $r^{}_{I} = r^{}_{o} = const.$,
and the coefficients of the second one are given by:
\ba
\alpha^{}_{2} & = & 2 r^{}_{o}\left(r^{2}_{o}+a^{2}\right) - a^{2}\left(r^{}_{o}-M^{}_{\MBH}\right)
\left(1-z^{}_{-}\right) \,, \\
\beta^{}_{2} & = & - a M^{}_{\MBH} \,, \\
\gamma^{}_{2} & = & -\frac{r^{}_{o}-M^{}_{\MBH}}{1-z^{}_{-}}\,, \\
\lambda^{}_{2} & = & - r^{}_{o}\Delta(r^{}_{o}) - \left(r^{}_{o}-M^{}_{\MBH}\right)
\left(r^{2}_{o} + a^{2}z^{}_{-}\right) \,,
\ea

One can then combine the two expressions in Eq.~\eqref{equationsforEandL} to eliminate one
of the two unknowns. For instance, we can eliminate $L^{}_{z}$ and obtain an equation
for $E$, which can be written in the following form:
\ba
& &\left([\alpha,\gamma]^{2}+4[\alpha,\beta][\gamma,\beta]\right) E^{4} \nonumber \\
&+& 2\left([\alpha,\gamma][\lambda,\gamma]+2[\gamma,\beta][\lambda,\beta]\right) E^{2}
+ [\lambda,\gamma]^{2} = 0\,, \label{equationforE}
\ea
where the notation $[\ast,\ast]$ denotes the following antisymmetric product:
\ba
[\Pi,\Omega]\equiv \Pi^{}_{\APO}\Omega^{}_{\PERI} - \Pi^{}_{\PERI}\Omega^{}_{\APO}\,,
\ea
and the subscripts denote at which radii we have to evaluate the quantity 
(e.g., $\Pi^{}_{\APO} = \Pi(r^{}_{\APO})$).  
Eq.~\eqref{equationforE} is bi-quadratic in $E$, which means that there are two solutions 
for $E^{2}$ and from each of these, there are two
values of $E$, one positive and one negative, that are related by time-inversion.
From the two solutions for $E^{2}$, the larger one corresponds to retrograde
orbits, while the smaller one corresponds to prograde ones.  Given $E^{2}$, we can
then find $L^{}_{z}$ through
\ba
L^{2}_{z} = \frac{1}{[\beta,\gamma]}\left([\alpha,\beta]E^{2} + [\lambda,\beta] \right)\,,
\ea
where the positive solution correspond to prograde orbits and the negative one
for retrograde orbits.  Finally, the Carter constant $C$ is given by Eq.~\eqref{Cexpression}
and $Q$ from Eq.~\eqref{c_carter}.
Once we know the constants of motion $(E,L^{}_{z},C)$ we can also find the inclination
angle $\iota$ from Eq.~\eqref{iota-angle}.

The next point is the computation of the extrema of the radial and polar motions. 
For the radial motion there are four extrema [see Eq.~\eqref{new-rhs-rdot}]: 
$r^{}_{\APO} > r^{}_{\PERI} > r^{}_{3} > r^{}_{4}$.   From Eqs.~\eqref{rdot-GR} and~\eqref{new-rhs-rdot} we
know that these extrema must satisfy the following relations:
\ba
&&r^{}_{\APO}+r^{}_{\PERI}+r^{}_{3}+r^{}_{4} = \frac{2M_{\MBH}}{1-E^{2}}\,, 
\label{rextrema1} \\
&&r^{}_{\APO}r^{}_{\PERI}+r^{}_{3}r^{}_{4}+(r^{}_{\APO}+r^{}_{\PERI})(r^{}_{3}+r^{}_{4}) 
\nonumber \\
&&= \frac{a^{2}(1-E^{2})+L^{2}_{z}+C}{1-E^{2}}\,, \label{rextrema2} \\
&&(r^{}_{\APO}+r^{}_{\PERI})r^{}_{3}r^{}_{4}+(r^{}_{3}+r^{}_{4})r^{}_{\APO}r^{}_{\PERI}=
\frac{2M_{\MBH}Q}{1-E^{2}}\,, \label{rextrema3} \qquad \\
&&r^{}_{\APO}r^{}_{\PERI}r^{}_{3}r^{}_{4} = \frac{a^{2}C}{1-E^{2}} \,. \label{rextrema4} 
\ea
From these relations we can find $(r^{}_{3},r^{}_{4})$ using, for instance, Eqs.~\eqref{rextrema1}
and~\eqref{rextrema4}, to get
\ba
r^{}_{3,4} = {\cal A} \pm \sqrt{{\cal A}^{2} - {\cal B}} \,,
\ea
where the plus sign corresponds to $r^{}_{3}$, and ${\cal A}$ and ${\cal B}$
are given by
\ba
{\cal A} & = & \frac{M_{\MBH}}{1-E^{2}} - \frac{r^{}_{\APO}+r^{}_{\PERI}}{2}\,,\\
{\cal B} & = & \frac{a^{2} C}{1-E^{2}}\frac{1}{r^{}_{\APO}r^{}_{\PERI}}\,.
\ea
For the polar motion, and using the variable $z=\cos^{2}\theta$, there are two extrema
[see Eq.~\eqref{new-rhs-thetadot}]: $z^{}_{+}$ and $z^{}_{-}$, with $z^{}_{+} > z^{}_{-}$.  
From Eqs.~\eqref{thetadot-GR} and~\eqref{new-rhs-thetadot}, these extrema satisfy
\ba
z^{}_{-} + z^{}_{+} & = & \frac{a^{2}(1-E^{2})+L^{2}_{z}+C}{a^{2}(1-E^{2})}\,, \\
z^{}_{-} z^{}_{+} & = & \frac{C}{a^{2}(1-E^{2})}\,.
\ea
Given $z^{}_{-}$, from Eq.~\eqref{thetamin}, we can find $z^{}_{+}$ from any of
these two.  

Let us now consider the inverse case in which we know the set $(E,L^{}_{z},C/Q)$ and we want
to find the orbital parameters $(p,e,\theta^{}_{\INC})$ and other important quantities.
We can start from the equations for the extrema for the radial and polar motions.  
In the first case, any extrema $r^{}_{\star}$ will satisfy the following quartic
[from Eqs.~\eqref{rdot-GR} and~\eqref{new-rhs-rdot}]:
\ba 
r^4_{\star} + a^{}_3 r^3_{\star} + a^{}_2 r^2_{\star} + a^{}_1 r^{}_{\star} + a^{}_0 = 0\,,
\ea
where the coefficients are given by:
\ba
a^{}_3 & = & - \frac{2 M_{\MBH}}{1 - E^2}\,,\quad 
a^{}_2  = \frac{a^2 ( 1 - E^2 ) + L^2_{z} + C}{1 - E^2}\,, \\
a^{}_1 & = & - \frac{2 M_{\MBH} Q}{1 - E^2}\,, \quad 
a^{}_0 =  \frac{a^2 C}{1 - E^2}\,.
\ea
We can solve this quartic equation by steps (see, e.g.~\cite{Abramowitz:1970as}).  First, let us consider 
the following cubic equation:                                                           
\ba
 y^3 + b^{}_2 y^2 + b^{}_1 y + b^{}_0 = 0\,,
\ea
where the coefficients are given by 
\ba
b^{}_2 & = & \frac{5}{2} \delta\,, \quad
b^{}_1 = 2 \delta^2 - \epsilon\,, \\
b^{}_0 & = &\frac{1}{2} \delta^3 - \frac{1}{2} \delta \epsilon - \frac{1}{8} \tau^2\,,
\ea
and where $\delta$, $\tau$, and $\epsilon$ are the coefficients of the {\em depressed} 
quartic $u^4 + \delta u^2 + \tau u + \epsilon = 0\,.$  This equation is associated with 
the initial quartic via the change of variable: $r^{}_{\star} = u - a^{}_3/4$.
Then, the relations between the coefficients of the depressed quartic and those  
of the initial quartic are:                                                        
\ba
\delta & = & - \frac{3}{8}a^{2}_3 + a^{}_2\,, \quad
\tau = \frac{1}{8}a^{3}_3 - \frac{1}{2}a^{}_2 a^{}_3 + a^{}_1\,, \\
\epsilon & = & - \frac{3}{256} a^{4}_3 + \frac{1}{16} a^{}_2 a^{2}_3 - \frac{1}{4}a^{}_1 a^{}_3 
+ a^{}_0\,.
\ea
Now, let us consider a real solution to the cubic equation above, namely $y^{}_1$ (a cubic equation with 
real coefficients always has at least one real root).  Then, the four solutions of the initial quartic can 
be written as follows:                
\ba
r^{}_{\star} &=& -\frac{1}{4}a^{}_3 + \frac{1}{2}\left\{ s^{}_1 \sqrt{\delta + 2 y^{}_1} 
\right. \nonumber \\
&+& \left. s^{}_2 \sqrt{ - \left( 3 \delta + 2 y^{}_1 + s^{}_1\frac{2\tau}{\sqrt{\delta + 2 y^{}_1}}\right)} \right\}\,,
\ea
where $s^{}_1$ and $s^{}_2$ are two independent signs, which lead to the four solutions.  
We can then immediately identify the extrema $r^{}_{\APO} > r^{}_{\PERI} > r^{}_{3} > r^{}_{4}$
and from Eq.~\eqref{peri-and-apo} we find $e$ and $p$.

In the case of the polar motion, things are a bit easier.  We need to find the     
roots of the following quadratic equation for $z^{}_{\star} = \cos^2\theta$
[from Eqs.~\eqref{thetadot-GR} and~\eqref{new-rhs-thetadot}]:                  
\ba
z^2_{\star} + c^{}_1 z^{}_{\star} + c^{}_0 = 0\,,
\ea
where the two coefficients are given by 
\ba
c^{}_1 & = & - \frac{a^2 ( 1 - E^2 ) + L^2_{z} + C}{a^2 ( 1 - E^2 )}\,,\\
c^{}_0 & = & \frac{C}{a^2 ( 1 - E^2 )}\,.
\ea
Then, the two extrema of polar motion are given by
\ba
z^{}_{\pm} = - \frac{c^{}_1}{2 c^{}_0} \pm \sqrt{ \left(\frac{c^{}_1}{2 c^{}_0}\right)^2 - c^{}_0 }\,,
\ea
and from Eqs.~\eqref{z-def} and~\eqref{theta-inc} we find the inclination angle $\theta^{}_{\INC}$.  
The inclination angle $\iota$ follows from Eq.~\eqref{iota-angle} as before.

\section{formulas for the Fundamental Frequencies and Periods}
\label{fundamental-frequencies-and-periods}

In order to give formulas for the fundamental frequencies and periods with respect to the
Boyer-Lindquist coordinate time, it is very convenient to start first considering the
frequencies and periods with respect to a new time (see~\cite{Mino:2003yg}): $d/d\lambda \equiv \rho^{2} d/d\tau$, 
which separates the radial and polar dependence
in the sense that it terms of $\lambda$ the equations for $r$ and $\theta$ become:
\be
\left(\frac{d {r}}{d\lambda} \right)^{2} = R(r)\,,\qquad
\left(\frac{d{\theta}}{d \lambda} \right)^{2} = \Theta(\theta)\,,
\ee
where $R(r)$ and $\Theta(\theta)$ denote the right-hand sides of Eqs.~\eqref{rdot-GR} and~\eqref{thetadot-GR}
respectively. The fundamental frequencies of the radial and polar motions associated 
with the $\lambda$ time are $\Upsilon^{}_{r} = 2\pi/\Lambda^{}_{r}$ and $\Upsilon^{}_{\theta} =  
2\pi/\Lambda^{}_{\theta}$, where the periods, denoted by $\Lambda^{}_{r}$ and $\Lambda^{}_{\theta}$, are given by 
\be
\Lambda^{}_{r} = 2 \int^{r^{}_{\APO}}_{r^{}_{\PERI}} \frac{dr}{\sqrt{R}}\,, ~~ 
\Lambda^{}_{\theta} = 2 \int^{\pi-\theta^{}_{\MIN}}_{\theta^{}_{\MIN}} 
\frac{d\theta}{\sqrt{\Theta}} \,.
\ee 
Then, following~\cite{Fujita:2009bp}, the radial and polar frequencies with respect to the
$\lambda$ time can be written as
\be
\Upsilon^{}_{r} = \frac{\pi\sqrt{(1-E^{2})(r^{}_{\APO}-r^{}_{3})(r^{}_{\PERI}-r^{}_{4})}}{2\,{\cal K}(k^{}_{r})}\,,
\ee
\be
\Lambda^{}_{\theta} = \frac{\pi\,a\sqrt{(1-E^{2})z^{}_{+}}}{2\,{\cal K}(k^{}_{\theta})}\,,
\ee
where
\be
k^{}_{r} = \sqrt{\frac{(r^{}_{\APO}-r^{}_{\PERI})(r^{}_{3}-r^{}_{4})}{(r^{}_{\APO}-r^{}_{3})(r^{}_{\PERI}-r^{}_{4})}}\,,~~
k^{}_{\theta} = \sqrt{\frac{z^{}_{-}}{z^{}_{+}}}\,,
\ee
and ${\cal K}$ denotes the complete elliptic integral of the first kind (we adopt
the definitions of~\cite{Abramowitz:1970as} for the elliptic functions).

The right-hand side of the equations for $t$ and $\phi$, Eqs.~\eqref{tdot-GR} and~\eqref{phidot-GR},
have the following structure in terms of the $\lambda$ time
\ba
\frac{dt}{d\lambda} & = &  T^{}_{r}(r) + T^{}_{\theta}(\theta) + a L^{}_{z}\,, \\
\frac{d\phi}{d\lambda} & = & \Phi^{}_{r}(r) + \Phi^{}_{\theta}(\theta) - a E\,,
\ea
and from here we can compute the azimuthal frequency with respect to the $\lambda$ time,r
$\Upsilon^{}_{\phi}$, and also the rate of change of the time $t$ with respect to the time $\lambda$,
$\Upsilon^{}_{t}$, as follows (see~\cite{Fujita:2009bp}, but note that there is a typo in the first
expression in equation 21, our Eq.~\eqref{upsilon-t}, there is a closing square bracket that is 
located in the wrong place, the one that has $E/2$ as coefficient):
\begin{widetext}
\ba
\Upsilon^{}_{\phi} & =  & \frac{2a\,\Upsilon^{}_{r}}{\pi(r^{}_{+}-r^{}_{-})\sqrt{(1-E^{2})
(r^{}_{\APO}-r^{}_{3})(r^{}_{\PERI}-r^{}_{4})}}\left\{ 
\frac{2M^{}_{\bullet}E\,r^{}_{+}-aL^{}_{z}}{r^{}_{3}-r^{}_{+}}
\left[ {\cal K}(k^{}_{r}) - \frac{r^{}_{\PERI}-r^{}_{3}}{r^{}_{\PERI}-r^{}_{+}}
\Pi(-h^{}_{+},k^{}_{r}) \right] \right. \nonumber \\
&- & \left. \;(\,+ \;\longleftrightarrow\; -\,)  \right\} + 
\frac{2 L^{}_{z}}{\pi a\,\sqrt{(1-E^{2})z^{}_{+}}}\Upsilon^{}_{\theta} \,, \label{upsilon-phi}
\ea
\ba
\Upsilon^{}_{t} & = & 4M^{2}_{\bullet}E + \frac{2a\,E\sqrt{z^{}_{+}}}{\pi\sqrt{1-E^{2}}}
\left[{\cal K}(k^{}_{\theta}) - {\cal E}(k^{}_{\theta}) \right]\Upsilon^{}_{\theta} \nonumber \\
& + &\frac{2\Upsilon^{}_{r}}{\pi\sqrt{(1-E^{2})(r^{}_{\APO}-r^{}_{3})(r^{}_{\PERI}-r^{}_{4})}}\left\{
\frac{E}{2}\left[ \left(r^{}_{3}(r^{}_{\APO} + r^{}_{\PERI} + r^{}_{3}) 
- r^{}_{\APO}r^{}_{\PERI}\right){\cal K}(k^{}_{r}) \right. \right. \nonumber \\
&+& \left. (r^{}_{\PERI}-r^{}_{3})(r^{}_{\APO} + r^{}_{\PERI} + r^{}_{3} + r^{}_{4})\Pi(-h^{}_{r},k^{}_{r})
+(r^{}_{\APO}-r^{}_{3})(r^{}_{\PERI}-r^{}_{4}){\cal E}(k^{}_{r}) \right] \nonumber \\
& + &  2M^{}_{\bullet}E\left[r^{}_{3}{\cal K}(k^{}_{r}) + (r^{}_{\PERI}-r^{}_{3})\Pi(-h^{}_{r},k^{}_{r}) \right] 
\nonumber \\
& + & \left. \frac{2M^{}_{\bullet}}{r^{}_{+} - r^{}_{-}}\left[ \frac{(4M^{2}_{\bullet}E-aL^{}_{z})r^{}_{+} 
-2M^{}_{\bullet}a^{2}E}{r^{}_{3}-r^{}_{+}}
\left( {\cal K}(k^{}_{r}) - \frac{r^{}_{\PERI}-r^{}_{3}}{r^{}_{\PERI}-r^{}_{+}}\Pi(-h^{}_{+},k^{}_{r}) \right)
\;- \;(\,+ \;\longleftrightarrow\; -\,)  \right]
\right\} \,,  \label{upsilon-t}
\ea
\end{widetext}
where ${\cal E}$ and $\Pi$ denote the complete elliptic integrals of the second and third kinds
respectively, and 
\be
h^{}_{r} = \frac{r^{}_{\APO} - r^{}_{\PERI}}{r^{}_{\APO}-r^{}_{3}}\,,
\ee
\be
h^{}_{\pm} = \frac{(r^{}_{\APO}-r^{}_{\PERI})(r^{}_{3}-r^{}_{\pm})}{(r^{}_{\APO}-r^{}_{3})(r^{}_{2}-r^{}_{\pm})}\,.
\ee

From these expressions for the fundamental frequencies and periods with respect to the $\lambda$
time we can compute the corresponding ones for the Boyer-Lindquist coordinate time $t$ by using
the following simple expressions:
\be
\Omega^{}_{r} = \frac{\Upsilon^{}_{r}}{\Upsilon^{}_{t}}\,, \quad
\Omega^{}_{\theta} = \frac{\Upsilon^{}_{\theta}}{\Upsilon^{}_{t}}\,, \quad
\Omega^{}_{\phi} = \frac{\Upsilon^{}_{\phi}}{\Upsilon^{}_{t}}\,,  \label{fundamentalfrequenciesBLtime}
\ee
and
\be
T^{}_{r} = \frac{2\pi}{\Omega^{}_{r}}\,, \quad 
T^{}_{\theta} = \frac{2\pi}{\Omega^{}_{\theta}}\,, \quad 
T^{}_{\phi} = \frac{2\pi}{\Omega^{}_{\phi}}\,. \label{fundamentalperiodsBLtime}
\ee
%

\section{formulas for the Evolution of Circular non-Equatorial Orbits}
\label{circular-nonequatorial}
Here we provide the expressions of the components of the matrix that determines
the evolution of the Carter constant and radius of circular nonequatorial orbits
in terms of the evolution of the energy and angular momentum component along
the spin axis [see Eq.~\eqref{Cdot-rdot}].  The expressions of
the $c^{}_{AB}$ ($A,B=1,2$) and $d$ quantities, as functions 
of $(M^{}_{\MBH},a;E,L^{}_{z},C,r^{}_{o})$ are:
\begin{widetext}
\ba
c^{}_{11} & = & 4E(1-E^{2})r^{6}_{o}
            - 12M^{}_{\MBH}Er^{5}_{o}
            + 2E\left[ a^{2}(1-E^{2})+3(C+L^{2}_{z}) \right]r^{4}_{o}
            + 8M^{}_{\MBH}\left[  2a(L^{}_{z}-aE) - E(C+L^{2}_{z}) +a^{2}E^{3}  \right]r^{3}_{o}  \nonumber \\
          & + & 2a\left[ a^{3}E(1-E^{2}) + aE(C+L^{2}_{z}) + 6 M^{2}_{\MBH}(aE-L^{}_{z}) \right]r^{2}_{o}     
            - 4M^{}_{\MBH}a^{2}E\left[ (aE-L^{}_{z})^{2} + C\right]r_{o} \nonumber \\
          & + & 4aM^{2}_{\MBH}(L^{}_{z}-aE)\left[  (aE-L^{}_{z})^{2} + C\right] \,,
\ea
\ba
c^{}_{12} & = & 4(1-E^{2})L^{}_{z}r^{4}_{o}
            - 16M^{}_{\MBH}(1-E^{2})(L^{}_{z}-aE)r^{3}_{o}
            +  2\left[ 6M^{2}_{\MBH}(L^{}_{z}-aE) - L^{}_{z}(C+L^{2}_{z} + a^{2}(1-E^{2})\right]r^{2}_{o} \nonumber \\
          & + & 4M^{}_{\MBH}L^{}_{z}\left[ (aE-L^{}_{z})^{2} + C\right]r^{}_{o}  
            - 4 M^{2}_{\MBH}(L^{}_{z}-aE)\left[  (aE-L^{}_{z})^{2} + C\right] \,,
\ea
\ba
c^{}_{21} & = & -2 \left[ E r^{5}_{o} -3M^{}_{\MBH}E r^{4}_{o} + 2 a^{2}E r^{3}_{o} + aM^{}_{\MBH}(L^{}_{z}-2aE) r^{2}_{o} \right. \nonumber \\
          & + & \left. a^4 E r^{}_{o}  - a^{3}M^{}_{\MBH}(L^{}_{z}-a E) \right] \,,
\ea
\ba
c^{}_{22} & = & -2a\left[ M^{}_{\MBH} E r^{2}_{o} - a L^{}_{z} r^{}_{o} - a M^{}_{\MBH}(aE-L^{}_{z}) \right] \,,
\ea
\ba
d & = & 2(1-E^{2})r^{4}_{o}
    - 4M^{}_{\MBH}(1-E^{2})r^{3}_{o}
    + \left[ 6M^{2}_{\MBH} - C - L^{2}_{z} + 5a^{2}(1-E^{2}) \right]r^{2}_{o} 
    + 2 M^{}_{\MBH}\left[ a^{2}(E^{2}-3) - 2 a L^{}_{z} E + C + L^{2}_{z}\right]r^{}_{o}  \nonumber \\
  & + & a^{4}(1-E^{2}) + a^{2}( C + L^{2}_{z} ) - 2M^{2}_{\MBH}\left[ (aE-L^{}_{z})^{2} + C  \right] \,.
\ea
\end{widetext}
%

%

\end{document}